\documentclass[conference]{IEEEtran}
\usepackage[english]{babel}
\usepackage{xcolor}
\usepackage{pifont}
\usepackage[caption=false]{subfig}%
\usepackage{booktabs} 
\usepackage{multirow}
\usepackage{xspace}
\usepackage{enumitem}
\usepackage{balance}
\usepackage[multiple]{footmisc}
\usepackage{graphicx}
\usepackage{float}
\usepackage{hyperref}
\usepackage{booktabs}
\usepackage{longtable} 
\usepackage{supertabular}
\usepackage{caption}
\usepackage[autostyle]{csquotes}
\usepackage{listings}
\usepackage{tikz}
\usepackage{amsmath,amssymb,amsfonts}
\usepackage{algorithmic}
\usepackage{soul}
\usepackage{amssymb}
\usepackage{pifont}
\usepackage{pgfplots} 
\usepackage[edges]{forest}
\usepackage{adjustbox} 
\usepackage{pifont}
\usepackage{textcomp}
\usepackage{adjustbox}
\usepackage{lscape}
\usepackage{tabularx}
\usepackage{color}
\usepackage[newcommands]{ragged2e}
\usepackage{colortbl}
\usepackage{array}
\usepackage{url}
\usepackage{rotating}
\newcommand{\cmark}{\ding{51}}%
\newcommand{\xmark}{\ding{55}}%

\newcolumntype{C}[1]{>{\centering\arraybackslash\hspace{0pt}}p{#1}}
\def\BibTeX{{\rm B\kern-.05em{\sc i\kern-.025em b}\kern-.08em
   T\kern-.1667em\lower.7ex\hbox{E}\kern-.125emX}}

\title{Intrusion Detection based on Federated Learning: a systematic review}

\author{Jos\'{e} L. Hern\'{a}ndez-Ramos, Georgios Karopoulos, Efstratios Chatzoglou, Vasileios Kouliaridis, \\Enrique Marmol, Aurora Gonzalez-Vidal and Georgios Kambourakis

\thanks{Jos\'{e} L. Hern\'{a}ndez-Ramos, Enrique Marmol and Aurora Gonzalez-Vidal are with the University of Murcia, Spain. E-mail: \{jluis.hernandez, enrique.marmol, aurora.gonzalez2\}@um.es}

\thanks{Georgios Karopoulos, and Vasileios Kouliaridis are with the European Commission, Joint Research Centre, Ispra 21027, Italy. E-mail: \{georgios.karopoulos, vasileios.kouliaridis\}@ec.europa.eu}

\thanks{Efstratios Chatzoglou and Georgios Kambourakis are with the Department of Information \& Communication Systems Engineering, University of the Aegean, Greece. E-mail: \{efchatzoglou, gkamb\}@aegean.gr}}

\newcolumntype{M}[1]{>{\centering\arraybackslash}m{#1}}
\begin{document}

\markboth{IEEE}{Intrusion Detection based on Federated Learning: a systematic review}
\maketitle

\begin{abstract}
The evolution of cybersecurity is undoubtedly associated and intertwined with the development and improvement of artificial intelligence (AI). As a key tool for realizing more cybersecure ecosystems, Intrusion Detection Systems (IDSs) have evolved tremendously in recent years by integrating machine learning (ML) techniques for the detection of increasingly sophisticated cybersecurity attacks hidden in big data. However, these approaches have traditionally been based on centralized learning architectures, in which data from end nodes are shared with data centers for analysis. Recently, the application of federated learning (FL) in this context has attracted great interest to come up with collaborative intrusion detection approaches where data does not need to be shared. Due to the recent rise of this field, this work presents a complete, contemporary taxonomy for FL-enabled IDS approaches that stems from a comprehensive survey of the literature in the time span from 2018 to 2022. Precisely, our discussion includes an analysis of the main ML models, datasets, aggregation functions, as well as implementation libraries, which are employed by the proposed FL-enabled IDS approaches. On top of everything else, we provide a critical view of the current state of the research around this topic, and describe the main challenges and future directions based on the analysis of the literature and our own experience in this area.
\end{abstract}

\begin{IEEEkeywords}
Federated Learning, Intrusion Detection Systems
\end{IEEEkeywords}
\IEEEpeerreviewmaketitle

\section{Introduction}
\label{sec:introduction}

Nowadays, it is clear that Artificial Intelligence (AI) and, in particular, the application of Machine Learning (ML) techniques, will play a key role in the development of the next digital age~\cite{jordan2015machine}. Indeed, several worldwide initiatives, including the EU AI Act~\cite{act2021proposal} and the Artificial Intelligence Strategy in the U.S., demonstrate the need for regulatory frameworks to foster the development of compliant AI systems considering ethical, social, and legal aspects. In the field of cybersecurity, the application of ML is often seen as a double-edged sword in which potential attackers can develop increasingly sophisticated attacks that need to be mitigated through effective and efficient approaches~\cite{taddeo2019trusting}. One of the most prominent advances in this field recently has been the definition of Federated Learning (FL)~\cite{mcmahan2017communication} as a decentralized learning approach where end nodes are not required to share their data to build an ML model. Instead, the training of a certain model is performed throughout different rounds in which such model is built from the end nodes' updates by using a certain aggregation function. This distributed scheme is crucial for the development of ML-enabled systems and applications, while end users' privacy is well preserved. From a practical point of view, the development of FL will be driven by the increasing processing capabilities of end nodes to execute ML algorithms, as well as the need to reduce latency in the decision-making processes by processing data at the network edge~\cite{lu2020low}.

While the deployment of FL approaches has been widely considered in recent years in a plethora of scenarios, including Internet of Things (IoT)~\cite{nguyen2021federated2}, edge computing~\cite{xia2021survey}, healthcare~\cite{nguyen2022federated}, smart cities and transportation systems~\cite{du2020federated}, the application of FL in the field of cybersecurity is intended to foster the development of collaborative approaches to realize more effective and efficient systems for the identification and mitigation of cyberattacks~\cite{alazab2021federated, ghimire2022recent}. That is, FL could play a prominent role in large-scale cybersecurity scenarios where organizations could share information about threats and security attacks without the need to share their actual data. This naturally works in favor of cyber threat intelligence as well. In particular, the well-known Intrusion Detection Systems (IDSs) are essential to identify potential security threats in any IT system. Over the last few decades, IDSs have evolved from signature-based systems to sophisticated ML-based approaches with the capability to detect slow-rate, deceptive, or even unknown attacks~\cite{khraisat2019survey, ring2019survey, thakkar2020review}. In this context, the application of FL is intended to mitigate privacy issues, otherwise present, due to the need of sharing network traffic data, which is normally required to identify cyberattacks.

\textbf{Our contribution:} The development of FL-driven IDSs has recently attracted immense interest from the research community, as widely recognized by recent surveys around this topic~\cite{lavaur2022evolution, agrawal2021federated}. However, as detailed in subsection~\ref{sS:Comparison:with:existing:surveys}, such survey works lack exhaustiveness since they either concentrate on a certain scenario (e.g., IoT) or their analysis neglects key aspects of FL, such as the aggregation function used or the dataset employed. To cover this literature gap, our analysis is founded on a proposed taxonomy around the main aspects of FL-enabled IDSs, including ML model, aggregation function, dataset, as well as implementation libraries, which are key to assess the current development of such systems. We scrutinize each of these aspects and elaborate on their impact on FL-enabled IDS. Moreover, our analysis encompasses key challenges and future paths that need to be considered for the development of FL-enabled IDSs. This analysis hinges in part on our previous experience in this area~\cite{campos2021evaluating} and includes the study of some of the recently proposed solutions to tackle FL challenges, including communication overhead, security and privacy issues, as well as the lack of a standardized set of evaluation metrics to ease the comparison of FL-enabled IDS approaches. While these challenges need to be considered in the coming years, it is rather straightforward that the future of IDS and other security systems can take advantage of FL's decentralized nature to realize sophisticated security approaches without the need to share private information~\cite{ijisec/KoliasKK17}.

\subsection{Comparison with existing surveys}
\label{sS:Comparison:with:existing:surveys}

FL has attracted a significant interest in the few recent years as an alternative to legacy centralized ML approaches. Indeed, several recent studies and surveys have been contributed towards providing a comprehensive overview about the current landscape of FL approaches. Table~\ref{table:surveys} provides a chronologically-ordered list of major past FL surveys, including also its relationship with the work at hand.


One of the first works in this topic~\cite{yang2019federated} provides a categorization of FL according to data partitioning, either horizontal and vertical, and describes the main privacy approaches for FL as well as potential applications. Furthermore,~\cite{zhang2021survey} analyzes the use of FL around five main aspects: data partitioning, privacy mechanism, ML model, communication architecture, and systems heterogeneity. The work in~\cite{abdulrahman2021survey} includes a taxonomy for FL applications, and concentrates on potential application areas, security and privacy, and resource management on the use of FL. Moreover,~\cite{wahab2021federated} comprises a tutorial highlighting the applications and future directions of FL in the domain of communications and networking. Precisely, the authors provide an analysis of the existing literature around five main axes, including statistical challenges, communication efficiency, client selection and scheduling, security and privacy concerns, as well as service pricing. The authors in~\cite{aledhari2020federated} analyze the main technologies and potential FL applications, as well as additional aspects, including implementation libraries, which are also considered in our analysis for FL-enabled IDS. A review of the main FL applications is also provided by~\cite{li2020review}, which describes some of the main challenges associated to FL, such as security and privacy concerns. More recently,~\cite{liu2022distributed} offers a description about the main architectures for FL settings, including a description of existing FL implementations, which are also considered in the present work.

Other recent surveys around FL are focused on specific settings. For instance, the use of FL in mobile edge networks is comprehensively analyzed by~\cite{lim2020federated}, which describes potential applications, as well as challenges and future trends around communication, resource allocation, and security/privacy. Also related to edge scenarios,~\cite{xia2021survey} analyzes the aspects of security/privacy and deployment of FL in such settings. Moreover, the deployment of FL techniques in IoT scenarios has been studied by~\cite{imteaj2021survey}. The authors describe the main challenges associated with the deployment of FL in the case of resource-constrained IoT devices and networks and propose potential solutions.
The work in~\cite{khan2021federated} provides a taxonomy of FL for IoT and analyzes the main research challenges relevant to such a scenario. Additionally,~\cite{nguyen2021federated2} describes comprehensively the potential IoT applications and services enabled through FL such as data sharing, attack detection, localization and mobile crowdsensing in several contexts, including smart healthcare and smart cities. Indeed,~\cite{nguyen2022federated} is focused on the use of FL for smart healthcare, which is analyzed around potential applications, as well as challenges and future directions. Moreover, other recent works are focused on the application of blockchain technology in FL settings. In this direction,~\cite{zhu2023blockchain} describes some of the issues associated to typical FL deployments with a single aggregator, and analyzes different system models for the integration of blockchain. Furthermore, the authors describe emerging applications in this field, as well as future research directions. In addition,~\cite{issa2023blockchain} concentrates on the application of blockchain to address the security and privacy challenges of FL settings, and in particular, in the scope of IoT deployments. 

Besides the above-mentioned works, other recent surveys around FL security and privacy aspects have attracted a significant interest. 
Specifically,~\cite{mothukuri2021survey} offers a comprehensive overview about security and privacy challenges in FL, as well as potential mitigation techniques. In the same direction,~\cite{blanco2021achieving} contributes an empirical analysis around security and privacy concerns in FL. More focused on privacy aspects,~\cite{yin2021comprehensive} proposed a taxonomy to systematically analyze potential privacy leakage risks in FL. Furthermore, a recent survey~\cite{tariq2023trustworthy} provides the fundamental principles of trustworthiness in FL, as well as a taxonomy including security/privacy, fairness and interpretability aspects. The authors also analyze the challenges and future directions of such perspective. With reference to Section~\ref{sec:challenges}, it should be noted that these qualities are also analyzed in our work in the context of FL-enabled IDSs.

\begin{table*}[]
\centering
\scriptsize
\begin{tabular}{C{1cm}C{1cm}C{5.5cm}C{7cm}C{1.5cm}}
\hline
\textbf{Reference} & \textbf{Year} & \textbf{Contributions} & \textbf{Relationship with our work} & \textbf{Citations in Scopus (as of 10/08/2023)} 
\\ \hline
\hline
\cite{yang2019federated} & 2019 & Categorization of FL architectures considering potential applications and related techniques & The proposed definitions and categorization are considered in our taxonomy in subsection~\ref{sec:taxonomy} to classify FL-enabled IDSs & 2293\\ \hline

\cite{lim2020federated} & 2020 & Analysis of FL around communication, resource allocation and security/privacy, including applications and challenges for multi-access edge computing (MEC) settings & The described challenges are considered in the scope of FL-enabled IDS. Cyberattack detection is regarded as one of the main FL applications that is the central topic of our work &  809 \\\hline

\cite{aledhari2020federated} & 2020 & Analysis of technologies and potential FL applications, including associated challenges & FL implementations are also considered in our analysis in Section~\ref{sec:implementations} alongside major challenges in the context of FL-enabled IDS & 232\\\hline
 
\cite{li2020review} & 2020 & Analysis of FL challenges around security and privacy, and description of potential applications & With reference to Section~\ref{sec:challenges}, some of the described challenges are considered in our work, particularly through the prism of FL-enabled IDS & 202\\\hline
 
\cite{abdulrahman2021survey} & 2020 & Analysis of FL challenges according to a proposed classification and definition of a taxonomy for FL applications & Some of these challenges are examined in our work, but in the scope of FL-enabled IDS. Moreover, in Section~\ref{sec:taxonomy}, the present work covers a number of taxonomy aspects, including aggregation functions and datasets & 177 \\\hline
 
\cite{wahab2021federated} & 2021 & Description of main FL aspects, including aggregation functions and analysis of existing literature in respect to five main FL challenges & Aggregation functions are also analyzed in our work and used in the proposed taxonomy in Section~\ref{sec:taxonomy}. The described challenges are considered in Section~\ref{sec:challenges} from an IDS viewpoint & 169\\\hline
 
\cite{zhang2021survey} & 2021 & Analysis of FL literature according to data partitioning, privacy mechanism, ML model, communication architecture, and system heterogeneity & All these aspects have been taken into account for the definition of (a) the proposed taxonomy in Section~\ref{sec:taxonomy} and (b) the challenges associated with FL-enabled IDSs in Section~\ref{sec:challenges} & 216\\\hline
 
\cite{nguyen2021federated2} & 2021 & Analysis of the use cases and applications of FL for IoT, as well as the main challenges and research directions & Our work concentrates on the application of FL for IDS, which is considered as a potential use-case. Additionally, in Section~\ref{sec:taxonomy}, we reckon with key research challenges and their impact on FL-enabled IDS approaches & 190\\\hline
 
\cite{xia2021survey} & 2021 & Analysis on the application of FL to edge computing by considering security/privacy, efficiency, and migration scheduling & In the current work, security/privacy and efficiency aspects are addressed in Section~\ref{sec:challenges} & 67\\\hline

\cite{mothukuri2021survey} & 2021 & A comprehensive overview around security and privacy in FL, and description of the main technologies in this domain & While security and privacy aspects are examined as challenges in our work, in Sections~\ref{sec:aggregationMethods} and~\ref{sec:implementations}, we additionally consider some of the aggregation functions and FL implementations in our analysis for FL-enabled IDS & 371\\\hline

\cite{blanco2021achieving} & 2021 & Analysis of security and privacy aspects affecting FL setting, including the description of potential attacks and countermeasures, as well as an evaluation of such methods & The described challenges and mitigation techniques are considered in our work in Section~\ref{sec:challenges} regarding their relationship with FL-enabled IDS & 30\\\hline

\cite{khan2021federated} & 2021 & Comprehensive analysis on the use of FL for IoT scenarios, including a proposed taxonomy, and open research challenges & While it is focused on IoT settings, some of the described challenges, e.g., those related to security and data heterogeneity, and taxonomy aspects are also considered in our work in the scope of FL-enabled IDSs in Section~\ref{sec:taxonomy} & 167 \\\hline

\cite{yin2021comprehensive} & 2021 & Comprehensive analysis of privacy concerns and mitigation techniques for FL, including a proposed taxonomy around those aspects & With reference to Section~\ref{sec:challenges}, privacy aspects are considered as one of the main challenges around the development of FL-enabled IDSs & 102\\\hline

\cite{imteaj2021survey} & 2022 & Analysis of the main challenges of FL when considering its application on IoT constrained scenarios, as well as potential solutions & Some of the described challenges are also taken into account in our work in Section~\ref{sec:challenges} regarding their impact on the development of FL-enabled IDSs & 117\\\hline

\cite{nguyen2022federated} & 2022 & Analysis of the requirements of using FL in smart healthcare, and overview of the main applications, as well as the main challenges and research directions & With reference to Section~\ref{sec:challenges} the description of several challenges around FL in this context, such as communication issues, or security/privacy aspects are considered in our work with reference to the application of FL for IDSs & 73 \\\hline

\cite{liu2022distributed} & 2022 & Description of the evolution of distributed ML and, in particular, FL architectures and techniques & Our work also provides a description of existing FL implementations (given in Section~\ref{sec:implementations}), and research directions, e.g., around data heterogeneity, that are considered in Section~\ref{sec:challenges} for FL-enabled IDSs & 41\\\hline

\cite{zhu2023blockchain} & 2023 & Analysis of blockchain as enabling technology in FL settings, as well as description of challenges and future directions & The application of blockchain in FL scenarios is addressed as a potential approach to mitigate the issues associated with typical FL settings where the aggregator could become a bottleneck (see Section~\ref{sec:challenges_aggregator_bottleneck}) & 4\\\hline

\cite{issa2023blockchain} & 2023 & Analysis of blockchain technology to address some of the main security and privacy challenges related to FL in IoT scenarios  & Security and privacy challenges in FL (and FL-enabled IDS in particular) are analyzed in Section \ref{sec:challenges}. Furthermore our proposed taxonomy in Section \ref{sec:taxonomy} considers some of the aspects of this work, such as data partitioning and aggregation function & 10\\\hline

\cite{tariq2023trustworthy} & 2023 & Description of the main principles of trustworthy FL, including security/privacy, fairness and interpretability & Our work also considers some of the main aspects related to security and privacy in the context of FL-enabled IDS in Section \ref{sec:challenges} & Not in Scopus\\\hline

\end{tabular}
\caption{Relationship of the present work with existing surveys on FL}
\label{table:surveys}
\end{table*}

With reference to the above analysis and Table~\ref{table:surveys}, despite the great number of surveys about FL, only a few recent works have partially analyzed the current landscape of approaches related to the application of FL specifically for the development of IDS approaches. To make this point clearer, Table~\ref{tab:comparison-surveys} compares the current work with existing surveys on this topic, based on seven distinct criteria, namely, the definition of a taxonomy, the analysis of ML models, the employed datasets, as well as the use of aggregation functions and specific implementation libraries. Based on our analysis, with reference to the same table, a survey addresses a certain aspect if it offers a comprehensive description of the different alternatives, e.g., it analyzes a significant number of IDS datasets and discusses their impact on the development of FL-enabled IDSs. If a certain criterion is not comprehensively analyzed, say, only the most used aggregation functions are described, we consider that aspect partially addressed. If an aspect is not examined, even if it is used to classify existing works, then it is deemed as not covered.

\begin{table*}[h]
\centering
\scriptsize
\begin{tabular}{p{1cm}p{1cm}p{1cm}p{1cm}p{1cm}p{1.5cm}p{1.5cm}p{1cm}p{1.5cm}p{1.5cm}}
\hline
\textbf{Reference} & \textbf{Year} & \textbf{Taxonomy} & \textbf{ML models} & \textbf{Datasets} & \textbf{Implementations} & \textbf{Aggregation functions} & \textbf{Challenges} & \textbf{Systematic} & \textbf{Analyzed works}   \\ \hline
\hline
\cite{agrawal2021federated} & 2021 & \xmark & \xmark & \xmark & \xmark & \xmark & \cmark  & \xmark & 15\\\hline

\cite{ferrag2021federated} & 2021 & \xmark & Partially & Partially & \xmark & \xmark & \cmark  & \xmark  & 14\\\hline

\cite{campos2021evaluating} & 2022 & \xmark & Partially & \xmark & Partially & \xmark &  Partially & \xmark & 12\\\hline

\cite{fedorchenko2022comparative} & 2022 & \xmark & \xmark  & \xmark & \xmark  & Partially & Partially & \xmark & 11\\\hline

\cite{belenguer2022review} & 2022 & \xmark & Partially  & Partially & \xmark  & Partially & Partially & \xmark & 18\\\hline

\cite{lavaur2022evolution} & 2022 & \cmark & Partially  & \xmark & \xmark  & Partially & Yes & \cmark  & 22\\\hline

\cite{arisdakessian2022survey} & 2022 & \cmark & Partially  & \xmark & \xmark  & \xmark & Partially & \cmark  & 4 \\\hline

\cite{venkatasubramanian2023iot} & 2023 & \cmark & Partially  & \xmark & \xmark  & \xmark & Partially & \xmark  & 11 \\\hline
 
Our work & 2023 & \cmark &  \cmark & \cmark & \cmark & \cmark & \cmark &  \cmark & 104 \\\hline

\end{tabular}
\caption{Comparison of this work based on seven distinct criteria with existing surveys on FL-enabled IDS approaches}
\label{tab:comparison-surveys}
\end{table*}

As recapitulated in Table~\ref{tab:comparison-surveys}, the following surveys either address a certain criterion partially or they are marginally in scope, or both. The survey on FL-enabled IDS given in~\cite{agrawal2021federated} analyzes relevant concepts, challenges, and future directions. However, it does not provide a comprehensive review of existing approaches, and it does not define a taxonomy to classify such solutions. Furthermore, it does not refer to existing IDS datasets and their potential suitability to be considered in FL scenarios. Recently, the authors in~\cite{fedorchenko2022comparative} offered a comparative review of several FL-enabled IDS approaches, and discussed the main challenges associated with them. Even though this contribution does consider some of the aspects put forward by the present work, e.g., dataset, implementation library, or ML technique, it does not provide a comprehensive analysis of these features and their impact on FL-enabled IDS. Moreover, the review scope is limited, only referring to 11 works. Additionally, the work in~\cite{lavaur2022evolution} defines a taxonomy to analyze the state of the art of FL-enabled IDS approaches from both a quantitative and qualitative perspective through a systematic literature review (SLR)~\cite{keele2007guidelines}. The authors also highlight some of the main open problems, as well as relevant future directions. Nevertheless, they only analyze 22 works without thoroughly covering several key aspects of FL-powered IDS. 

In addition to the above-mentioned contributions some others concentrated on the use of FL-enabled IDS approaches in the context of IoT. Particularly, the authors in~\cite{belenguer2022review} provided a review based on the different ML techniques used by such works. Furthermore,~\cite{ferrag2021federated} offered an overview about the use of FL to strengthen cybersecurity in IoT, considering different applications, as well as integration with other technologies, such as blockchain. The authors also provided a performance analysis on the application of FL to specific datasets, as well as a description of the main challenges of FL-enabled IDS in IoT. In a similar direction, our previous work~\cite{campos2021evaluating} evaluated the use of FL in IDS for IoT with different data distributions, and provided an analysis of the main challenges and future perspectives in this ecosystem.
Furthermore, \cite{arisdakessian2022survey} analyzed different aspects related to IDS in IoT scenarios, including an attack taxonomy and deployment options in FL settings. Indeed, FL is not the focus of such work as the authors only analyze four works proposing an FL-enabled IDS approach. In addition, \cite{venkatasubramanian2023iot} is focused on the use of FL for malware analysis in IoT, even if several works on IDS are also analyzed according to the employed dataset and ML model. Unlike these studies, the work at hand offers a comprehensive analysis of the main aspects that characterize FL-enabled IDS approaches that are used to compare more than 100 published works on this topic so far. This analysis is then used to provide an overview of the key challenges for the development of FL-enabled IDSs that will serve as a reference for future research in this domain.

\subsection{Literature search and article selection process}
\label{sS:Methodology:of:Survey}

For searching the literature for relevant works, we first consider the set of references exploited in~\cite{campos2021evaluating, ruzafa2021intrusion}. Then, to dig deeper and find the most recent relevant works, we took advantage of the Scopus database. A first coarse search was done using the query ``TITLE-ABS-KEY ((``federated learning'' OR ``federated machine learning'') AND (``intrusion detection'')). After filtering the obtained results, we realized that some references that we had considered in our previous works were not obtained using the previous query. This is because some works use terms related to intrusion detection interchangeably; for instance, several articles use relevant to IDS terms like ``malware'' or ``attack detection''. As a result, we amended the query with the expression: TITLE-ABS-KEY ((``federated learning'' OR ``federated machine learning'') AND (``intrusion detection'' OR ``network anomaly detection'' OR ``attack detection'' OR ``malware detection'')), and removed some references, which were not actually related to intrusion detection. This for instance applies to works proposing anomaly detection schemes and evaluate them by means of non-IDS datasets.

\begin{figure}[h]
\begin{tikzpicture}
\begin{axis}[
ybar,
ymin=0,
bar width=20pt, 
/pgf/number format/.cd,
set thousands separator={},
height=7cm,
width=9cm]
\addplot coordinates {
	(2018,2) 
	(2019,6) 
	(2020,24) 
	(2021,65) 
	(2022,156)
};
\end{axis}
\end{tikzpicture}
\caption{Number of FL-enabled IDS publications through time}
\label{fig:fl:pubs:by:year}
\end{figure}
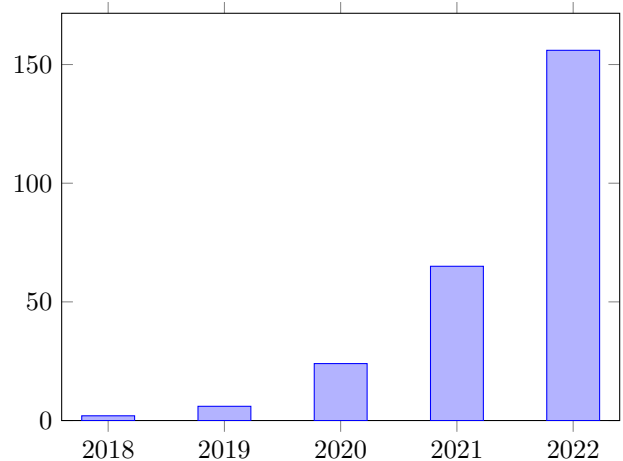

Figure~\ref{fig:fl:pubs:by:year} illustrates the evolution in the number of published works related to the previous search, which demonstrates the increasing interest by the research community in the development of FL-enabled IDS approaches. Moreover, in the case of the identification of relevant surveys, we realized that their majority were not yet in Scopus database because they were very recent. Therefore, in this case, we used the Google Scholar database with the search (``federated learning'' ``intrusion detection'' ``survey'').


\subsection{Structure of the survey}
\label{SS:Structure:of:survey}

The rest of the paper is organized as follows. The next section presents a background on FL and IDS, as well as an overarching taxonomy that is used to classify the existing literature. Next, the subsequent sections focus on the main aspects of the proposed taxonomy. Section~\ref{sec:datasets} elaborates on existing IDS datasets for FL and elaborates on their use in the literature. Furthermore, Section~\ref{sec:models} offers an overview of the ML/DL models that have been considered for FL-enabled IDS approaches. Section~\ref{sec:aggregationMethods} discusses the different aggregation techniques used in FL, and their use in existing literature. Section~{\ref{sec:evaluation}} describes the primary factors influencing the evaluation of FL-based systems, i.e., evaluation metrics and FL implementations. Section~\ref{sec:challenges} discusses key challenges in this field, while the last section concludes. 





\topcaption{List of acronyms}
\label{tab:acronyms}
\tablehead{\hline\textbf{Acronym} & \textbf{Definition} \\\hline \hline}
\begin{supertabular}{cc}

AE & Autoencoder \\ 
ARP & Address Resolution Protocol \\ 
AUC & Area Under the Curve \\ 
BNN & Binarized Neural Network \\ 
C\&C & Command-and-Control \\ 
CAN & Controller Area Network \\ 
CKS & Cohen Kappa Score \\
CM & Coordinate Median \\
CMFL & Communication-Mitigated Federated Learning \\
CNN & Convolutional Neural Network \\ 
CoAP & Constrained Application Protocol \\ 
CSV & comma-separated values \\ 
DBN & Deep Belief Network \\ 
DDoS & Distributed Denial of Service \\ 
DDQN & Double Deep Q-Network \\ 
DL & Deep Learning \\ 
DLC & Data Length Code \\
DNN & Deep Neural Network \\
DNP & Distributed Network Protocol \\ 
DNS & Domain Name System \\ 
DoS & Denial of Service \\ 
DP & Differential Privacy \\ 
DQN & Deep Q-Network \\ 
DT & Decision Tree \\ 
EAP & Extensible Authentication Protocol \\ 
ENN & Edited Nearest Neighbors \\ 
FAR & False Acceptance Rate \\ 
FD & Federated Distillation \\ 
FL & Federated Learning \\ 
FNN & Feedforward Neural Network  \\ 
FN & False Negative \\
FNR & False Negative Rate \\
FOR & False Omission Rate \\ 
FP & False Positive \\
FPR & False Positive Rate \\ 
GAN & Generative Adversarial Network \\ 
GBDT & Gradient Boosting Decision Tree \\ 
GBRM & Gaussian Binary Restricted Boltzmann Machine \\ 
GMM & Gaussian Mixture Model \\ 
GPS & Global Positioning System\\
GRU & Gated recurrent unit \\ 
HE & Homomorphic Encryption \\ 
IDS & Intrusion Detection System \\ 
IIoT & Industrial Internet of Things \\
IoT & Internet of Things \\ 
IRL & Inverse Reinforcement Learning \\ 
KNN & K-Nearest Neighbors \\ 
LAN & Local Area Network \\ 
LDAP & Lightweight Directory Access Protocol \\ 
LLM & Large Language Model \\ 
LR & Logistic Regression \\
LSTM & Long short-term memory \\ 
MAE & Mean Absolute Error \\
MAPE & Mean Absolute Percentage Error \\
MCC & Matthews correlation coefficient \\ 
MEC & Mobile Edge Computing \\ 
MSSQL & Microsoft SQL Server \\ 
ML & Machine Learning \\ 
MLP & Multilayer perceptron \\ 
MQTT & Message Queuing Telemetry Transport \\ 
MSE & Mean Square Error \\ 
MTL & Multi-Task Learning \\ 
MTNN & Multi-Task Neural Network \\ 
NB & Naive Bayesian\\
NetBIOS & NETwork Basic Input Output System \\ 
NLP & Natural Language Processing \\ 
NN & Neural Networks \\
NTP & Network Time Protocol \\ 
OBD & On-Board Diagnostics \\
OS-ELM & Online Sequential Extreme Learning Machine \\
PFNM & Probabilistic Federated Neural Matching \\ 
RaNN & Random Neural Network \\ 
RL & Reinforcement Learning \\ 
RF & Random Forest \\ 
RMSE & Root Mean Squared Error \\
RNN & Recurrent Neural Network \\ 
ROC & Receiver operating characteristic \\ 
RPM & revolutions per minute \\
RTSP & Real Time Streaming Protocol \\ 
SACN & Subgraph Aggregated Capsule Network \\ 
SAFA & Semi-Asynchronous Federated Averaging \\
SDN & Software-Defined Networking \\ 
SGD & Stochastic Gradient Descent \\ 
SMC & Secure Multi-party Computation \\ 
SMOTE & Synthetic Minority Over-sampling Technique \\ 
SMPC & Secure Multi-Party Computation \\
SNMP & Simple Network Management Protocol \\ 
SOC & Security Operation Centers \\ 
SOM & Self-organizing map \\ 
SSDP & Simple Service Discovery Protocol \\ 
SVM & Support-vector machine \\ 
TEE & Trusted Execution Environment \\ 
TFF & TensorFlow Federated \\ 
TFTP & Trivial File Transfer Protocol \\ 
TL & Transfer Learning \\ 
TN & True Negative \\
TNR & True Negative Rate \\
TP & True Positive \\
TPR & True Positive Rate \\ 
UAV & Unmanned Aerial Vehicle \\ 
UDP & User Datagram Protocol \\ 
VAE & Variational Autoencoder \\ 
VPN & Virtual Private Network \\ 
WCSS & Within Cluster Sum of Squares \\ 
WSN & Wireless Sensor Network \\ 
XSS & Cross-site scripting \\ \hline
\end{supertabular}
\section{Background}
\label{S:Background}

For aiding the reader in assessing the contents of the subsequent sections, the current section provides an overview of the basic concepts around IDS approaches and FL. Furthermore, it describes the taxonomy that is used in our work to classify the current literature around FL-enabled IDS.

\subsection{Intrusion Detection Systems}
\label{SS:IDS}

According to NIST SP 800-94, intrusion detection refers to \textit{``the process of monitoring the events occurring in a computer system or network and analyzing them for signs of possible incidents''}. Therefore, an  IDS is usually considered as a \textit{``software that automates the intrusion detection process''}~\cite{nist80094}. An IDS represents a widely deployed security mechanism to identify potential malicious entities and actions targeting a certain IT or operational technology (OT) system. Although the initial development of IDSs was focused on \textit{signature-based} approaches in which network traffic was compared with previously stored network patterns, \textit{anomaly-based} IDSs quickly attracted significant interest, especially with the application of ML techniques~\cite{da2019internet, khraisat2019survey, karop22}. That is, an anomaly-based IDS builds a normal behavior model of a certain system so that any detected deviation from that behavior is likely to be considered an intrusion. Compared to signature-based IDSs, one of the main advantages of anomaly-based approaches is that they can detect previously unseen attacks. Additionally, an IDS can also be classified depending on the source of information for intrusion detection, for example, if the data originates from end nodes or network traffic. However, in practice, it is becoming commonplace to use hybrid approaches where information from both the network and end nodes is used to identify possible intrusions related to network and application layers.

The application of ML in the development of anomaly-based IDS has represented a hot topic in cybersecurity research by using both supervised and unsupervised approaches~\cite{liu2019machine}. In the former case, the identification of cybersecurity attacks relies on labeled data and a set of features that characterize the information (usually network traffic) for the training process. That is, each network flow or other type of information is already marked for the training as an attack or benign traffic, depending on whether it corresponds to an intrusion or not. In the scope of IDS, several different supervised techniques have been considered, including decision trees~\cite{mahbooba2021explainable, ahmim2019novel, panigrahi2021consolidated}, SVM~\cite{mohammadi2021comprehensive}, as well as different types of neural networks~\cite{drewek2021survey}. However, it is not always feasible find or create such labelled information in real-world scenarios. Therefore, the use of unsupervised approaches has been proposed for identifying potential cybersecurity attacks using unlabeled data~\cite{nisioti2018intrusion}. In this direction, typical approaches based on clustering~\cite{borkar2019novel} are employed to group samples from the same type (attack or benign traffic) in a certain cluster. Additionally, there are hybrid or semi-supervised approaches~\cite{li2020enhancing, gao2018novel}, in which supervised and unsupervised learning techniques are used alongside. That is, as described by~\cite{nisioti2018intrusion}, hybrid approaches can take advantage of the performance of supervised approaches for known attacks, as well as the ability of unsupervised approaches to detect new attacks.

The performance of the ML technique is key in the operation of an ML-enabled IDS; a poor performance could have severe consequences on the system where the IDS is deployed. While a \textit{false negative} would represent an attack instance that is not detected by ML-enabled IDS, a \textit{false positive} could lead the system to deploy countermeasures against a false attack, which would cause unnecessary use of resources that could be exploited by opponents to trigger a real attack (e.g., a  DoS attack). Furthermore, as widely recognized in the literature~\cite{kim2020collaborative}, the time required to detect a certain attack by an IDS is crucial to reducing the eventual damage caused to the system. 

The development of ML-enabled IDSs has experienced a strong interest in recent years through centralized environments where the information is shared by end nodes to build the corresponding model. As previously mentioned, this approach raises privacy concerns that need to be addressed through decentralized settings. The next subsection provides an overview about FL, which is intended to mitigate these issues for the next generation of IDSs.

\subsection{Federated Learning}
\label{SS:FL}

\begin{figure*} 
	\centering
		\includegraphics[width=1\textwidth]{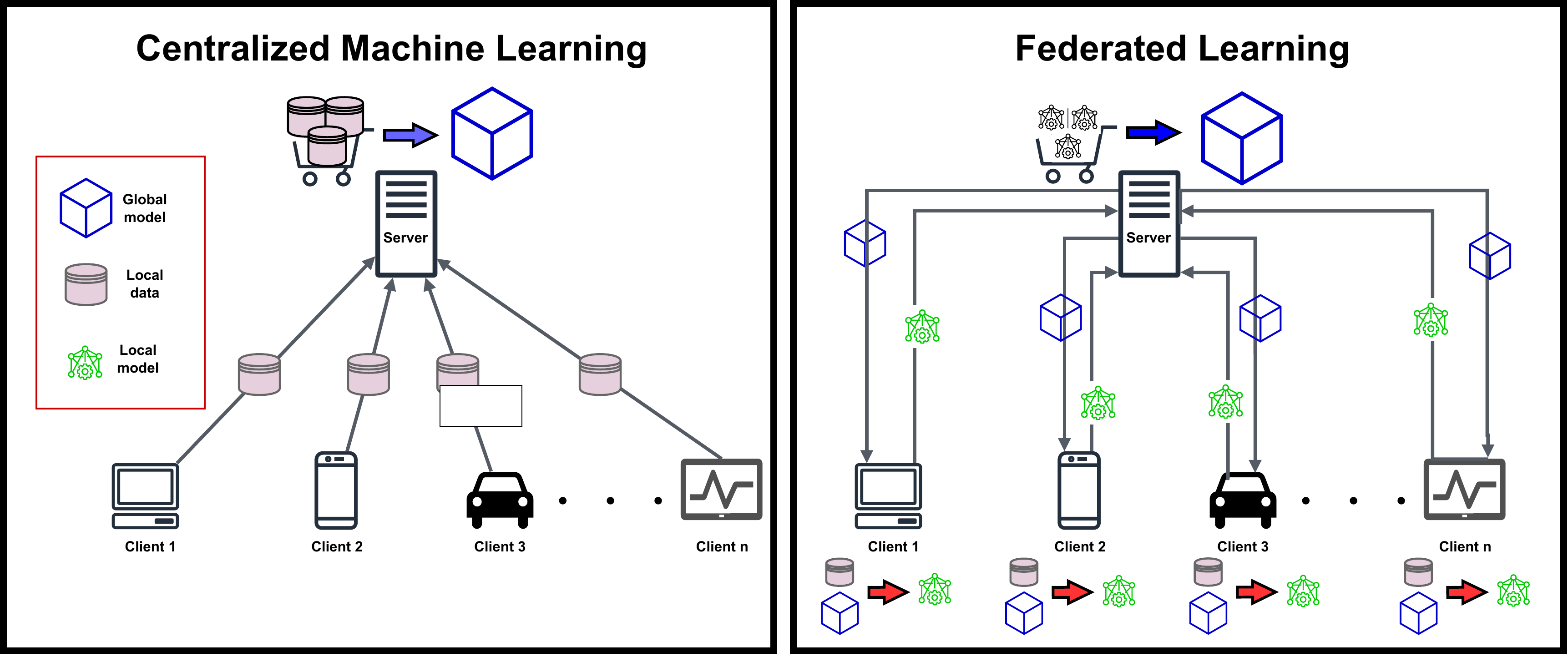}
	\caption{A bird's-eye view of centralized ML vs. FL approach}
	\label{fig:federated}
\end{figure*}

Federated Learning (FL) was coined in 2016~\cite{mcmahan2017communication} as a distributed and collaborative approach to ML, in which a set of entities is trained over their local data to build a global model. Unlike traditional ML approaches, end nodes do not need to share their data with typical data centers to be further processed. Figure~\ref{fig:federated} depicts an overview of the centralized ML and FL approaches. Generally, the FL training process is divided into a set of \textit{rounds} where each party trains a global model by using its local data until a certain number of rounds or the model converges. Each party is called an \textit{FL client} and can be represented by different entities depending on the scenario, e.g., IoT device, smartphone, or vehicle. Furthermore, this local training can be based on a wide variety of ML techniques, including supervised and unsupervised approaches. The local processing is normally called \textit{epoch} and can be repeated several times in each training round.

In a typical FL scenario, the local updates computed by the different FL clients are then shared with a central entity called \textit{aggregator}, which is responsible for generating a new model based on the received updates using an \textit{aggregation function}. While the commonest function is based on averaging (FedAvg~\cite{mcmahan2017communication}), as discussed in Section~\ref{sec:aggregationMethods}, other more sophisticated functions have also been proposed in recent years to cope with non-independent and identically distributed (non-IID) data distributions, which are common in FL settings (see Section~\ref{sec:challenges}). The new model generated by the aggregator is then again shared with FL clients to start a new round of training. It should be noted that, in each round, the aggregator can select a subset of the clients to participate, considering different aspects, such as connection/device status or impact of the updates on the model's convergence~\cite{nishio2019client, cho2020client}.

As mentioned earlier, the role of FL in the development of the next generation of IDS is decisive to realizing approaches detecting cybersecurity attacks without the need to share private data. The application of FL in the evolution of IDS goes beyond small-scale scenarios where specific attacks need to be detected; it can foster a collaborative sharing approach of cybersecurity information among companies and institutions from different countries without the need to share sensitive information~\cite{alazab2021federated, sarhan2021cyber}. Indeed, in an increasingly hyperconnected society~\cite{hernandez2019toward}, the sharing of cybersecurity information, e.g., related to Information and Communications Technology (ICT) products' vulnerabilities or attacks on critical infrastructures, is essential to foster the cooperation for the detection of cyberattacks~\cite{hernandez2021challenges, neisse2020interledger}. Nevertheless, several stakeholders are still reluctant to share cybersecurity information because it may contain sensitive data that could be exploited by potential attackers. Therefore, the use of FL can play a key role in addressing these challenges. Furthermore, the integration of FL in the development of IDSs is expected to offer benefits in terms of reduced latency in the cyberattack detection process since the data does not need to be shared with traditional cloud data centers to be processed. Actually, the aggregation process could be carried out by intermediate processing nodes at the edge of the network for speeding up the decision-making process. Besides the expected benefits, the development of FL still needs to address several challenges in the coming years, including communication requirements and security aspects, which are detailed in Section~\ref{sec:challenges}.
\subsection{A taxonomy for FL-enabled IDS approaches}
\label{sec:taxonomy}

\begin{figure*}
   \centering
\begin{forest}
for tree={
        draw,
        font=\small\linespread{0.84}\selectfont,
        calign=edge midpoint,
        align=center,
if level =1{draw,
            minimum height=5.6ex,
            edge path={\noexpand\path[\forestoption{edge}]
            (!u.south) -- ++ (0,-3mm) -| (.child anchor);}
            }{},
if level>=1{grow'=0,
           folder,
           folder indent=4mm,
           l sep=7mm,
           s sep=1mm}{},
if level =2{draw, align=left}{},
if level>=3{draw=none,
           text width=11em,
           align=left}{},
              }
[FL-based IDS
   [Network architecture
     [Centralized]
     [Decentralized]
   ]
  [Data distribution
      [Data availability
          [Cross-silo]
          [Cross-device]
      ]
      [Data partitioning
          [Horizontal FL]
          [Vertical FL]
          [Transfer FL]
      ]
    ]
     [Dataset \\ (Section \ref{sec:datasets})]
    [ML model \\ (Section \ref{sec:models})
      [Supervised \\ (Section \ref{sec:models_supervised}]
      [Unsupervised\\ (Section \ref{sec:models_unsupervised}]
      [Semi-supervised\\ (Section \ref{sec:models_semisupervised}]
      [Reinforcement\\ (Section \ref{sec:models_reinforcement}]
     ]
    [Aggregation function \\ (Section \ref{sec:aggregationMethods})][Evaluation \\ (Section \ref{sec:evaluation})
       [Metrics\\ (Section \ref{sec:metrics}]
       [FL implementation\\ (Section \ref{sec:implementations}]
          ]
     [Challenges (Section \ref{sec:challenges})]
    ]
\end{forest}
  \caption{Taxonomy of FL-based IDS}
 \label{fig:taxonomy}
\end{figure*}
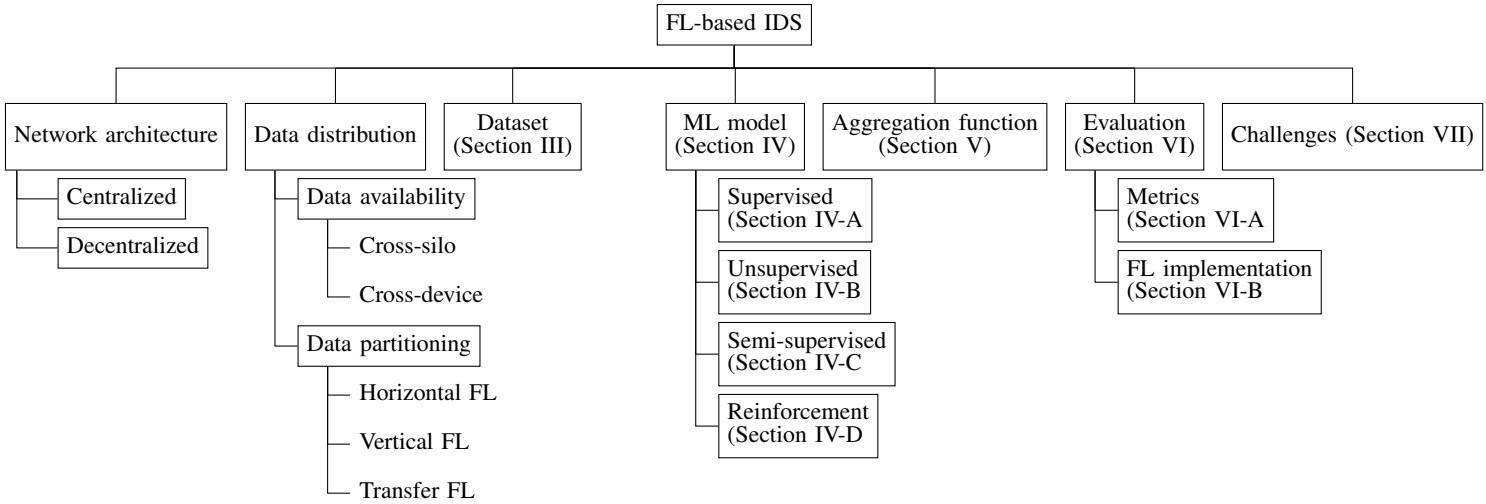

With regard to general classifications on IDS systems~\cite{nisioti2018intrusion,karop22}, an FL-based IDS is an anomaly-based system, more particularly belonging to the ML subclass. Therefore, any FL-enabled IDS taxonomy can be seen as a sub-tree under the ML/anomaly-based leaf of the general IDS taxonomy. Under this assumption, Figure~\ref{fig:taxonomy} presents a taxonomy of FL-based IDSs, taking also into account relevant taxonomies found in related work. Also with reference to table~\ref{tab:comparison-surveys}, it is to be noted that only one recent work proposes a taxonomy for FL-based IDS systems~\cite{lavaur2022evolution}, whereas other relevant surveys provide taxonomies for FL systems in general~\cite{belenguer2022review,li2021survey,imteaj2021survey,JATAIN2021,mothukuri2021survey,nguyen2021federated2}. 
We argue here that an FL-based IDS is a special case of an IDS, which is obviously clearly impacted by the aspects around FL; thus, the proposed taxonomy shown in Figure~\ref{fig:taxonomy} is highly influenced by the aforementioned general FL taxonomies rather than trying to come up with a completely different classification. Moreover, we elaborated on the existing taxonomies around IDS to incorporate any IDS-specific characteristics. As shown in Figure~\ref{fig:taxonomy}, the proposed taxonomy builds around seven criteria detailed below.

The \textit{network architecture} criterion shows where the aggregation takes place. Specifically, an IDS can be categorized as centralized, when a central server acts as the aggregator of the model parameters, or decentralized, when the model parameters are exchanged point-to-point among a subset of the system nodes. The next classification criterion concerns \textit{data availability}, which is affected by the number and size of the client nodes. In the cross-silo category we find small client numbers that own a large amount of data and typically are data centers belonging to large corporations or organizations. In the cross-device case, there is a large number of low-end devices, such as smartphones or IoT devices, with a rather limited data volume and processing power. \textit{Data partitioning} concerns the way data are split among participants on the sample and feature space. In horizontal FL, the clients have the same feature space but different sample space, whereas in vertical FL the sample space is the same and the feature space differs. Federated transfer learning (FTL) can be considered a hybrid category between the previous two, where both the sample and the feature spaces are different. 
The \textit{ML model} aspect considers the type of ML algorithm used, dividing them into four categories: supervised, unsupervised, semi-supervised, and reinforcement learning (RL). An IDS can further be classified depending on the employed ML algorithm, using existing taxonomies, such as those given in~\cite{sultana2019survey,liu2019machine,gamage2020deep}. The \textit{aggregation method} aspect classifies FL-based IDS systems according to the algorithm used for aggregating the received model updates. The next criterion, namely, \textit{implementation framework}, concerns the framework used for implementing  the FL solution. The \textit{dataset} aspect is not actually a characteristic of the IDS, but it demonstrates which of the available datasets has been used for testing a proposed FL-based IDS and conducting the relevant experiments.


For each taxonomy's criterion, we provide an analysis in the following sections, considering the existing literature on FL-enabled IDS approaches. Based on the proposed taxonomy in Figure~\ref{fig:taxonomy}, appendix~\ref{sec:appendix} classifies the FL-enabled IDS approaches analyzed in this work. The next sections dig into the main aspects of the proposed taxonomy, including datasets, ML models, aggregation functions, as well as evaluation considerations. 


\forestset{
  direction switch/.style={
   for tree={edge+=thick},
    where level>=1{folder, grow'=0}{for children=forked edge},
   where level=3{}{draw}
  },
}
\section{Datasets} \label{sec:datasets}
The development of FL-enabled IDS approaches is closely related to the use of appropriate datasets for evaluation purposes. This section offers a general description and analysis of 36 pertinent network datasets that have been proposed in the literature during the time span between 2017 and 2022. For the analysis of the different corpora, we consider diverse aspects that are typically employed to compare IDS datasets~\cite{ring2019survey, thakkar2020review}. Precisely, with reference to Table~\ref{table:comparison1}, the datasets are compared based on several aspects, including the number of features and samples, the type of attacks, as well as if the corpus was obtained through a realistic testbed. Furthermore, we consider if the dataset is directly applicable to simulate an FL scenario in which each client makes use of its own data for training. Indeed, based on previous analysis~\cite{campos2021evaluating, rey2022federated}, we notice that a significant amount of research proposals around FL-enabled IDS propose artificial divisions of existing datasets, for example without considering network traffic's IP address. The basic characteristics of each one of the considered datasets are sketched below, while details per corpus are provided in Table~\ref{table:comparison1}. Furthermore, we show which existing FL-enabled IDS works exploited each dataset. As will be detailed in Section~\ref{sec:datasets_analysis}, it should be noted that a significant portion of existing FL-enabled IDS approaches were proposed by using obsolete datasets previous to 2017.


\subsection{Internet datasets}
\label{SS:Internet datasets}

The \textbf{CIC-IDS2017} dataset~\cite{sharafaldin2018toward} is one of the most utilized datasets for IDS purposes. It was created in an emulated networking environment and comprises five days of network traffic. This labelled dataset covers a broad range of attacks, including brute force, DoS, network infiltration, and botnets. It contains 80 features associated with the network traffic that are extracted by using the CICFlowMeter tool~\cite{lashkari2017cicflowmeter}, and provides information such as timestamp, source/destination IP address and port. Furthermore, it can be downloaded in either pcap or CSV format.

The \textbf{CSE-CIC-IDS2018} dataset~\cite{sharafaldin2018toward} is an extension of the CIC-IDS2017 one, that is, it contains the same methodology of exporting and labelling flow-based features, but with different attacks, including brute-force, TLS Heartbleed, botnet, DoS, distributed DoS (DDoS), Web attacks, and network infiltration. Additionally, the network topology is bigger, i.e., the attacker controls 50 machines and the targeted organization comprises five departments, where each one contains 420 client machines and 30 servers.

The \textbf{CIDDS-001} dataset~\cite{ring2017flow} capitalizes on OpenStack as a tool to generate labelled and realistic benchmark datasets for IDS. This dataset is flow-based, anonymized, labelled, and the captured network traffic stems from a realistic testbed. It contains different assaults, including DoS, brute force, and port scanning. The \textbf{CIDDS-002} dataset~\cite{ring2017creation} extends CIDDS-001~\cite{ring2017flow} for specific port scanning attacks, including SYN, ACK, UDP, FIN and Ping cases. Like its predecessor, it is provided in a flow-based, anonymized and labelled format based on a similar deployment. 

The \textbf{CIC DoS}~\cite{jazi2017detecting} dataset contains eight different application-layer DoS attacks launched against 10 web servers. Particularly, the focus is on DoS slow-rate attacks, like the well-known Slowloris. The dataset also contains attack-free traffic obtained from the ISCXIDS2012 dataset~\cite{shiravi2012toward}, and is offered in a packet-based format with a size of 4.6 GB.

The \textbf{CIC DDoS}~\cite{sharafaldin2019developing} dataset considers DDoS attacks by using different well-known protocols, such as NTP, DNS, LDAP or SNMP. Such attacks were triggered during two days over different machines with diverse Windows OS versions, and more than 80 features were obtained through the CICFlowMeter tool. The dataset is available in both pcap and CSV format containing labeled flows.

The \textbf{PUF} dataset~\cite{sharma2018new} captures flow-based traffic. It focuses on DNS-based anomalies by labelling different flows, including DNS-over-TCP, DNS zone transfer, and DNS2TCP (a tool for relaying TCP connections over DNS).

The \textbf{TRAbID} dataset~\cite{viegas2017toward} concentrates on legacy assaults, including SYN flood, UDP flood, and ICMP flood. Ubuntu 16.04 was used for client machines and Ubuntu 14.04 for vulnerable servers. The dataset was also evaluated by the authors in terms of anomaly detection.

The \textbf{Unified host and network} dataset~\cite{turcotte2019unified} comprises MS Windows-based event logs, which can be analysed through ML techniques. Different host-based events were captured from the log files, including the creation and use of Kerberos authentication tickets, user log in or log out events, and if a workstation was locked or not.

The \textbf{AWID3} dataset~\cite{chatzoglou2021empirical} contains attacks exercised over an enterprise Wi-Fi network. It extends the well-known AWID2 dataset~\cite{kolias2015intrusion} by considering 21 different assaults, which are grouped into 801.11-specific and higher layer ones mounted against the local network or external targets. It was created from a physical lab that realistically emulates a typical enterprise network infrastructure, and it is offered in both CSV and pcap formats.

The \textbf{H23Q} dataset~\cite{H23Q} comprises HTTP/2, HTTP/3 and QUIC protocol traffic involving six diverse HTTP/3-enabled servers residing on Azure cloud infrastructure. Ten different attacks were recorded, including DDoS and Slowloris-alike ones. The testbed also included 12 legitimate clients who were accessing these servers from different public networks. Packet capture took place on each server, therefore each of them can be considered as a client under an FL setting. For more information about the QUIC assaults, the interested reader can refer to~\cite{QUIC:2022}.

\begin{table*}[]
\tiny
\centering
\begin{tabular}{C{1.5cm}C{0.5cm}C{0.6cm}C{1cm}C{1cm}C{5.5cm}C{0.8cm}C{0.8cm}C{0.8cm}C{2cm}}
\hline
\textbf{Dataset}& \textbf{Year}  & \textbf{\# features} & \textbf{\# samples (packets/flows)} & \textbf{Attack/benign traffic ratio} & \textbf{Attacks} & \textbf{Data labelled?} & \textbf{Realistic testbed?} & \textbf{Division} & \textbf{FL-enabled IDS works using the dataset}\\
 \hline  \hline

CIC-IDS2017 \cite{sharafaldin2018toward}& 2017 & 80 & $\approx$28M f  & 0.17:1 &\vspace{-1.3em}\flushleft{Brute force, DoS, DDoS, Heartbleed, XSS, SQL Injection, Data exfiltration, Botnet, Scan} & \checkmark & \checkmark & IP address & \cite{preuveneers2018chained, zhao2019multi, chen2020intrusion, qin2020line, fan2020iotdefender, qin2021fnel, kwon2022anomaly, otoum2021federated, yadav2021unsupervised, ayed2021federated, hao2021secure, wettlaufer2021property, mcosker2021architecture, wei2021federated, otoum2022feasibility, zhang2021flddos, otoum2021federatedreinforcement, tang2022federated, markovic2022random, chen2022privacy}\\
 \hline
 
TRAbID \cite{viegas2017toward}& 2017 & 50 & $\approx$
207M p & 0.15:1 & \vspace{-1.3em}\flushleft{DoS (SYN flood, UDP flood, ICMP flood, TCP keepalive, SMTP flood, HTTP flood), Port scan, Vulnerability scan} & \checkmark & \checkmark & N/A & -\\
\hline

Unified host and network \cite{turcotte2019unified} & 2017 & 21 & N/A & N/A &  \vspace{-1.3em}\flushleft{Privilege escalation}  & \xmark & \checkmark & N/A & -\\
\hline

CIDDS-001 \cite{ring2017flow}& 2017 & 14 & $\approx$32M f & 0.12:1 &\vspace{-1.3em}\flushleft{DoS, Brute force, Port scan}  & \checkmark & \checkmark & IP address & -\\
\hline

CIDDS-002 \cite{ring2017creation}& 2017 & 14 & $\approx$16M f& 0.036:1 & \vspace{-1.3em}\flushleft{Port scan (SYN, ACK, UDP, FIN, Ping)} &  \checkmark & \checkmark & IP address & -\\
\hline

CSE-CIC-IDS2018 \cite{sharafaldin2018toward}& 2018 & 80 & $\approx$4,5M f & 0.58:1 &\vspace{-1.3em}\flushleft{DoS, Heartbleed, DoS Slowloris, DDoS, Bruteforce (SSH, FTP), Scan, Injection} & \checkmark & \checkmark & IP address & \cite{popoola2021federated, shi2021data, chen2022privacy, friha2022felids, zhang2022secfednids}\\
\hline

PUF \cite{sharma2018new}& 2018 & 12 & $\approx$0.3M f & 0.14:1 & \vspace{-1.3em}\flushleft{DNS attacks} &  \checkmark & \checkmark & N/A & -\\
\hline

CIC-DoS \cite{jazi2017detecting}& 2017 & 2 & $\approx$4,6GB p & N/A &\vspace{-1.3em}\flushleft{High-volume HTTP attacks: DoS improved GET (Goldeneye), DDoS GET(ddossim), DoS GET (hulk). Low-volume HTTP attacks: slow-send body (Slowhttptest), slow send body (RUDY]), slow-send headers (Slowhttptest), slow send headers (Slowloris), slow-read (Slowhttptest)} & \checkmark & \checkmark & N/A & -\\
\hline

CIC-DDoS2019 \cite{sharafaldin2019developing}& 2019 & 80 & $\approx$50M f & 879.4:1 &\vspace{-1.3em}\flushleft{DDoS attacks over the following protocols: NTP (DDoS\_NTP), DNS (DDoS\_DNS), LDAP (DDoS\_LDAP), MSSQL (DDoS\_MSSQL), NetBIOS (DDoS\_NetBIOS), SNMP (DDoS\_SNMP), SSDP (DDoS\_SSDP), UDP (DDoS\_UDP), UDP-Lag, WebDDoS, SYN and TFTP} & \checkmark & \checkmark & IP address & \cite{qin2021fnel, zhang2021flddos, li2021fids, yuan2021towards, duy2021federated, dong2021towards, lv2022ddos}\\  \hline

AWID3~\cite{chatzoglou2021empirical}& 2021 & 254 & $\approx$37M p & 0.18:1 & \vspace{-1.3em}\flushleft{Deauthentication flooding, Deauthentication flooding broadcast, Disassociation flooding, Disassociation flooding broadcast, Amok, Association Request flooding, Reassociation Request flooding, Beacon flooding, Rogue AP, Channel MitM attack, Key Reinstallation (Krack), TK Reinstallation (Kr00k), SSH Brute force, Botnet, WannaCry plus (Malware), TeslaCrypt (Malware), SQL Injection, SSDP Amplification, Captive portal (Evil Twin), Eaphammer (Evil Twin), ARP spoofing, DNS spoofing} & \checkmark~\cite{BestFeaturesAWID3,AWID3:BestFeatureSelection2} & \checkmark & IP address & -\\
\hline

H23Q \cite{H23Q}& 2022  & 200 & $\approx$10M p & 0.08:1 & \vspace{-1.3em}\flushleft{HTTP3-flood, Fuzzing, HTTP3-loris(DDoS), HTTP3-stream, QUIC-flood(DDoS), QUIC-loris(DDoS), QUIC-enc, HTTP-smuggle, HTTP2-concurrent, HTTP2-pause} & \checkmark & \checkmark & Device & -\\
\hline
\end{tabular}
\caption{Comparison of contemporary Internet datasets which can be used for assessing FL-enabled IDS.}.
\end{table*}

\subsection{IoT datasets}\label{sec:iot_datasets}

The \textbf{N-BaIoT} dataset~\cite{meidan2018n} focuses on botnet attacks on IoT networks. Specifically, it simulates two different botnets, namely, Bashlite and Mirai~\cite{Mirai-Kolias}. Both these botnets are capable of launching TCP and UDP flooding attacks, along with other ones, including port scanning. The dataset was evaluated by the authors by means of autoencoder models.

The \textbf{Kitsune} dataset~\cite{mirsky2018kitsune} was obtained based on the traffic provided by a video surveillance network with eight IP cameras, as well as an IoT deployment comprising nine devices, including a thermostat, a baby monitor, and two doorbells. More than 100 features were identified, and the attacks include MiTM, reconnaissance, and DoS. The dataset is available in both pcap and CSV formats.

The \textbf{Bot-IoT} dataset~\cite{koroniotis2019towards} also targets botnet attacks against IoT networks. It contains mostly DoS and DDoS assaults based on network flow features originating from each participating device. In this respect, the dataset focuses on IoT-specific protocols like MQTT having several devices residing on Amazon Web Services (AWS).

The \textbf{WUSTL-IIoT} dataset \cite{zolanvari2019machine} concentrates on an IIoT scenario considering 41 features to characterize network traffic and different attacks related to command injection, DoS, reconnaissance, and backdoor traffic. The data was obtained over 53 hours by using a testbed in an industrial plant with several IIoT systems. 

The \textbf{IoT network intrusion dataset}~\cite{q70p-q449-19} contains various types of network attacks against an IoT infrastructure. For capturing the traffic, two typical smart home devices were used, namely, SKT NUGU (NU 100) and EZVIZ Wi-Fi Camera. The dataset comprises 42 raw network packet files (pcap) collected at different time points. The packet files were captured by using the monitor mode of the wireless network adapter. This dataset contains zero wireless headers as these were removed through the Aircrack-ng tool. The \textbf{IoTID20}~\cite{ullah2020scheme} is built from the previous dataset by applying CICFlowMeter to extract 83 features from the network traffic, so that it is released in CSV format. 

The \textbf{IoT-23} dataset~\cite{iot23} is released through 20 malware captures and three captures related to benign network traffic. In particular, malicious traffic was obtained by using a Raspberry Pi device, while benign traffic was associated to three different IoT devices. The attacks in this dataset include traffic related to different botnets, such as Mirai and Okiru, as well as port scanning and DDoS. 

The \textbf{IoT DoS and DDoS Attack Dataset}~\cite{0s0p-s959-21} was created by applying data preprocessing techniques over the CIC DDoS dataset (described in subsection~\ref{SS:Internet datasets}). Although the original dataset was in CSV format, to efficiently utilize the potential of CNNs for DoS and DDoS attack detection, the network traffic data has been converted to an image format.

The \textbf{MQTT-IoT-IDS2020} dataset~\cite{bhxy-ep04-20} is derived from a simulated MQTT-based network. The simulated network comprised twelve sensors, a broker, a simulated camera, and an attacker. Five scenarios were recorded in an equal number of pcap files: normal operation, aggressive scan, UDP scan, Sparta SSH brute-force, and MQTT brute-force attack. Three types of features were extracted from the raw pcap files: packet features, unidirectional flow features, and bidirectional flow features. The dataset is also available in CSV format.

The \textbf{MQTTset} dataset~\cite{vaccari2020mqttset} is another dataset focused on the MQTT protocol created from the traffic generated by different IoT devices, such as door locks, temperature sensors, motion sensors, and fan speed controllers. It uses 33 features and provides data from different attacks, including flooding DoS, MQTT publish flood, and malformed data. 

The \textbf{IoT Healthcare Security Dataset}~\cite{9w13-2t13-21} was generated with the assist of IoT-Flock, an open-source tool for IoT traffic generation. This dataset contains one use-case corresponding to an IoT-based intensive care unit with the capacity of two beds, where each bed is equipped with nine patient monitoring devices (sensors) and one control unit called Bedx-Control-Unit. The dataset includes attacks related to MQTT and CoAP protocols. 

The \textbf{CCD-INID-V1} dataset~\cite{liu2021using} was compiled and assessed on a hybrid setting, that is, a combination of fog and cloud computing. Specifically, authors generated features from the fog layer and utilized training and testing phases at the cloud layer. The data was collected in smart lab and smart home environments using Rainbow HAT sensor boards installed on Raspberry Pi. The dataset was evaluated by means of legacy ML algorithms, including KNN, NB, LR, and SVM.

The \textbf{X-IIoTID}~\cite{al2021x} dataset captures recent attacker behavior targeted against industrial control loops devices, and edge, mobile, and cloud network domains. It contains device-agnostic features, making it suitable for industrial IoT (IIoT) systems. The dataset has more than 820K records and 68 features. These features were extracted from various sources, including network traffic, system logs, application logs, device resources, and commercial IDS logs.

The \textbf{Edge-IIoTset} dataset~\cite{ferrag2022edge} can be used by ML-based IDS in two different modes, namely, centralized and FL. The IoT data were generated from an assortment of IoT devices, such as low-cost digital sensors, ultrasonic sensors, and others. The authors analyzed 14 attacks related to IoT and IIoT protocols, which are categorized into five categories, namely DoS/DDoS, information gathering, MitM, injection, and malware. Specifically, the testbed comprised seven layers, i.e., cloud computing, network functions virtualization, a blockchain network, fog computing, software-defined networking, edge computing, and IoT/IIoT perception. 

The \textbf{LATAM-DDoS-IoT dataset}~\cite{almaraz2022toward} is a dataset centers on DoS and DDoS attacks against IoT devices. In particular, it contains 20 features and was created by using a production network with physical IoT devices, such as smart light bulbs and voice assistants. 

The \textbf{CIC IoT dataset} \cite{dadkhah2022towards} was recently created by using 81 IoT devices, such as cameras, smart lamps, speakers, doorbells, motion sensors, or weather stations which communicate through Wi-Fi, Zigbee and Z-Wave technologies. While the dataset is focused on device identification/profiling aspects, the authors carried out six different experiments, including the simulation of DoS and RTSP brute force attacks.

The \textbf{Ton\_IoT dataset} \cite{alsaedi2020ton_iot} was obtained through a deployment combining fog/edge and cloud components to simulate an IoT/IIoT production environment, considering different types of sensors. It provides network traffic information, as well as data related to sensor readings and telemetry using a set of 83 features. The data contains information regarding different types of attacks, including DoS/DDoS, ransomware, and XSS. Different parts of the dataset (e.g., related to network traffic or sensor readins) can be downloaded in CSV and pcap formats.

The \textbf{MedBIoT} dataset \cite{guerra2020medbiot} is focused on IoT botnets, including Mirai, BashLite, and Torii. It was obtained through a network comprised of 83 real and emulated devices, such as locks, switches, fans, and lights. Authors used real malware, considering the mentioned botnets. The dataset employs 100 features to characterize network traffic and can be downloaded in the form of pcap files.

\begin{table*}[]
\tiny
\centering
\begin{tabular}
{C{1.5cm}C{0.5cm}C{0.6cm}C{1cm}C{1cm}C{5.5cm}C{0.8cm}C{0.8cm}C{0.8cm}C{2cm}}
\hline
\textbf{Dataset}& \textbf{Year}  & \textbf{\# features} & \textbf{\# samples (packets/flows)} & \textbf{Attack/benign traffic ratio} & \textbf{Attacks} & \textbf{Data labelled?} & \textbf{Realistic testbed?} & \textbf{Division} & \textbf{FL-enabled IDS works using the dataset}\\
 \hline  \hline
N-BaIoT \cite{meidan2018n} & 2018  & 115 & $\approx$7M p & 11.7:1 & \vspace{-1.3em}\flushleft{Mirai Bot, BashLite Bot}  & \checkmark & \checkmark& Device & \cite{zhao2022semi, rey2022federated, khoa2020collaborative, popoola2021federatedzero, zakariyya2021memory} \\
 \hline

Kitsune \cite{mirsky2018kitsune} & 2018 & 115 & $\approx$21M p & 0.9:1 & \vspace{-1.3em}\flushleft{Reconnaissance, MitM, DoS, and botnet} & \checkmark & \checkmark & IP address & \cite{vy2021federated, zakariyya2021memory}\\
 \hline

Bot-IoT~\cite{koroniotis2019towards}& 2019 &  46 & $\approx$73M p & 9725.7:1 & \vspace{-1.3em}\flushleft{PortScan, OS Fingerprinting, DoS/DDoS, data theft, keylogging} & \checkmark & \checkmark & IP address & \cite{LocKedge, popoola2021federatedzero, popoola2021federated}\\
 \hline

WUSTL-IIoT~\cite{zolanvari2019machine} & 2019 & 41 & $\approx$1.2M p & 0.08:1 & \vspace{-1.3em}\flushleft{command injection, DoS, reconnaissance and backdoor}  & \checkmark & \checkmark & IP address & -\\
\hline

IoT network intrusion dataset~\cite{q70p-q449-19}& 2019 & N/A & $\approx$1.3M p & 9:1 & \vspace{-1.3em}\flushleft{Mirai ACK flooding, Mirai brut force, Mirai HTTP flooding, Mirai UDP flooding, MitM, portscan, OS Fingerprinting}  & \xmark & \checkmark & N/A & - \\
\hline

IoTID20 \cite{ullah2020scheme} & 2020 & 83 & 625,784 p & 14.6:1 & \vspace{-1.3em}\flushleft{Mirai ACK flooding, Mirai brut force, Mirai HTTP flooding, Mirai UDP flooding, MitM, portscan, OS Fingerprinting} & \checkmark & \checkmark & IP address & -\\
 \hline
 
IoT-23 \cite{iot23} & 2020 & 21 & $\approx$325M f & 9.5:1 & \vspace{-1.3em}\flushleft{PortScan, Okiru botnet, DDoS, C\&C, C\&C-HeartBeat, C\&C-File download, C\&C-Heart-Beat-FileDownload, C\&C-Tori, C\&C-Mirai} & \checkmark & \xmark & IP address & -\\
 \hline

MQTT-IoT-IDS2020~\cite{bhxy-ep04-20}& 2020 & 44 & $\approx$22M p & 8.7:1 & \vspace{-1.3em}\flushleft{Aggressive scan, UDP scan, Sparta SSH brute force, and MQTT brute-force attack} & \checkmark & \checkmark & IP address & \cite{attota2021ensemble}\\
\hline

MQTTset~\cite{vaccari2020mqttset}& 2020 & 33 & $\approx$12M p & 0.014:1 & \vspace{-1.3em}\flushleft{Flooding DoS, MQTT Publish flood, SlowITe, malformed data, brute force authentication} & \checkmark & \xmark & IP address & \cite{friha2022felids}\\
\hline

MedBIoT \cite{guerra2020medbiot} & 2020 & 100 & $\approx$18M p & 0.42:1 & \vspace{-1.3em}\flushleft{Mirai botnet, BashLite botnet, Torii botnet}  & \checkmark & \checkmark & IP address & - \\
 \hline

IoT Healthcare Security Dataset~\cite{9w13-2t13-21}& 2021 & 10 & 56,609 p & 0.7:1 & \vspace{-1.3em}\flushleft{MQTT Publish Flood, MQTT Authentication Bypass Attack, MQTT Packet Crafting Attack, and COAP Replay Attack} & \checkmark & \checkmark & N/A & -\\
\hline

ToN\_IoT \cite{alsaedi2020ton_iot}& 2021 &  83 & $\approx$5.3M p & 0.88:1 & \vspace{-1.3em}\flushleft{Backdoor, DoS, DDoS, Injection, MITM, Password, Ransomware, Scanning, XSS}  & \checkmark & \checkmark & IP address & \cite{ruzafa2021intrusion, kumar2021pefl, campos2021evaluating, de2022improving, kwon2022anomaly, popoola2021federated, singh2022dew}\\
 \hline

CCD-INID-V1~\cite{liu2021using}& 2021 &  83 & N/A & N/A & \vspace{-1.3em}\flushleft{ARP poisoning, APR DoS, UDP Flood, Brute force, Slowloris} & \checkmark & \xmark & IP address & -\\
\hline

IoT DoS and DDoS Attack Dataset~\cite{0s0p-s959-21}& 2021 &  80 & N/A & N/A & \vspace{-1.3em}\flushleft{CIC DDoS attacks}  & \checkmark & \checkmark &IP address  & -\\
\hline

X-IIoTID~\cite{al2021x} & 2021 & 68 & $\approx$0.8M p & 0.5:1 & \vspace{-1.3em}\flushleft{Generic scanning, scanning vulnerabilities, WebSocket fuzzing, discovering CoAP resources, brute force, dictionary attack, malicious insider, reverse shell, MitM, MQTT cloud broker-subscription, Modbus-register reading, TCP relay attack, data exfiltration, poisoning of cloud data (i.e., false data injections), fake notification, ransomware, and Ransom DoS}  & \checkmark & \checkmark & IP address & -\\
\hline

Edge-IIoTset~\cite{ferrag2022edge} & 2022 &  61 & $\approx$21M f & 1.15:1 & \vspace{-1.3em}\flushleft{TCP SYN DDoS, UDP flood DDoS, HTTP flood DDoS, ICMP flood DDoS, Port scanning, OS Fingerprinting, Vulnerability scanning, DNS spoofing, ARP spoofing, XSS, SQL Injection, Remote file inclusion, Backdoor, Password cracking, Ransomware}  & \checkmark & \checkmark & IP address & \cite{ferrag2022edge} \\ \hline

LATAM-DDoS-IoT~\cite{almaraz2022toward}& 2022 & 20 & $\approx$30M f (DoS), $\approx$50M f (DDoS) & 50.25:1 & \vspace{-1.3em}\flushleft{DoS, DDoS}  & \checkmark & \checkmark & IP address & -\\
\hline

CIC IoT ~\cite{dadkhah2022towards}& 2022 & 48 & 30k & N/A & \vspace{-1.3em}\flushleft{DoS, RTSP brute force}  & \xmark & \checkmark & IP address & -\\
\hline
\end{tabular}
\caption{Comparison of contemporary IoT datasets which can be used for assessing FL-enabled IDS.}.
\end{table*}

\subsection{Other datasets}
\label{sec:other_datasets}

The \textbf{InSDN} dataset~\cite{elsayed2020insdn} considers software-defined networking (SDN) scenarios by using 83 features to characterize network traffic in such environments. It contains information about different attacks, such as DoS, web attacks, and exploitation that was obtained through several tools, including Metasploit and Nmap. 

The \textbf{OTIDS} IDS dataset~\cite{lee2017otids} released back in 2017 targets in-vehicle networking (IVN), and specifically the CAN bus. Along with CAN normal traffic, it contains three attack categories, namely, DoS, fuzzy, and impersonation. The two datasets of normal traffic have a total size of around 354 MB ($\approx$4.6M messages), whereas the sizes of the three attack datasets are 60 MB (DoS), 50 MB (Fuzzy), and 84 MB (impersonation). The dataset was extracted from a KIA SOUL car by logging the CAN traffic via the On-Board Diagnostics (OBD) port. Additionally, the dataset is partially labeled, that is, only for the DoS attack, and is offered in text format.

The \textbf{Vehicular Reference Misbehavior Dataset (VeReMi)} dataset~\cite{heijden2018veremi} was built through a simulated vehicular environment considering five different attacks in which the vehicle's position is forged. The dataset also contains information about several scenarios in which vehicle density (i.e., number of vehicles) and attacker density are changed. It is a labelled dataset in which 17 features are used to characterize the vehicles' messages. 

The \textbf{UAV Attack Dataset}~\cite{00dg-0d12-20} is one of the few datasets containing ad-hoc network traffic. It targets Unmanned Aerial Vehicle (UAV) environments and was created with the aid of five different simulation platforms. The dataset contains two attack scenarios, namely, Global Positioning System (GPS) spoofing and jamming. Although it seems to be the only one available for studying UAV IDS, it has a rather small size of less than 19 MB.

The \textbf{Intrusion Detection in CAN bus}~\cite{24m9-a446-19} dataset comprises a fusion of three other datasets. The first is the Car Hacking Dataset v1~\cite{lee2017otids}, from which the authors picked the ``fuzzy'' attack. The other two datasets were new collections, utilizing the ML350 and CL2000 Mercedes models. Overall, the dataset contains two attacks, namely DoS and ``fuzzy'', and has an unzipped size of 43 MB.

The \textbf{Car Hacking} dataset~\cite{song2020vehicle} focuses on different attacks on CAN bus, including DoS, fuzzy and spoofing (RPM/gear). It considers five features, namely timestamp, CAN ID, DLC, Data, and flag, and was built by logging CAN traffic via the OBD-II port from a real vehicle.

The \textbf{Car Hacking: Attack \& Defense Challenge} dataset~\cite{kang2021car} was created through a competition in Korea to develop attack and detection techniques of CAN by using Hyundai Avante CN7 as target vehicle. The dataset is composed of different attacks, including data about flooding, spoofing, replay, and fuzzing attacks on CAN messages, and similar features compared to the Car Hacking dataset.

 \begin{table*}[]
\tiny
\centering
\begin{tabular}{C{1.5cm}C{0.5cm}C{0.6cm}C{1cm}C{1cm}C{5.5cm}C{0.8cm}C{0.8cm}C{0.8cm}C{2cm}}
\hline
\textbf{Dataset}& \textbf{Year}  & \textbf{\# features} & \textbf{\# samples (packets/flows)} & \textbf{Attack/benign traffic ratio} & \textbf{Attacks} & \textbf{Data labelled?} & \textbf{Realistic testbed?} & \textbf{Division} & \textbf{FL-enabled IDS works using the dataset}\\
 \hline  \hline

InSDN \cite{elsayed2020insdn}& 2020 & 83 & $\approx$0.3M f & 4.03:1 &\vspace{-1.3em}\flushleft{Botnet, DoS, DDoS, web attacks, password brute-forcing, probe, exploitation} & \checkmark & \checkmark & IP address & \cite{friha2022felids}\\
\hline

OTIDS \cite{lee2017otids}& 2017 & 11 & $\approx$4.6M p & 0.9:1 & \vspace{-1.3em}\flushleft{DoS, Fuzzy, impersonation} & \checkmark & \checkmark & Vehicle & \cite{aliyu2021blockchain}\\
\hline

VeReMi \cite{heijden2018veremi}& 2018 & 17 & $\approx$2.2M p & 0.35:1 & \vspace{-1.3em}\flushleft{Misbehavior (position forging) attacks} & \checkmark & \checkmark & Vehicle & \cite{uprety2021privacy}\\
\hline

Intrusion Detection in CAN bus~\cite{24m9-a446-19}& 2019 & N/A & N/A & N/A & \vspace{-1.3em}\flushleft{DoS, fuzzy} & \xmark & \checkmark & Device & - \\
\hline

UAV Attack Dataset~\cite{00dg-0d12-20, whelan2020novelty}& 2020  & 85 & 786,728 (sensor values) & 0.02:1 & \vspace{-1.3em}\flushleft{GPS spoofing, and jamming}  &  \checkmark  & \checkmark & Device & - \\
\hline

Car Hacking~\cite{song2020vehicle}& 2020 & 5 & $\approx$17.5M messages & 0.17:1 & \vspace{-1.3em}\flushleft{Flooding, spoofing, replay, fuzzing} & \checkmark & \checkmark & Vehicle & \cite{abdel2021federated} \\
\hline

Car Hacking A\&D Challenge~\cite{kang2021car}& 2021 & 6 & $\approx$1.2M messages & 16.8:1 & \vspace{-1.3em}\flushleft{DoS, fizzy, spoofing (drive gear), spoofing (RPM gauze)} & \checkmark & \checkmark & Vehicle & \cite{driss2022federated} \\
\hline

\end{tabular}
\caption{Comparison of other contemporary datasets which can be used for assessing FL-enabled IDS.}.
\label{table:comparison1}
\end{table*}

\subsection{Analysis}
\label{sec:datasets_analysis}

Based on the previous description and the information provided in Table~\ref{table:comparison1}, several conclusions can be drawn. First, in recent years there is a clear growing trend in the generation of IoT-related datasets that demonstrates an increase on the use and deployment of IoT technologies. Indeed, half of the IDS datasets analyzed (18 out of 36) are related to IoT settings. Second, there is also an increase in proposed datasets related to vehicular environments, especially around CAN-based in-vehicle IDS~\cite{karop22}. Third, regarding the benign/attack ratio, even though it is widely recognized that real-life scenarios are characterized by a higher proportion of benign traffic, we verified that there is a significant disparity between the different datasets, with some of them (e.g., Bot-IoT or IoT network intrusion datasets) containing more samples related to attack traffic. A fourth point is that most of the datasets do not contain modern attacks, while a significant portion of them embrace a small number of attacks. Another significant observation is that, according to our analysis, most of the recent datasets can be realistically divided (e.g., using the IP address) to be used in a FL scenario. However, many of them (e.g., Kitsune or IoT-23) are not directly applicable, but must be previously pre-processed so that the network traffic from the same IP address can be considered as a device acting as a FL client. In the case of datasets related to in-vehicle networks, we verified that most of the datasets are based on a single vehicle from which the data was obtained; therefore it is rather infeasible to create an FL scenario with a realistic division. Furthermore, some datasets only consider a tiny number of devices that could be considered as FL clients. For example, IoT network intrusion and IoTID20 datasets are based on network traffic from only two devices. In the case of IoT-23, while benign traffic originates from three devices, the attack traffic is obtained by a single device.

\begin{figure*}
\begin{adjustbox}{width=\textwidth}
\begin{tikzpicture}
 \small
\begin{axis} [
    xbar,
    label style={font=\large},
    xmin=0,
    height=15cm,
    width=18cm,
    bar width=5pt,
    ylabel={Dataset}, 
    ylabel near ticks,
    xlabel={\# of references},
     nodes near coords, 
     enlarge y limits=0.02,
     ytick={1,2,3,4,5,6,7,8,9,10,11,12,13,14,15,16,17,18,19,20,21,22,23,24,25,26,27,28,29,30,31,32,33,34,35,36,37,38,39,40,41,42},
     yticklabels={
        NREL dataset \cite{nreldataset},
        InSDN \cite{elsayed2020insdn},
        CTU-13 \cite{garcia2014empirical},
        Car Hacking \cite{song2020vehicle},
        ISCXVPN2016 \cite{draper2016characterization},
        ISCXTor2016 \cite{lashkari2017characterization},
        Microsoft Malware \cite{ronen2018microsoft},
        Genome \cite{zhou2012dissecting},
        Contagio \cite{contagiodataset},
        WSN-DS \cite{almomani2016wsn},
        CRAWDAD \cite{crawdad},
        MQTT-IoT-IDS2020 \cite{bhxy-ep04-20},
        MQTTset \cite{vaccari2020mqttset},
        SEA \cite{schonlau2001computer},
        \cite{igbe2017deterministic},
        IoT-DDoS \cite{iot-ddos},
        WUSTL \cite{teixeira2018scada},
        VeReMi \cite{heijden2018veremi},
        VirusShare \cite{virusshare},
        Car Hacking A\&D Challenge \cite{kang2021car},
        \cite{javed2020alphalogger},
        Edge-IIoTset \cite{ferrag2022edge},
        \cite{frazao2018denial},
        AndroZoo \cite{allix2016androzoo},
        OTIDS \cite{lee2017otids},
        Maldroid \cite{mahdavifar2020dynamic},
        SWAT \cite{goh2016dataset},
        ISCXBotnet2014 \cite{beigi2014towards},
        Kitsune \cite{mirsky2018kitsune},
        AWID3 \cite{chatzoglou2021empirical},
        Bot-IoT \cite{koroniotis2019towards},
        Industrial dataset \cite{morris2014industrial},
        Drebin \cite{arp2014drebin},
        N-BaIoT \cite{meidan2018n},
        CSE-CIC-IDS2018 \cite{sharafaldin2018toward},
        Ton\_IoT \cite{alsaedi2020ton_iot},
        CIC-DDoS2019 \cite{sharafaldin2019developing},
        KDDCup99 \cite{kddcup_dataset},
        UNSW-NB15 \cite{moustafa2015unsw},
        Generated,
        CIC-IDS2017 \cite{sharafaldin2018toward},
        NSL-KDD \cite{tavallaee2009detailed}
        }
   ]
\addplot coordinates {
    (23,42)
   (20,41)
   (12,40)
  (11,39)
  (11,38)
  (7,37)
  (7,36)
  (5,35)
 (5,34)
 (3,33)
  (3,32)
  (3,31)
  (2,30)
  (2,29)
  (2,28)
  (2,27)
  (2,26)
   (2,25)
   (2,24)
  (1,23)
 (1,22)
  (1,21)
  (1,20)
   (1,19)
  (1,18)
   (1,17)
  (1,16)
   (1,15)
   (1,14)
   (1,13)
   (1,12)
(1,11)
   (1,10)
 (1,9)
   (1,8)
  (1,7)
   (1,6)
   (1,5)
   (1,4)
   (1,3)
   (1,2)
    (1,1)
};
\end{axis}
\end{tikzpicture}
\end{adjustbox}
\caption{Datasets employed by existing FL-enabled IDS proposals}
\label{fig:datasets}
\end{figure*}
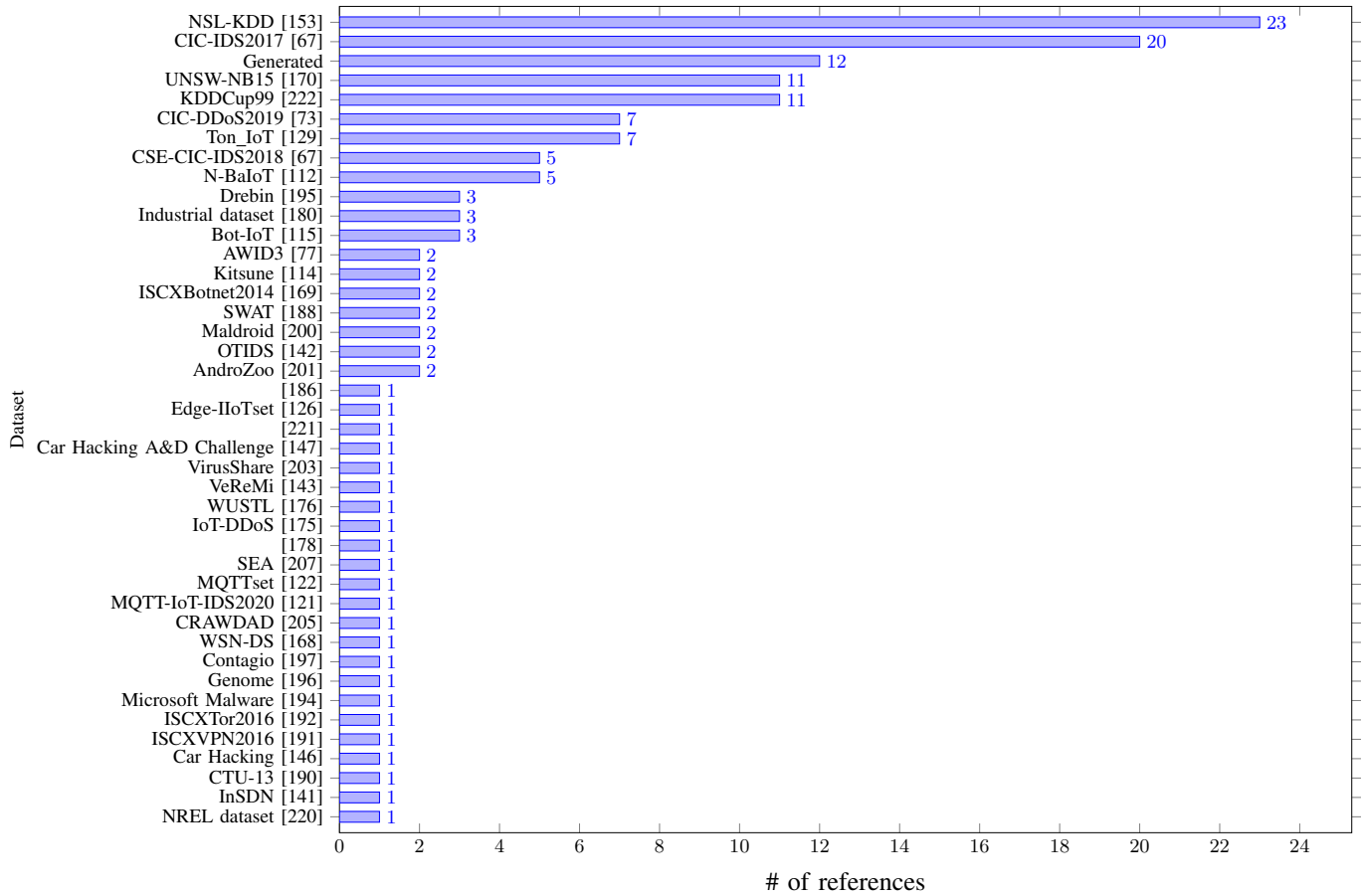

In addition to the analysis of the described datasets, Table~\ref{table:comparison1} shows which datasets are used by proposed FL-enabled IDS. We verified that only 56 works out of the 104 analyzed (see Appendix~\ref{sec:appendix} for a summary of such analysis)  utilized recent datasets. Namely, Figure~\ref{fig:datasets} provides the usage statistics of the datasets employed by the existing literature around FL-enabled IDS approaches. From the figure, we see that the most used dataset is NSL-KDD~\cite{tavallaee2009detailed} (used in 23 works so far), which was created in 2009 as an improved version (after removing duplicate packets/flows) of the KDDCup99~\cite{bolon2011feature} dataset, which, although deprecated, is still widely used (11). Both these datasets focus on attacks related to DoS, privilege escalation, and probing. However, according to~\cite{ring2019survey}, these datasets do not have IP address information, which could be exploited for a realistic division to be used in a FL scenario. That is, according to our analysis, we verified that the proposed works use different and artificial divisions of the dataset. For example, NSL-KDD is used by~\cite{al2020federated} dividing a subset of the dataset among 10 clients, as well as by~\cite{rahman2020internet}, which uses four different nodes. In the case of~\cite{mirzaee2021fids}, the authors split the dataset randomly among 16 nodes and~\cite{wang2022feco} uses 50 clients. Other references consider a variable number of nodes~\cite{shahid2021detecting, saadat2021hierarchical, liu2022intrusion}, and in some works (e.g.,~\cite{luo2022federation, yang2022federation}) the number of nodes is not clearly stated. With reference to KDDCup99,~\cite{gaussian2020network} uses two clients, while~\cite{hei2020trusted} divided the dataset into four devices in an IoT environment where intermediate nodes act as clients. A variable number of nodes is considered by~\cite{liu2021blockchain}, while this number remains unspecified in~\cite{su2022detection}.

Besides KDDCup99 and NSL-KDD, one of the most widely used datasets in the literature around FL-enabled IDS is the CIC-IDS2017 one (20). For this dataset, some of the proposed works leverage the information (e.g., IP address) provided about each of the 12 victim devices. Specifically, the authors in~\cite{preuveneers2018chained} consider up to 12 nodes acting as FL clients in their experiments. However, in other cases, including~\cite{zhao2019multi, otoum2021federated}), it is unclear how the dataset is divided. Furthermore, the authors in~\cite{tang2022federated} only considered two nodes, and~\cite{wei2021federated} uses a variable number of devices (1, 5 and 10) acting as FL clients. In the case of~\cite{otoum2022feasibility}, the same dataset is distributed among a variable number of vehicles. Additionally, certain works~\cite{qin2021fnel} only consider a subset of the attacks contained in the dataset. CIC-IDS2017, KDDCup99, and the WSN-DS dataset~\cite{almomani2016wsn} (focused on DoS attacks on WSNs) are used in~\cite{chen2020intrusion} by taking into account different numbers of clients. Moreover, the approach proposed by~\cite{qin2020line} is evaluated using two datasets independently: CIC-IDS2017 and ISCXBotnet2014~\cite{beigi2014towards}, which are divided considering a variable number of clients (2 to 8). It should be noted that ISCXBotnet2014 concentrates on botnets and is built based on other datasets, including the ISCX2012IDS one~\cite{shiravi2012toward}. On the other hand, the work in~\cite{fan2020iotdefender} uses Transfer Learning (TL) to obtain personalized models. This is done by utilizing CIC-IDS2017 as a public dataset, and NSL-KDD, Kitsune and IoT network intrusion as private datasets using four clients. 

Another widely used dataset is UNSW-NB15~\cite{moustafa2015unsw} (11), which was proposed in 2015 with 49 features to characterize network traffic considering 45 IP addresses, which can be used for FL purposes. For instance, the work in~\cite{zhao2020network} splits this dataset between two clients to simulate a federated TL scenario. In the case of~\cite{aouedi2022fluids}, the authors divide the dataset among 100 nodes, but only 10\% are used in each training round, and~\cite{lalouani2022robust} uses 10 nodes. Moreover, the authors in~\cite{markovic2022random} exploit various datasets, including KDDCup99, NSL-KDD, UNSW-NB15, and CIC-IDS2017, by modifying the number of nodes. In the case of~\cite{cheng2022federated}, the authors employ 10 nodes considering NSL-KDD and UNSW-NB15 datasets.

Other more recent datasets, such as CIC-DDoS2019 and CSE-CIC-IDS2018 were used in a significant mass of works, that is, 7 and 5, respectively. For example,~\cite{li2021fids} divides the CIC-DDoS2019 among five nodes, and~\cite{yuan2021towards} considers a variable number of clients. In the case of~\cite{duy2021federated}, the same dataset is split into three clients, although the authors also consider additional nodes to provide a comparison in terms of communication overhead. Furthermore, the authors in~\cite{lv2022ddos} used 50 clients in which CIC-DDoS2019 was distributed, albeit only a certain proportion of nodes (0.8) was selected in each training round. In addition, CIC-IDS2017, NSL-KDD and CIC-DDoS2019 are used by~\cite{zhang2021flddos} distributing the datasets among 21 nodes. Moreover, CSE-CIC-IDS2018 is used by~\cite{chen2022privacy}, which considers two clients to distribute other datasets, including KDDCup99, NSL-KDD and CIC-IDS2017. The authors in~\cite{shi2021data}, used a subset of the CSE-CIC-IDS2018 and UNSW-NB15 datasets that was divided among 10 and 15 clients, respectively. Furthermore, CSE-CIC-IDS2018, InSDN~\cite{elsayed2020insdn} and MQTTset~\cite{vaccari2020mqttset} were used by~\cite{friha2022felids} considering a variable number of nodes (5 to 15).

Based on the information provided in Figure~\ref{fig:datasets}, we verified that despite some of the IoT datasets are quite recent, they are also widely considered in the development of FL-enabled IDS approaches. In the case of Ton\_IoT (7),~\cite{campos2021evaluating} divides the dataset considering the IP address of the nodes and in particular they select a subset of the dataset made up of 10 clients. The work in~\cite{de2022improving} also divides the dataset by IP address considering the network traffic of 13 nodes. However, the work in~\cite{kumar2021pefl} only uses two clients to split the dataset, and~\cite{singh2022dew} proposes a hierarchical approach making use of the NSL-KDD and Ton\_IoT datasets, although the number of the participating devices is not mentioned. For N-BaIoT (5),~\cite{zhao2022semi} considers different data distributions using 27 and 89 nodes to partition the dataset. The work in~\cite{rey2022federated}, on the other hand, splits the dataset across eight nodes using the division of the network traffic as provided in the dataset.

N-BaIoT was also used by~\cite{khoa2020collaborative}, which splits additional datasets (KDDCup99, NSL-KDD, UNSW-NB15) between two and three clients. Several datasets related to IoT environments (N-BaIoT, Kitsune, IoT-DDoS~\cite{iot-ddos} and WUSTL~\cite{teixeira2018scada}) are used by~\cite{zakariyya2021memory} independently considering three FL clients. Interestingly, the WUSTL dataset is a previous version of WUST-IIoT without IP address information, while the IoT-DDoS dataset is a subset of BoT-IoT made up of traffic only related to DDoS attacks. Moreover, the work in~\cite{LocKedge} uses the BoT-IoT dataset considering the attackers' IP address, so that each attacker in the dataset (i.e., four nodes) is considered as a FL client. In the case of~\cite{popoola2021federated}, the authors divided the BoT-IoT and N-BaIoT datasets among five nodes. Additionally, in~\cite{ferrag2022edge}, the authors proposed the Edge-IIoTset dataset, which is described in Section~\ref{sec:iot_datasets}, and used for the evaluation of different FL-enabled scenarios.

Some works focus on vehicular scenarios making use of some of the datasets described in subsection~\ref{sec:other_datasets}. For instance,~\cite{aliyu2021blockchain} employs the OTIDS dataset, which is distributed among a variable number of nodes. However, as already mentioned, this dataset comprises information generated by a single vehicle. This is also the case of~\cite{driss2022federated} and the Car Hacking A\&D Challenge dataset. Furthermore,~\cite{abdel2021federated} uses the Car Hacking dataset distributed among 20 nodes, while~\cite{uprety2021privacy} centers on the detection of vehicle misbehavior using the VeReMi dataset, although it is artificially distributed among 10 nodes.

Beyond the vehicular context, industrial environments have been also considered for the development of FL-enabled IDS approaches. In this direction,~\cite{zhou2020federated} uses two different datasets considering a variable number of nodes. The first~\cite{igbe2017deterministic} focuses on the ModBus protocol~\cite{fovino2009design} and the second~\cite{morris2014industrial} on the DNP3 protocol~\cite{east2009taxonomy}. Both these datasets are offered in the form of pcap files. The work in~\cite{morris2014industrial} is also used by~\cite{li2020deepfed} and ~\cite{aouedi2022federated}. In the first, the authors consider a variable number of clients, while in the second the authors distribute the dataset among 100 nodes with 10\% of them training in each round. In~\cite{kelli2021ids}, the authors generate a dataset considering the DNP3 protocol. Furthermore,~\cite{mothukuri2021federated} exploits another dataset containing information about DoS attacks~\cite{frazao2018denial} associated with the ModBus protocol. The authors in~\cite{novikova2022federated} focus on a water treatment scenario using the SecureWater Treatment (SWaT) dataset~\cite{goh2016dataset}. The peculiarity of this work is that, unlike most of the proposed approaches, it considers a vertical FL scenario where clients do not share the same feature space.

Other datasets are scarcely used in the current literature. For instance, the authors in~\cite{ibitoye2022differentially} use the CTU-13 dataset~\cite{garcia2014empirical}, which concentrates on different botnets and is distributed among 200 clients. Furthermore,~\cite{zhao2019multi} uses the ISCXVPN2016~\cite{draper2016characterization} and ISCXTor2016~\cite{lashkari2017characterization} datasets. While the former dataset is split into two clients, it is unclear if this division can be done for the latter. A similar case is presented in~\cite{singh2020collaborative}, which uses the Microsoft Malware dataset~\cite{ronen2018microsoft}, which is randomly divided into six nodes of equal size. Other barely considered datasets are Drebin~\cite{arp2014drebin}, Genome~\cite{zhou2012dissecting}, and Contagio~\cite{contagiodataset}, which are used by~\cite{taheri2020fed} through modifying the number of clients. In the case of~\cite{pasdar2022train}, the authors use the Maldroid~\cite{mahdavifar2020dynamic}, Drebin and Androzoo~\cite{allix2016androzoo} datasets considering three clients. Furthermore, \cite{pei2022knowledge} uses the same datasets and the VirusShare repository \cite{virusshare} with a variable number of clients. 

Moreover,~\cite{mowla2019federated} uses the CRAWDAD~\cite{crawdad} dataset (divided into six clients), which contains 802.11p packets with information related to jamming attacks. In ~\cite{zhao2020intelligent}, the authors use the SEA dataset~\cite{schonlau2001computer}, which provides the logs of different UNIX users. 

Other works generate their own datasets~\cite{nguyen2019diot, li2020distributed, sun2020intrusion} in different scenarios, including detection of attacks on Android devices~\cite{hsu2020privacy, rehman2022federated}, or even against solar farms~\cite{zhao2021federated}. Moreover, the work in~\cite{zhang2022grid} considers false data injection attacks on smart grids, while~\cite{shukla2021device} uses the VirusTotal database~\cite{peng2019opening}, which also seems to be used by the same authors in~\cite{shukla2022rafel}. On the downside, some of the aforementioned datasets are not publicly available, therefore it is difficult to compare the proposed approaches with the existing literature. Last but not least, there exist few other works, including~\cite{nguyen2021federated1} and~\cite{alazzam2022federated}, which do not provide details of the dataset being used.

Overall, based on the previous analysis about the datasets used in the existing FL-enabled IDS proposals, a number of key points arise. First off, as already mentioned, we verified that a significant number of proposed works are based on datasets prior to 2017 that contain obsolete attacks, which do not reflect modern communication networks. Second, some of the proposed works employ datasets that do not even provide network traffic information (e.g., the IP address) that can be used to realistically divide the dataset into several nodes acting as FL clients. Third, although most of the works utilize datasets that can be realistically divided, the approaches are usually based on random divisions among different numbers of clients. Fourth, and related to the previous point, most contributions do not make the divisions of the datasets used in their evaluation publicly available. Therefore, it is almost impossible to reproduce the results obtained or provide realistic comparisons between similar works, even if they are based on the same dataset. In many cases, the divisions are made so that the nodes have the same number of samples or have samples of all classes, i.e, type of attacks. This circumstance goes against the nature of a FL scenario that is characterized by non-IID data distributions. Fifth, the approaches do not reflect heterogeneity, that is, considering scenarios where nodes employ dissimilar features to characterize network traffic. Actually, only one of the works analyzed~\cite{novikova2022federated} explicitly considers a vertical FL scenario. 



\section{ML/DL models for FL-enabled IDS}
\label{sec:models}
\begin{figure} [!ht]
	\centering
		\includegraphics[width=0.5\textwidth]{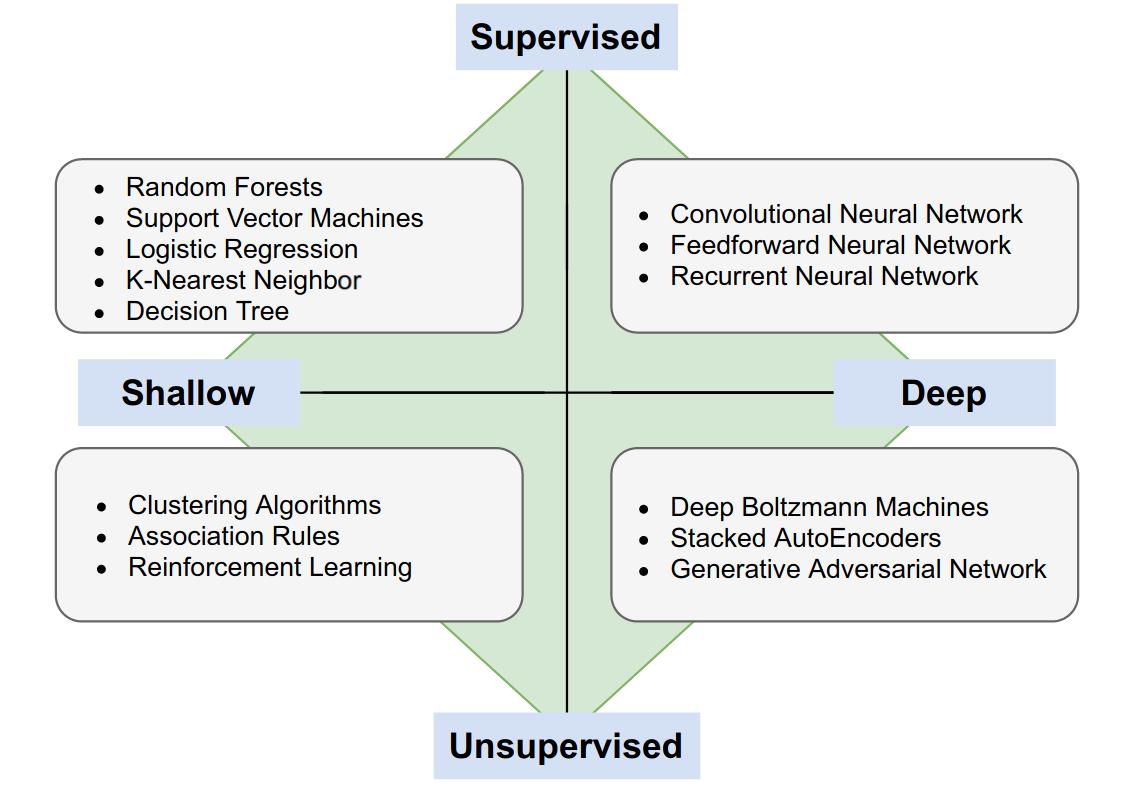}
	\caption{Typical ML algorithms in Cybersecurity \cite{apruzzese2022role}.}
	\label{fig:tml}
\end{figure}

This section alongside Table~\ref{tab:super_taxonomy_long_table} in Appendix~\ref{sec:appendix} offers a comprehensive overview and analysis of the ML techniques used by existing FL-enabled IDS in the literature. Some of the most prevalent ML algorithms in cybersecurity~\cite{apruzzese2022role} are shown in Figure~\ref{fig:tml}. Specifically, in the following subsections, we briefly refer to four major categories, namely supervised, unsupervised, semi-supervised and reinforcement learning techniques. An outline of the surveyed works per type of learning is given in Table~\ref{tab:super_taxonomy_long_table} in Appendix~\ref{sec:appendix}.

\subsection{Supervised learning}\label{sec:models_supervised}

Supervised learning techniques rely on the use of labeled data to train a model. The training dataset consists of inputs and correct outputs, which enable the model to learn by adjusting its hyperparameters to the data. The dataset guides the model so that it can learn the relationship between inputs and outputs and make predictions on new data. Some of the most commonly used supervised learning techniques are Logistic Regression (LR)~\cite{wright1995logistic}, Decision Trees (DT)~\cite{kotsiantis2013decision}, Random Forest (RF)~\cite{breiman2001random}, Neural Networks (NN)~\cite{anderson1995introduction}, and Support Vector Machines (SVM)~\cite{hearst1998support}.

Based on the information provided in Annex~\ref{sec:appendix}, we conclude that most of the proposed FL-enabled IDS approaches are based on the use of different types of NN. A NN consists primarily of units referred to as neurons, which are typically organized into layers and connected to each other with an associated weight that determines the input's influence on the neuron's output. Typically, the NNs used in IDS employ a number of features in the dataset as the number of neurons in the input layer, and depending on whether it is binary or multiclass classification, the number of classes (i.e., benign and types of attack) of traffic represented in the dataset. There is a plethora of NNs, including feedforward (FNN)~\cite{svozil1997introduction}, recurrent (RNN)~\cite{salehinejad2017recent} or convolutional (CNN)~\cite{gu2018recent} ones, which are widely considered for the development of FL-enabled IDS, as described below. 

In the case of FNNs, signals only flow from input to output. The most well-known example of this type of network is represented by a Multilayer peceptron (MLP)~\cite{murtagh1991multilayer}, which is characterized by one or several hidden layers between the input and output layers. Furthermore, an MLP is a fully connected network; each neuron in a hidden layer receives inputs from all the neurons in the previous layer. Based on MLP, the work in~\cite{al2020federated} builds a simple MLP structure with two hidden layers of 256 neurons to develop a federated mimic learning system, where a master model is trained on private data and label a public dataset to train a student model. In the case of~\cite{liu2021blockchain}, the authors use an MLP in a blockchain-enabled FL architecture to detect attacks in vehicular environments, although details of the MLP layers are not provided. Moreover, the authors in~\cite{rey2022federated} use four MLPs with different configurations in terms of number of hidden layers and number of neurons per layer.

As an alternative to FNNs, RNNs allow signals to flow in both directions; that is, signals feed back to neurons in the different layers. One of the main examples of RNN is the Long short-term memory (LSTM), which maintains an internal memory that is updated with new inputs. Specifically, this memory is maintained by each neuron using three different gates: input, output, and forget. Based on this model, the work in~\cite{singh2022dew} uses a hierarchical LSTM approach for medical environments where FL nodes are medical institutions, such as hospitals, providing better results compared to naive LSTM. In the case of~\cite{zhao2021federated}, the authors use a 5-layer LSTM for false data injection attack detection in solar farms. Another well-known example of RNN is represented by Gated recurrent units (GRU), which are computationally more efficient using fewer parameters than LSTMs. Additionally, GRUs aim to solve the vanishing gradient problem which comes with a standard RNN using the update gate and reset gate~\cite{gonzalez2022parking}. Based on the use of GRUs,~\cite{li2021fids} uses a simple GRU structure comprising two fully connected layers and two identical GRU layers. In the case of~\cite{kumar2021pefl}, the authors use a GRU model consisting of two hidden layers with 16 and 8 neurons. Additionally, they previously employ an LSTM-autoencoder (AE) approach to transform the original data into a new representation to further improve privacy. The study in ~\cite{tang2022federated} uses a simple GRU architecture composed of 256 nodes in the hidden layer. In the case of~\cite{mothukuri2021federated}, the authors capitalize on a combination of LSTM and four GRU models with different parameters. Also, based on GRU,~\cite{nguyen2019diot} presents the DIoT system, which builds communication profiles associated with a type of device to detect attacks. In particular, several gateways are responsible for generating GRU models from the connected IoT devices, which are then aggregated by a service.

Another well-known example of neural networks are CNNs. These are based on the shared-weight architecture of the convolution kernels or filters to extract relevant features from the input data. They can be considered as regularized versions of MLPs intended to reduce overfitting by leveraging the hierarchical structure of data. CNNs build patterns of increasing complexity using smaller and simpler patterns embedded in their filters. According to our analysis, the use of CNNs is widely considered in the development of FL-enabled IDSs. In particular, this approach is used by~\cite{li2020distributed} for intrusion detection in satellite-terrestrial integrated networks using a dataset generated with different attacks against such networks. In ~\cite{sun2020intrusion}, the authors employ a four-layer CNN to detect attacks in a dataset generated in a Local Area Network (LAN). 

Also based on a CNN, the work in~\cite{zhou2020federated} proposes an approach for payload classification as a basis for detecting potential attacks in an edge-enabled industrial environment. In a similar direction, the authors in~\cite{fan2020iotdefender} use a CNN architecture deployed on different Mobile Edge Computing (MEC) nodes. They additionally use TL to obtain customized models in different IoT networks. Moreover, in~\cite{duy2021federated}, a 16-layer CNN is utilized in combination with a simple 7-layer structure~\cite{shi2021data}. In the case of~\cite{pasdar2022train}, the authors present a two-phase system based on a CNN, which additionally employs TL to create customized models and FL for collaborative learning. Another approach based on CNN is proposed by~\cite{lv2022ddos} with two convolutional layers and three fully connected layers to detect DDoS attacks. For improving effectiveness, the authors group several traffic flows related to a certain attack to be used by the CNN.

A germinal CNN-based work is presented in~\cite{luo2022federation}; the same study was improved in~\cite{yang2022federation} to consider every attack in the employed dataset. The study in~\cite{cheng2022federated} also uses a three-layer CNN to build a model which is then transferred to another domain for accelerating model training. The authors also rely on Reinforcement Learning (RL) to build a client selection mechanism to improve system performance. Additionally,~\cite{ibitoye2022differentially} exploits a CNN-based approach that includes a noise layer to implement a Differential Privacy (DP) mechanism for reinforcing privacy.

Other recent approaches consider the use of CNNs in combination with other models. For example, in~\cite{liu2022intrusion}, the authors combine CNNs and MLPs; CNNs are used to extract data features that are then used by an MLP model for classification. The study in~\cite{zhang2022grid} uses a combination of CNNs and LSTMs for false data injection attack detection in the context of smart grids, although further details about the model are not provided. Additionally,~\cite{zhao2020intelligent} compares the performance between a federated version of CNNs and LSTMs, resulting that the latter offers better performance over a dataset consisting of UNIX user logs~\cite{schonlau2001computer}. The use of a simple RNN is considered by~\cite{zhang2021flddos} as a more effective alternative compared to MLPs and CNNs to detect DoS attacks using multiple datasets. In~\cite{li2020deepfed}, the authors propose a model consisting of a CNN and a GRU, whose outputs are then passed to an MLP for intrusion detection in industrial environments. In particular, the input features are a one-dimensional vector representing the numerical features of network traffic data. This vector is processed by three convolutional blocks and two identical GRU layers. The output is fed to the MLP module, which has two fully connected layers and a dropout layer. Ultimately, a softmax layer provides the final classification result.

In addition, some approaches do not provide details on the type of NN being used. For instance,~\cite{mowla2019federated} proposes a security architecture for detecting jamming attacks in a UAV ecosystem. The authors additionally use a technique based on the Dempster-Shafer theory~\cite{shafer1992dempster} for prioritizing clients in each training round. Moreover,~\cite{rahman2020internet} uses a simple NN with a hidden layer to analyze the impact of different data distributions. Also based on a (fully connected) NN, the work in~\cite{zakariyya2021memory} builds an efficient memory-based model based on the backpropagation algorithm. 

The authors in~\cite{qin2021fnel} exploit a three-layer NN to design an attack detection approach based on the principles of never ending learning~\cite{mitchell2018never}; this is done to obtain knowledge about new threats. A NN-based model is also used by~\cite{uprety2021privacy}, although no further details about the used model are provided. Another NN is used by~\cite{saadat2021hierarchical} based on a simple architecture with two hidden layers in an FL-enabled hierarchical IDS approach where edge nodes perform partial aggregations. Additionally,~\cite{LocKedge} proposes an NN model by reducing the number of layers and evaluating the impact of different numbers of neurons used in the hidden layer.

The authors in~\cite{wang2022feco} rely on a lightweight NN with two hidden layers to propose a federated approach based on contrastive learning~\cite{khosla2020supervised}, with the aim of converting the original traffic instances into a new representation. The contribution in~\cite{lalouani2022robust} considers a simple NN structure comprising two hidden layers with 64 neurons. The use of an NN is also considered by~\cite{otoum2022feasibility}, which compares the performance and privacy properties between FL, TL, and Split Learning~\cite{singh2019detailed}. Also based on NNs, other works make use of Deep NNs (DNNs), which are based on the use of a greater number of hidden layers. For instance, the use of DNN combined with TL is proposed by~\cite{zhao2020network} to tackle the problem of training data scarcity. In particular, first, FL is used to build a global model, which is reconstructed and re-trained in a second phase to improve detection effectiveness. In a similar approach,~\cite{otoum2021federated} uses a 6-layer DNN to pre-train a model on the network edge that is then transferred and customized on different end nodes. Also through a DNN,~\cite{popoola2021federated} implements a model composed of 4 hidden layers with 100 neurons each. In ~\cite{kelli2021ids}, the authors rely on a 6-layer architecture to detect attacks in industrial environments. Moreover,~\cite{rehman2022federated} uses a DNN composed of five dense layers and five dropout layers for the detection of side channel attacks on Android devices. In~\cite{friha2022felids}, the authors compare federated approaches based on CNN, RNN, and DNN models with different configurations.

Other recent works consider propose certain variations of traditional NN approaches. For instance, Binarized NNs (BNNs)~\cite{hubara2016binarized} are characterized by using binary (+1 or -1) representations for weights and activations. Therefore, the sign function becomes the most suitable activation function under this setting~\cite{xu2021learning}. In particular, targeting IoT scenarios where gateways act as FL clients,~\cite{qin2020line} proposed a BNN as  an efficient approach for federated training in terms of memory and communication costs. Furthermore, the authors in~\cite{zhao2019multi} propose a multi-task learning (MTL)~\cite{zhang2021surveyMTL} approach called MT-DNN-FL, which simultaneously detects network anomalies, recognizes VPN traffic, and classifies traffic using publicly accessible datasets. Recall that a Multi-task NN (MTNN)~\cite{crawshaw2020multi} is an example of MTL techniques, which are typically used to train a model for different purposes simultaneously; therefore, this approach is especially useful in environments where there is a scarcity of data for each specific task.

In addition to NNs, other techniques have also been considered in the development of FL-enabled IDS approaches. For example, DTs~\cite{kotsiantis2013decision} are one of the simplest ML techniques based on a hierarchical structure where each node represents a decision (on the basis of a certain variable) and each leaf represents a prediction. Based on DTs, the work in~\cite{novikova2022federated} uses the Gradient Boosting Decision Tree (GBDT) algorithm~\cite{zhang2019incremental}, which combines multiple DTs for creating a more accurate prediction model. Additionally,~\cite{yuan2021towards} uses the LightGBM classifier~\cite{ke2017lightgbm}, which significantly accelerates the training process compared to GBDT. However, according to the results presented, LightGBM is not deployed in a federated scenario. The same classifier is used in~\cite{dong2021towards} by adjusting its main hyperparameters to mitigate overfitting . Unlike GBDT that builds trees one after another, RF~\cite{breiman2001random} creates several completely independent trees that are then combined to improve the model's accuracy. In this direction,~\cite{aliyu2021blockchain} proposes an RF model for intrusion detection in a vehicular scenario combined with the Fourier transform~\cite{bracewell1986fourier} to extract and transform CAN ID sequences. In~\cite{markovic2022random}, the authors propose an RF approach on which the optimal number of DTs to be used is analyzed considering different datasets. Furthermore,~\cite{driss2022federated} uses RF to ensemble local models that are trained using GRUs.

Other popular ML techniques are represented by LR~\cite{wright1995logistic} and SVM~\cite{hearst1998support}. LR uses a set of variables and a logistic function to model the probability that a sample belongs to a certain class. Based on LR,~\cite{campos2021evaluating} proposes an approach for creating a multi-class classification system using the softmax function on the Ton\_IoT dataset (see Section~\ref{sec:datasets}). A similar approach is used by~\cite{ruzafa2021intrusion}, which adds DP techniques using the same configuration. Also based on a similar environment, the authors in ~\cite{de2022improving} employ LR for binary classification on a different subset of the same dataset. Regarding \cite{shahid2021detecting}, the authors compare a simple LR model with a DNN-based approach , obtaining different results depending on the data distribution between participating nodes. 

For SVM, the goal is to find a hyperplane to separate different classes with the largest possible margin, so that the samples of each class are as far as possible. In this direction,~\cite{hsu2020privacy} proposes the use of a federated version of SVM to detect malware on Android devices. Additionally,~\cite{chen2020intrusion} uses GRU in combination with SVM considering different datasets. In a smart metering environment, ~\cite{mirzaee2021fids} employs a simple DNN architecture that is compared with SVM, KNN, and LR. A comparison of different techniques is also given in~\cite{hei2020trusted}. Specifically, the authors use MLP, DT, SVM and RF in a federated environment where client updates are shared through a blockchain. In~\cite{ferrag2022edge}, the authors propose the Edge-IIoTset dataset (see Section~\ref{sec:datasets}), which is evaluated using different models, including DT, RF, SVM, DNN and K-Nearest Neighbors (KNN). Moreover, the work in~\cite{shukla2021device} compares the use of different supervised approaches, including CNN, RF, LR, and KNN. The same work is extended in~\cite{shukla2022rafel} considering additional models, such as Mobile-Net~\cite{howard2017mobilenets} (a type of CNN for computer vision) and MLP, as well as an approach to reduce the communication cost. 

Finally, one of the most recent ML techniques is represented by transformers~\cite{lin2022survey}, which are the basis of popular tools such as Chat-GPT. A transformer is based on an encoder-decoder architecture and uses a self-attention mechanism to give assign different weights to each part of the input data. Although they are mostly known for their results in Natural Language Processing (NLP), such models can also be considered for intrusion detection. In this direction,~\cite{abdel2021federated} uses transformers in a supervised approach to detect attacks in a vehicular environment through a training process using labelled data. 

\subsection{Unsupervised learning}\label{sec:models_unsupervised}
Based on our analysis, the use of unsupervised learning techniques is not widely considered in the development of FL-enabled IDS. Recall that this type of learning mainly consists of finding patterns from unlabeled data, while some well-known techniques are clustering~\cite{xu2015comprehensive} and dimensionality reduction~\cite{sorzano2014survey}. Clustering groups the data into clusters based on the similarities among samples. One of the main clustering techniques is k-means~\cite{ahmed2020k}, which is an iterative algorithm that partitions the dataset into K distinct non-overlapping subgroups (clusters) where each observation/subject belongs to only one group. The objective is to minimise the intra-cluster distance and maximize the inter-clusters one. It assigns data points to a cluster such that the sum of the squared distance between the data points and the cluster’s centroid is minimized. The less the variation within clusters, the more homogeneous the data points are within the same cluster. In this direction,~\cite{xie2021improved} uses k-means clustering based on cosine distance~\cite{senoussaoui2013study} (instead of the commonest Euclidean distance) to measure the distance between different data. Additionally, the authors use a three-way clustering approach in which clusters are represented by three regions~\cite{wang2018ce3}. The same technique is applied in~\cite{kwon2022anomaly} over a set of benign data originated from different datasets to create several classes of normal traffic to be distributed among the different nodes. Furthermore,~\cite{yadav2021clustering} uses the same approach to detect nodes sending false gradients during the training process.

Beyond clustering, one of the most widely used unsupervised techniques for IDS is autoencoders (AEs)~\cite{bank2020autoencoders}. An AE is an example of an dimensionality reduction technique and is represented as a specific case of NN composed by an input layer (encoder), several hidden layers and an output layer (decoder), where the objective is to learn a compressed representation of certain input data. While the encoder is used for mapping the input data into a hidden representation, the decoder is intended to reconstruct the input data from such representation. With reference to IDS, an AE is trained with benign data to learn the underlying representation of normal traffic. Then, during the intrusion detection phase, a certain threshold is established so that in case the reconstruction error exceeds the threshold, the sample is considered as anomalous. Based on this technique, the goal of~\cite{preuveneers2018chained} was to evaluate different AE configurations by modifying the number of hidden layers and neurons on the CIC-IDS2017 dataset. The authors also integrated a blockchain architecture to support the auditing of models throughout the training process. The use of AE is also considered by~\cite{gaussian2020network}, which describes a system based on the Gaussian Mixture Model (GMM). The proposed scheme represents a federated version of the work described in~\cite{zong2018deep}. Related to this work, the study in~\cite{chen2022privacy} proposes an AE- and GMM-based approach considering data from different domains. Additionally, the authors propose an IDS optimization strategy to cope with the impact of different client data distributions. The contribution in~\cite{yadav2021unsupervised} relies on a combination of AEs and NNs. First, an AE is used to learn from unlabeled data and produce labeled data that is then used as input for a NN. Furthermore,~\cite{rey2022federated} assesses three different configurations of AEs, considering different numbers of hidden layers and neurons per layer on the N-BaIoT dataset. In this case, the configuration with the best results is based on a single hidden layer. Moreover, while the authors in~\cite{alazzam2022federated} use AEs, the configuration of the testbed is not provided.

In addition to the use of simple AEs, a few works employ variations of such technique for the development of FL-enabled IDS. The concept of \textit{stacked AE} is represented as several layers of AEs, where each layer's output is fed as input to the next layer. This AE scheme is considered in~\cite{cetin2019federated}, where the proposed architecture exploits the AWID dataset; therefore, the input layer has a number of neurons equal to the number of attributes in the dataset (74), and four neurons in the output layer correspond to the equal number of classes in AWID. Furthermore,~\cite{jahromi2021deep} proposes the use of stacked AE to detect attacks in industrial environments, although the authors do not provide details about the evaluation carried out. On the other hand,~\cite{wu2022fl} proposes the use of variational AEs (VAE) for anomaly detection in different datasets, including NSL-KDD. A VAE is an extension of the traditional AE that includes a probabilistic model to allow the generation of new data.

Other unsupervised techniques have also been considered for building FL-powered IDSs. Specifically, a recent trend is based on the use of GANs~\cite{gui2021review}. A GAN consists of two NNs called generator and discriminator, which compete during the training process. Specifically, the generator creates synthetic data to deceive the discriminator so that the latter learns to distinguish between synthetic and real data. In this direction,~\cite{taheri2020fed} uses a GAN to generate adversarial samples by considering a CNN architecture for generator and discriminator functions. The authors propose two different algorithms to mitigate the impact of such adversarial samples during the training process. The proposed approach in \cite{khoa2020collaborative} relies on DBN, which can be considered as a structure composed of several layers or processing units, such as AEs. In particular, the authors use Gaussian Binary Restricted Boltzmann Machine (GBRM)~\cite{melchior2017gaussian} to construct a Deep Belief Network (DBN), which is deployed on different gateways acting as FL clients. Also based on DBN,~\cite{xia2022fed_adbn} introduces an attention mechanism for attack detection in smart grid scenarios. Moreover,~\cite{mcosker2021architecture} contributes a rather preliminary work on the possibility of implementing Self-Organizing Maps (SOMs)~\cite{qu2021survey} in a federated architecture. A SOM is represented by a NN in which the neuron weights are adjusted so that data with similar characteristics are grouped in specific areas of the map. Another primal work is is given in~\cite{wettlaufer2021property}, which briefly compares isolation forest~\cite{hariri2019extended} and elliptic envelope~\cite{ashrafuzzaman2020elliptic} techniques on a subset of the CIC-IDS2017 dataset. The former technique is based on DTs representing a subset of the input data, while the latter fits an ellipse to the dataset and evaluates the distance of each sample to the ellipse’s centroid. 

\subsection{Semi-supervised learning}\label{sec:models_semisupervised}
Semi-supervised learning is based on the use of both labeled and unlabeled data to build a certain model. According to our analysis, the use of semi-supervised techniques is scarcely considered in the development of FL-enabled IDS. The work in~\cite{aouedi2022fluids} combines a 3-hidden-layer AE model and a simple NN. In particular, FL nodes (represented by IoT gateways) use an AE on unlabeled data, and the aggregator combines these models by adding layers to the model using labeled data. The same authors proposed a similar approach in~\cite{aouedi2022federated}, which is evaluated considering an industrial environment. Based on the integration of AEs and NNs,~\cite{qin2021federated} employs an approach called ONLAD, which is based on Online Sequential Extreme Learning Machine (OS-ELM)~\cite{liang2006fast}. They evaluated their approach through the NSL-KDD dataset. 

Moreover,~\cite{zhao2022semi} propose a semi-supervised approach based on a CNN model. It is assumed that each node initially is trained with labeled data, so that the traffic information obtained from other unlabeled data is classified using a discriminator considering the previous training. An additional approach is proposed by~\cite{singh2020collaborative}, which makes use of a lightweight CNN architecture and a GAN; the latter is used to generate unobserved data for training purposes. Also based on semi-supervised learning,~\cite{pei2022knowledge} capitalizes on a TL-based approach, where a cloud model is trained with labeled data and this information is used by the end nodes for attack classification on unlabeled data. Particularly, the authors use a graph-based representation called subgraph aggregated capsule network (SACN) to produce a more precise representation of complex attacks.

\subsection{Reinforcement learning} \label{sec:models_reinforcement}
Reinforcement Learning (RL)~\cite{kaelbling1996reinforcement} represents the branch of ML in which an agent learns to maximize a certain numerical reward by interacting with the environment. The approach is based on trial and error so that the agent learns which actions lead to greater reward over time. RL algorithms are mainly classified into: \textit{value-based} techniques, in which a certain value function indicates the quality of an action for a specific state, \textit{policy-based} techniques, in which a policy determines the best action in each state, and \textit{model-based} techniques, where the agent uses a model of the specific environment for the learning process.

Although RL has been widely used in different scenarios, such as the development of robots and control systems, its application in the development of IDS is not so widely considered~\cite{lopez2020application}. In the case of FL, our analysis reveals that only a few works are based on RL techniques. In particular,~\cite{otoum2021federatedreinforcement} uses a value-based approach using Q-learning~\cite{clifton2020q}, in which a table of values is maintained to choose the best action in each state. This table of ``q-values'' is updated using the rewards obtained from the actions taken by the agent. The term \textit{Q} refers to the function used to calculate the reward, considering a specific action and state. The authors also compare their approach with SVM on the CIC-IDS2017 dataset. Furthermore,~\cite{wei2021federated} uses a deep RL technique called Deep Q-Network (DQN)~\cite{mnih2015human}. This technique can be considered a variant of Q-learning, where a NN is used to approximate the Q function. Additionally, it incorporates prior knowledge using the \textit{replay experience} technique. The authors deploy this approach in a 5G network architecture where edge nodes cooperate with a cloud architecture to achieve a federated attack detection scheme. The work in~\cite{nguyen2021federated1} presents a monitoring framework called DeepMonitor, which includes intrusion detection capabilities for DDoS attacks based on Double Deep Q-Network (DDQN)~\cite{van2016deep}. DDQN uses two NNs to select the best action and evaluate the quality of that action respectively in order to reduce the possible overestimation of the reward in DQN.

\subsection{Analysis}
\label{sec:models_analysis}
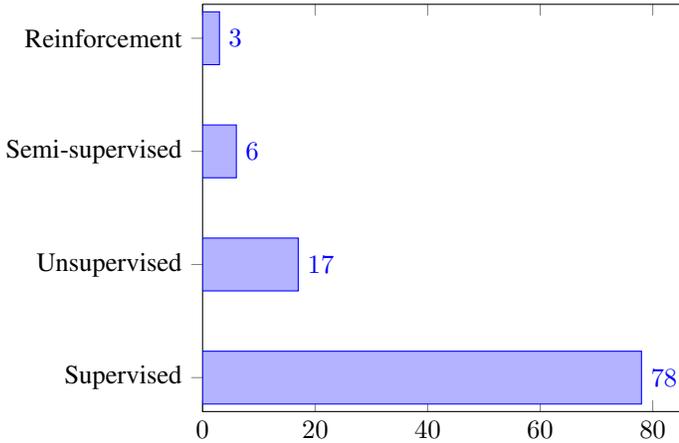
\begin{figure}[H]
\begin{tikzpicture}
\begin{axis}[
xbar,
xmin=0,
bar width=20pt, 
/pgf/number format/.cd,
symbolic y coords={ 
    Supervised,
    Unsupervised,
    Semi-supervised,
    Reinforcement},
height=7cm,
width=8cm,
ytick=data,
     nodes near coords]
\addplot coordinates {
	(78,Supervised) 
	(17,Unsupervised) 
	(6,Semi-supervised) 
	(3,Reinforcement) 
};
\end{axis}
\end{tikzpicture}
\caption{Number of publications according to the ML type}
\label{fig:fl:pubs:by:mltype}
\end{figure}

Based on the previous description, Table \ref{tab:models} provides an analysis on the use of ML techniques in existing FL-enabled IDS approaches. Furthermore, a more summarized perspective can be shown in Figure~\ref{fig:fl:pubs:by:mltype}. According to our analysis, 78 out of the 104 analyzed works (75\%) employ a supervised learning approach. While most of these works provide high performance in detecting different cyberattacks, they are characterized by the need for labeled data, which represents a significant limitation in real-life scenarios. Indeed, considering the dynamism, scale, and heterogeneity of current deployments, this could be impractical in many settings (e.g., IoT) where a large number of devices would need to be equipped with this information to detect potential threats. Furthermore, supervised learning techniques cannot be used to detect previously unknown attacks. This represents yet another significant limitation that is, however, inherent in most FL-enabled IDS approaches. 

Regarding the considered techniques, CNN represents the most widely used approach (33 out of 78) either uniquely or in combination with other techniques, such as MLP~\cite{liu2022intrusion} or GRU~\cite{li2020deepfed}. An additional important aspect to highlight is that a significant portion of the analyzed works describes the use of NNs (16) but not the specific type, even though some of them specify the number of layers and activation functions employed. Another notable observation is that while a substantial number of articles employ datasets related to IoT, most works use models without considering the inherent constraints of IoT devices and networks. Indeed, only one work~\cite{qin2020line} uses BNNs, which could be considered a promising approach for IoT scenarios. Additionally, we found that other promising options in terms of lightweightness and low complexity, such as RaNN~\cite{nakip2021mirai}, are not considered in the analyzed works.

Furthermore, while unsupervised and semi-supervised schemes mitigate some of the limitations of supervised learning, we found that only a small portion of the surveyed works considers these techniques. Specifically, FL-enabled IDS using unsupervised techniques represent 16\% (17 out of 104), while only a 5.8\% of them consider semi-supervised approaches (6 out of 104). 

In the case of unsupervised techniques, the most widely used approach is AE (9). Although AE is a widely known technique, it is worth noting that most of the surveyed works use simple or stacked AEs, and only one exploits VAEs~\cite{wu2022fl}, which are considered a promising approach for intrusion detection. Furthermore, we noticed that the use of GANs has been barely considered so far for generating attack data that could help improve the effectiveness of developed approaches. Due to the recent advances in generative models, it is very likely that the use of these techniques will expand in the coming years for the development of IDS.

Finally, only 2.9\% (3 out of 104) of the analyzed works employ an RL-based approach. Despite the potential advantages of RL in terms of adapting to possible environmental changes for attack detection, one of the commonly accepted challenges is the definition of a suitable reward function in the context of IDS~\cite{bertino2021ai}. In this regard, some recent works~\cite{lopez2020application} propose replacing the environment with a sampling function of recorded training intrusions. We believe that the use of RL approaches needs to be further investigated in the field of IDS, as well as variants based on Inverse Reinforcement Learning (IRL)~\cite{arora2021survey} that can aid in better understanding the cybersecurity behavior of devices and their users. As it is widely known, in many real-world scenarios, the combination of different approaches and techniques is likely to enhance the effectiveness of intrusion detection. One noteworthy aspect of our analysis of FL-enabled IDS development is the lack of schemes that consider different models to be used at each FL client. Although this may be feasible in small-scale controlled settings, enforcing the use of the same model for all FL clients is not a realistic approach for environments where different organizations, such as companies or even Security Operation Centers (SOCs) could be acting as FL clients during the training process. 

\begin{table}[ht]
\centering
\begin{tabular}{C{1.8cm}C{2cm}C{3.7cm}}
\hline
\textbf{ML approach} & \textbf{Technique} & \textbf{Works using this technique} \\
\hline \hline

\multirow{31}{*}{\textbf{Supervised}} 
& Based on NN   & \cite{zhao2019multi,zhao2020network, mowla2019federated, rahman2020internet, mirzaee2021fids, zakariyya2021memory, qin2021fnel, lalouani2021robust, otoum2021federated, shahid2021detecting, hao2021secure, popoola2021federated, agrawal2021temporal, attota2021ensemble, LocKedge, uprety2021privacy, saadat2021hierarchical, kelli2021ids, wang2022feco, lalouani2022robust, rehman2022federated, otoum2022feasibility, ferrag2022edge,popoola2021federatedzero,friha2022felids} \\

\cline{2-3}

& GRU           & \cite{nguyen2019diot, chen2020intrusion, li2021fids, vy2021federated, li2020deepfed, driss2022federated, kumar2021pefl,mothukuri2021federated,tang2022federated,tahir2021experience} \\

\cline{2-3}

& CNN & \cite{li2020distributed, sun2020intrusion, zhao2020intelligent, zhou2020federated, fan2020iotdefender, duy2021federated, chen2021trust, ayed2021federated, hao2021secure, vy2021federated, man2021intelligent, sun2020adaptive, shi2021data, li2020deepfed,shukla2021device, pasdar2022train, lv2022ddos, zhang2022grid, luo2022federation, yang2022federation, liu2022intrusion,ibitoye2022differentially,shukla2022rafel,friha2022felids,tabassum2022fedgan,zhang2022secfednids,sun2022hierarchical} \\

\cline{2-3}

& SVM & \cite{chen2020intrusion, hsu2020privacy, hei2020trusted, kumar2021security,ferrag2022edge} \\

\cline{2-3}

& MLP & \cite{al2020federated, hei2020trusted, dong2021towards, liu2021blockchain, li2020deepfed, liu2022intrusion,shukla2022rafel,sun2022hierarchical,rey2022federated} \\

\cline{2-3}

& BNN & \cite{qin2020line} \\

\cline{2-3}

& LSTM & \cite{zhao2020intelligent, hao2021secure, zhang2022grid, su2022detection, singh2022dew,mothukuri2021federated, zhao2021federated} \\

\cline{2-3}

& DT & \cite{hei2020trusted, yuan2021towards, dong2021towards,ferrag2022edge,novikova2022federated} \\

\cline{2-3}

& RF & \cite{hei2020trusted, aliyu2021blockchain, kumar2021security, shukla2021device, markovic2022random, driss2022federated,ferrag2022edge,shukla2022rafel}\\

\cline{2-3}

& RNN & \cite{zhang2021flddos,friha2022felids} \\

\cline{2-3}

& LR & \cite{shahid2021detecting, ruzafa2021intrusion, shukla2021device,campos2021evaluating,de2022improving, cheng2022federated,shukla2022rafel} \\

\cline{2-3}

& NB & \cite{kumar2021security} \\

\cline{2-3}

& KNN & \cite{shukla2021device,ferrag2022edge} \\

\cline{2-3}

& Transformers & \cite{abdel2021federated} \\

\hline

\multirow{8}{*}{\textbf{Unsupervised}} 
& AE & \cite{preuveneers2018chained,cetin2019federated,gaussian2020network,jahromi2021deep,yadav2021unsupervised,alazzam2022federated,rey2022federated,chen2022privacy,wu2022fl} \\
\cline{2-3}
& GAN & \cite{taheri2020fed} \\
\cline{2-3}
& DBN & \cite{khoa2020collaborative,xia2022fed_adbn} \\
\cline{2-3}
& Isolation Forest & \cite{wettlaufer2021property} \\
\cline{2-3}
& Elliptic Envelope & \cite{wettlaufer2021property} \\
\cline{2-3}
& SOM & \cite{mcosker2021architecture} \\
\cline{2-3}
& K-means & \cite{xie2021improved, yadav2021clustering,kwon2022anomaly} \\
\hline

\multirow{4}{*}{\textbf{Semi-supervised}} 
& AE-NN & \cite{qin2021federated,aouedi2022fluids,aouedi2022federated} \\ \cline{2-3}
& GAN-CNN & \cite{singh2020collaborative} \\
\cline{2-3}
& SACN & \cite{pei2022knowledge} \\
\cline{2-3}
& CNN & \cite{zhao2022semi}\\
\hline

\multirow{3}{*}{\textbf{Reinforcement}} 
& Q-learning & \cite{otoum2021federatedreinforcement}\\
\cline{2-3}
& DQN/DDQN & \cite{wei2021federated,nguyen2021federated1} \\
\cline{2-3}
& SOM & \cite{nguyen2021federated1}\\
\hline
\end{tabular}
\caption{Analysis on the use of ML techniques in FL-enabled IDS approaches}
\label{tab:models}
\end{table}

\section{Aggregation methods}
\label{sec:aggregationMethods}

As already described in Section~\ref{S:Background}, aggregation methods are a core part of FL. They determine how the local model updates provided by FL clients are combined, resulting in a new aggregated model in each training round. This section details the main aggregation functions used in current literature, indicating the advantages and disadvantages of each approach, and the use of such techniques in existing FL-enabled IDS approaches

\subsection{FedAvg}
\label{S:FedAvg}

FedAvg was initially proposed in the original FL paper~\cite{mcmahan2017communication}. This is the most straightforward aggregation function, in which each client's local update is aggregated on the server by calculating the average, depending on their data length. Specifically, the FedAvg objective function is calculated using equation~\ref{eq:Fed+1}, where $W = (w_i)_{i=1}^n$ are the weights of the server's model, $W^k= (w_i^k)_{i=1}^n$ the weights of client $k$,  $f_k$ is the loss function of client $k$, and $D_k$ and $D$ represent the length of the client $k$ dataset and the total length of the clients, respectively.

\begin{equation} \label{eq:Fed+1}
    \text{min }\left[F(W) = \frac{D_k}{D}\sum f_k(W^k)\right] \text{ subject to } w_i = w_i^k
\end{equation}

Based on this function, the weights are updated according to equation~\ref{eq:FedAvg}.

\begin{equation} \label{eq:FedAvg}
    W = \frac{D_k}{D}\sum W^k
\end{equation}

The operation of FedAvg is determined by the number of participating clients $C$ in each training round, the number of local training epochs $E$ that each client performs on its data, as well as the size of the local minibatch $B$ used for local training. Actually, FedAvg can be considered a generalization of FedSGD, which is based on Stochastic Gradient Descent (SGD), where $B = \infty$ and $E = 1$. Therefore, in FedSGD, each client is trained with the complete dataset and only performs one local training epoch in each round.

\subsection{FedPAQ}

FedPAQ~\cite{reisizadeh2020fedpaq}  follows a similar aggregation approach to FedAvg, while providing a higher degree of efficiency in environments with certain computing and network constraints. In particular, FedPAQ adds three main features: periodic averaging, partial device participation, and quantized message-passing. First, periodic averaging means that aggregation is performed at certain intervals, allowing each client to perform multiple epochs of local training before sending its updates to the server. Second, it allows only a subset of nodes to participate in each training round, reducing network overhead and improving scalability. Third, FedPAQ uses quantization to reduce the size of models sent to the server in each training round. Specifically, each node employs quantized operators~\cite{gholami2021survey} on the difference between the local model created by the client and the global model received during a training round. Overall, the server weights are updated based on equation~\ref{eq:AF2} where $W_r$ and $W_r^k$ represent the weights of the global model and client $k$ at round $r$, respectively, $S_n$ is the subset of the $n$ clients chosen in that round, and $Q$ is the quantized function. One example of this type of function $Q$ can be found in~\cite{alistarh2017qsgd}. 

\begin{equation} \label{eq:AF2}
    W_{r+1} = W_r + \frac{1}{|S_n|} \sum_{k\in S_n} Q(W^k_r-W_r)
\end{equation}

\subsection{FedMA}

FedMA~\cite{wang2020federated} is based on the principles of Probabilistic Federated Neural Matching (PFNM)~\cite{yurochkin2019bayesian} to address the issue of permutation invariance of NN parameters. Unlike FedAvg, which performs coordinate-wise averaging, FedMA conducts a matching process of the clients' NNs to perform layer-wise averaging. In addition to improving performance, this reduces the required communication rounds. In FedMA, after each client trains its model, the server sets the first layer's weights using one-layer matching, i.e., finding the permutations $\Pi_{k,l, \cdot}$, where $k$ is the client, and $l$ represents the layer, which minimizes equation~\ref{eq:minPerm}. Note that $\theta_i$ are the global model's weights, and $c(\cdot,\cdot)$ is a similarity function, such as the Euclidean distance. 

\begin{equation}
    \sum_{i=1}^L \sum_{k,l} min_{\theta_i} \Pi_{k,l,i}c(w_{kl}, \theta_i)
    \label{eq:minPerm}
\end{equation}

The weights are set layer by layer; once the optimized permutations are calculated, the server averages the weights of the first layer. The server then sends these weights to the clients, which in turn train all consecutive layers on their datasets, keeping the matched federated layers frozen. This procedure is then repeated up to the last layer, in which a weighted averaging is calculated. This procedure requires a number of communication rounds equal to the number of layers in the network. The equation of FedMA is given in equation~\ref{eq:4}, where $\Pi^T_k$ are the solution of the matched average of equation~\ref{eq:minPerm}.

\begin{equation}
    W_n = \frac{1}{K}\sum_k W_{k,n}\Pi^T_k
    \label{eq:4}
\end{equation}

\subsection{FedProx}

FedProx~\cite{fedprox} is a variant of FedAvg, whose main goal to address system and statistical heterogeneity of common FL settings. Especially, FedProx adds a \textit{proximal term} to restrict the local models to be closer to the central one by limiting the impact of local updates. Furthermore, it also addresses device heterogeneity to deal with scenarios involving devices with different computation capabilities. In particular, the approach is intended to address the problem of \textit{stragglers}, which represent devices with low capabilities that could have a negative impact on system convergence. The objective function of FedProx is calculated as shown in equation~\ref{eq:fedprox}, where $F_k$ is the loss function of client $k$, $W^k$ and $W$ are the client’s weights and aggregated weights, respectively, and $\mu$ is the proximal term chosen by the user. 

\begin{equation}
    min\, F_k(W^k) + \frac{\mu}{2}||W^k - W||^2
    \label{eq:fedprox}
\end{equation}

\subsection{Fed+}
\label{S:Fed+}

Fed+~\cite{yu2021fed} (or Fedplus) is a set of algorithms developed to handle data heterogeneity, and in particular, non-IID data. Fed+ takes the loss function of FedAvg and adds a penalty function on each client to remove the restriction that all clients' weights converge to the same point. That is, with reference to equation~\ref{eq:Fed+1}, the Fed+ objective function is calculated as in equation~\ref{eq:Fed+2}, where $\mu>0$ is a penalty constant chosen by the user, $A$ is an aggregation function (e.g., the average) of the clients' weights, and $B(\cdot,\cdot)$ is a distance function that penalizes the deviation of a local model $W^k$ from the output of $A(\cdot)$.

\begin{equation}
    \text{min }\left[F_\mu(W) = \frac{1}{D}\sum [f_k(W^k) +\mu B(W^k,A(W))]\right],
    \label{eq:Fed+2}
\end{equation}

Furthermore, with reference to equation~\ref{eq:Fed+2}, the corresponding weights are updated by using equation~\ref{eq:Fed+weights}, where $r$ is the round, $\nu$ the learning rate, $\theta = \frac{1}{1+\nu\mu}$ a constant that controls the degree of regularization, i.e., how the local weights are influenced by the distant function, and $Z^k$ the result of $B(W^k,\Tilde{W})$. This weighted average between their own weights and the one provided by function $Z$ shown in equation~\ref{eq:Fed+weights} leads to each client having their own model through the federated process. This removes the restriction that all clients' weights converge to the same point.

\begin{equation}
    W^k_{r+1} \longleftarrow \theta[W^k_r - \nu \nabla f_k(W^k_r)] + (1-\theta)Z^k,
    \label{eq:Fed+weights}
\end{equation}

\subsection{Others}

While previous approaches are widely considered in FL settings for different use cases and scenarios, a significant number of other aggregation functions have been also defined in the literature. The following subsections, provide a brief overview of these techniques for reasons of completeness. 

\subsubsection{Turbo-Aggregate}
\label{SS:Turbo-Aggregate}

Turbo-Aggregate~\cite{so2021turbo} stems from the idea of securing the communication of equation~\ref{eq:Fed+1} between clients and server through secure aggregation techniques used in~\cite{bonawitz2017practical}. The clients are separated into several groups, and model updates are shared among groups by using an additive secret sharing approach. Secret sharing is used to mask each local model by adding randomness in a way that can be cancelled out once the models are aggregated. In particular, the weights of each cluster are encoded and sent to the next cluster, where it will be decoded and aggregated with the weights of that cluster. The process is repeated for all the clusters. Finally, the last cluster sends the final aggregated model to the server, where it reconstructs the encoded model for obtaining the actual weights and sends them to the clients, where a new round starts. During the process, Turbo-Aggregate uses the Lagrange coding~\cite{yu2019lagrange} for adding redundancies to recover the data of dropped or delayed clients. Precisely, this is achieved by injecting redundancy via Lagrange polynomial so that the added redundancy can be exploited to reconstruct the aggregated model amidst potential dropouts. Turbo-Aggregate can be used both in a centralized (through a server) and decentralized (through direct interaction among devices) communication architecture.

\subsubsection{FedCD}

Similar to FedAvg, FedCD~\cite{kopparapu2020fedcd} begins with a global model on a centralized server that all devices are updated with. In each round, clients train different models and calculate a score related to the performance of these models. To face non-IID scenarios, FedCD, starting with one model in certain rounds, clones the models with high performance based on its score, deleting the ones with low scores. Finally, the model with the highest score is selected. However, this cloning can overload the devices' resources, thus delaying the whole process.

\subsubsection{Krum}
\label{SS:Krum}

Krum aggregation~\cite{blanchard2017machine} is intended to deal with scenarios including $f$ malicious peers. It is defined by $Kr(W_1,\dots,W_n)=W_k$, where certain weights $W_k$ are selected from the clients' weights $W_i$ to minimize their distance between its $n-f-2$ closest neighbours, where $n$ is the number of clients. Such weights $W_k$ are sent to the server at every round. However, the number of malicious clients should be known in advance, which makes Krum aggregation impractical.

\subsubsection{SAFA}

Semi-Asynchronous Federated Averaging (SAFA)~\cite{wu2020safa} was intended to cope with scenarios characterized by extremely weak communication conditions, where some clients may be disconnected depending on their capabilities. To this end, it is based on an asynchronous aggregation approach to mitigate the impact of stragglers. Furthermore, the authors propose a cache structure in the cloud to reduce the cost of communication, as well as a post-training client selection approach to improve convergence.

\subsubsection{CDA-FedAvg}

Concept-Drift-Aware Federated Averaging (CDA-FedAvg)~\cite{casado2022concept} applies FedAvg to concept drifts, that is, situations where data could change due to a certain event or circumstance~\cite{lu2018learning}. In this approach, the server aggregates weights asynchronously, namely, there is no predefined sequence to upload the global model. On the other hand, the clients have two different data storages (short and long term) to check if a drift occurs. If so, additional training rounds are performed to reflect the changes derived from the concept drift. 

\subsubsection{SecureBoost}
\label{SS:SecureBoost}

SecureBoost~\cite{cheng2021secureboost} is a framework for GBDT on vertically-partitioned data. SecureBoost enhances clients' privacy by sharing a GBDT model through multi-party data under privacy constraints. Specifically, SecureBoost determines inter-database intersections with a privacy-preserving protocol based on entity alignment technique~\cite{liang2004privacy}, XGBoost framework~\cite{chen2016xgboost}, and Paillier encryption~\cite{paillier1999public}. These three techniques are employed by clients to calculate the weights at every round prior to sending them to the server.

\subsubsection{CMFL}

Communication-Mitigated Federated Learning (CMFL)~\cite{luping2019cmfl} follows the idea of removing irrelevant updates from certain clients. To calculate the level of relevance of a certain model update, each client is forced to compare its local update with the global model by calculating the percentage of parameters with a different sign (positive or negative). The higher the percentage , the more irrelevant the update, so it can be discarded by considering a certain threshold. 

\subsubsection{DW-FedAvg}

Dynamic weighted federated averaging (DW-FedAvg)~\cite{chaudhuri2023dynamic} is based on FedAvg, with the goal of rewarding best performing clients. In particular, DW-FedAvg updates the weights following the equation $W = \sum \beta_k \cdot W^k$, where $\beta_k$ represents the adjusting factor associated with clients' weights. These $\beta_k$ are updated based on the local performance of each client. Initially, $\beta_k$ are initialized as the $\frac{1}{D_k}$, where $D_k$ is the length of the dataset of client $k$. Then, if client's accuracy is higher than the average accuracy, $\beta_k=\beta_k + \beta_k\cdot\alpha$, where $\alpha$ is a reward/penalty factor chosen by the user. In case the user's accuracy is lower than the average accuracy, $\beta_k=\beta_k - \beta_k\cdot\alpha$. Once $\beta_k$ are updated, these values are rescaled by $\beta_k = \frac{\beta_k}{\sum_k \beta_k}$. Finally, this process is repeated throughout all training rounds.

\subsection{Analysis}

After overviewing the current landscape of aggregation approaches in FL, Table~\ref{tab:agg-fun} provides a concise description of the key characteristics associated to the most relevant approaches, as well as the use of such techniques in existing FL-enabled IDS approaches. As previously described, the choice of the FL aggregation approach is crucial for the development of a robust and secure FL-enabled IDS. Although the initially proposed aggregation approach (i.e., FedAvg) relies on averaging client updates in each training round, subsequent studies have demonstrated the limitations of this approach in scenarios with a high degree of data and device heterogeneity, as well as in terms of security and efficiency~\cite{nilsson2018performance, zhu2021federated}. As a result, additional aggregation approaches were proposed to address (at least, partially) these limitations. For example, approaches based on FedCD, CMFL, FedMA, FedProx, and Fed+ deal with data and device heterogeneity for mitigating the impact derived from non-IID data distributions and stragglers. While FedMA seems to be oriented towards specific ML models, Fed+ offers a significantly flexible approach that can be considered in different aggregation functions, such as mean or median. Moreover, security aspects are addressed by approaches like Turbo-Aggregate, SecureBoost, and Krum, given in subsections~\ref{SS:Turbo-Aggregate},~\ref{SS:SecureBoost}, and~\ref{SS:Krum}, respectively. While the first two are more related to privacy aspects, Krum presents an approach that may be of interest for designing more robust FL systems, where some updates are discarded and therefore not aggregated. Additionally, depending on the number of clients involved, as well as the number of training rounds and their frequency, the computation and communication overhead of the FL aggregation process can be unaffordable under certain circumstances. In this regard, some of the described approaches, including FedPAQ or SAFA, are intended to improve the performance of the training process.

\begin{figure}[!htb]
\begin{tikzpicture}
\begin{axis}[
xbar,
xmin=0,
bar width=12pt, 
/pgf/number format/.cd,
symbolic y coords={ 
    Other,
    DW-FedAvg,
    FedPAQ,
    FedMA,
    Turbo-Aggregate,
    FedCD,
    SAFA,
    CDA-FedAvg,
    SecureBoost,
    CMFL,
    Krum,
    FedProx,
    Fed+,
    FedAvg},
height=10cm,
width=7cm,
ytick=data,
     nodes near coords]
\addplot coordinates {
	(0,FedPAQ) 
	(0,FedMA) 
	(3,FedProx)
    (3,Fed+) 
    (0,Turbo-Aggregate)
    (0,FedCD)
    (1,Krum)
    (0,SAFA)
    (0,CDA-FedAvg)
    (1,SecureBoost)
    (1,CMFL)
    (0,DW-FedAvg)
    (10,Other)
    (87,FedAvg)
};
\end{axis}
\end{tikzpicture}
\caption{Number of publications according to the use of a certain aggregation method}
\label{fig:fl:aggregation:usage}
\end{figure}
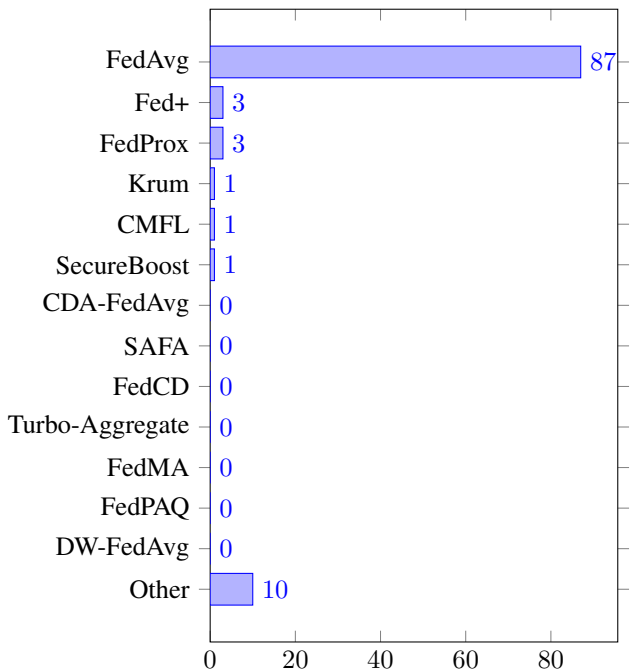

Beyond the general analysis of aggregation approaches, we examine the use of different methods in the context of FL-enabled IDS. To this end, Figure~\ref{fig:fl:aggregation:usage} offers a summarized overview of the usage of each aggregation function in existing literature. Based on our literature review, we found that the vast majority of approaches (87/104 or 83.7\%) use FedAvg or they are mostly based on it as the aggregation mechanism. One notable observation is that some works do not explicitly mention which aggregation function is employed, indicating that this aspect may currently be sufficiently considered in the current literature. In cases where it is not explicitly indicated, we assume that FedAvg is the employed method 

The convergence issues stemming from the use of FedAvg have been partially addressed by some approaches using alternative methods. In this regard, three works (2.88\%) consider FedProx-based approaches. In particular,~\cite{su2022detection} proposes a FedProx-based aggregation mechanism where nodes are grouped such that, in each training round, certain nodes are selected according to the descending gradient order. Additionally,~\cite{shukla2021device} and~\cite{shukla2022rafel} propose a robust version of FedProx where nodes suspected of producing fake updates are removed from the process. The use of Fed+ is also considered by three works (2.88\%); ~\cite{popoola2021federated} compares the Fed+ approach with other commonly used aggregation functions, such as FedAvg and Coordinate Median (CM). In the case of~\cite{campos2021evaluating}, the authors provide an evaluation of the impact of Fed+ compared to FedAvg, considering a non-IID scenario. Similarly,~\cite{ruzafa2021intrusion} demonstrates that the use of Fed+ substantially improves FedAvg in a scenario with non-IID data distribution where DP techniques are also employed. Furthermore, the Krum approach is used by~\cite{taheri2020fed} to evaluate the impact of different poisoning attacks, considering various aggregation functions. Particularly, the authors use Krum to assess the robustness of a malware detection system considering a GAN that generates adversarial samples. In the same direction,~\cite{rey2022federated} analyzes the impact of different aggregation functions, considering data and model poisoning attacks in their approach. These functions include versions of mean and median (coordinate-wise and trimmed mean~\cite{yin2018byzantine}) specifically designed to discard potential outliers. Furthermore, the work in~\cite{novikova2022federated} uses SecureBoost as an aggregation approach to provide additional privacy features, considering local models based on GBDT and vertical FL scenarios.

No less important, as shown in Figure~\ref{fig:fl:aggregation:usage}, a few works make use of ``other'' aggregation approaches. This category includes works proposing an aggregation scheme as part of their contributions. Among them,~\cite{chen2020intrusion} proposes FedAGRU as an improvement to FedAvg to reduce communication rounds while ensuring convergence. Inspired by the previous approach,~\cite{man2021intelligent} proposes an aggregation function (FedACNN) where the updates from each client are given different weights depending on their contribution to the global model convergence. In the case of~\cite{li2021fids}, the authors design an aggregation function based on FedAvg where certain parameters are dynamically adjusted. The work in~\cite{zhang2021flddos} introduces a hierarchical aggregation approach based on k-means to alleviate the data imbalance problem. Moreover,~\cite{agrawal2021temporal} proposes an aggregation function (i.e., a temporal weighted averaging) that takes into account the heterogeneity of clients participating in the FL training process. Specifically, it considers the time required by clients to perform local training. The authors in~\cite{liu2022intrusion} propose FedBatch, an adaptive aggregation approach where the aggregator does not need to wait to receive the aggregation from all clients in each training round. This proposal intends to address the straggler problem, especially in scenarios with interconnection issues between FL clients and the aggregator. Finally,~\cite{zhao2022semi} proposes an aggregation function based on the concepts of Federated Distillation (FD) to reduce communication overhead and improve performance in non-IID data scenarios. In spite of such efforts, we note that the existing literature around FL-enabled IDS lacks of suitable analysis about the implications associated to data/device heterogeneity and robustness in the development and deployment of such systems. 
 
\begin{table*}[]
\centering
\begin{tabular}{l C{3.5cm} C{3.5cm} C{3.5cm} C{5.5cm}}
\hline
\textbf{Technique} & \textbf{Core}  & \textbf{Advantages} & \textbf{Disadvantages} & \textbf{Works using this technique} \\ \hline \hline

FedAvg & Based on the weighted average of the updated weights provided by clients & It is widely used due to its low level of complexity  & It poses convergence issues in FL settings with non-IID data distributions & \cite{nguyen2019diot, zhao2019multi, li2020distributed, chen2020intrusion, zhao2020network, al2020federated, mowla2019federated, sun2020intrusion, hsu2020privacy, zhao2020intelligent, rahman2020internet, zhou2020federated, fan2020iotdefender, hei2020trusted, mirzaee2021fids, zakariyya2021memory, qin2021fnel, zhang2021flddos, li2021fids,  yuan2021towards, lalouani2021robust, duy2021federated, otoum2021federated, shahid2021detecting, dong2021towards, chen2021trust, ayed2021federated, hao2021secure, vy2021federated, popoola2021federated, ruzafa2021intrusion, man2021intelligent, attota2021ensemble, sun2020adaptive, aliyu2021blockchain, LocKedge, uprety2021privacy, shi2021data, liu2021blockchain,saadat2021hierarchical, li2020deepfed, kelli2021ids, kumar2021security, pasdar2022train, lv2022ddos, zhang2022grid, su2022detection, singh2022dew, luo2022federation, yang2022federation, markovic2022random, wang2022feco, driss2022federated, lalouani2022robust, rehman2022federated,otoum2022feasibility,ferrag2022edge,kumar2021pefl,campos2021evaluating,mothukuri2021federated,de2022improving,abdel2021federated,zhao2021federated,popoola2021federatedzero,ibitoye2022differentially,tang2022federated,friha2022felids,tabassum2022fedgan,zhang2022secfednids,tahir2021experience,sun2022hierarchical,preuveneers2018chained,cetin2019federated,taheri2020fed,gaussian2020network, khoa2020collaborative,jahromi2021deep,yadav2021unsupervised,wettlaufer2021property,mcosker2021architecture, xie2021improved,yadav2021clustering,alazzam2022federated,xia2022fed_adbn,rey2022federated,kwon2022anomaly,chen2022privacy,wu2022fl,qin2021federated,singh2020collaborative,aouedi2022fluids,aouedi2022federated,pei2022knowledge,zhao2022semi, otoum2021federatedreinforcement, wei2021federated,nguyen2021federated1}\\ \hline

FedPAQ  & It is  based on averaging model updates but providing a higher degree of efficiency through periodic averaging, partial device participation and quantized message-passing & It reduces communication and computation overhead  & It requires additional functionality on both aggregator and clients, and still poses issues with non-IID settings & -\\ \hline

FedMA  & It is based on the neurons' permutation invariance through a matching process of the clients' NNs to perform layer-wise averaging & It adapts to global model size and data heterogeneity and improves convergence, while requiring fewer training rounds & The computation of the permutation matrix may increase the computation time, and it is specific for CNN and LSTM & -\\ \hline

FedProx & It adds a proximal term to limit the impact of the different local updates & It addresses both data heterogeneity considering non-IID settings and device heterogeneity & It increases computation requirements and its performance largely depends on the appropriate election of the  proximal term & \cite{shukla2021device, su2022detection,shukla2022rafel}\\ \hline

Fed+ & Similar to the previous proximal term, a penalty constant is also added but different aggregation functions can be employed (beyond average) & It adds an increasing level of flexibility by considering different aggregation functions beyond average  & A higher flexibility comes with a cost in complexity, and the performance also depends on the right choice of the penalty constant & \cite{popoola2021federated, ruzafa2021intrusion,campos2021evaluating} \\ \hline









\end{tabular}
\caption{Comparison among main aggregation methods given in subsections ~\ref{S:FedAvg} to~\ref{S:Fed+}}
\label{tab:agg-fun}
\end{table*}

\section{Evaluation of FL-enabled IDS approaches}
\label{sec:evaluation}

This section describes the main aspects affecting the evaluation of FL-based systems, as well as analyzes their impact in the scope of FL-enabled IDS. For this purpose, we describe the most commonly used evaluation metrics in ML and FL, and provide an overview of the main existing FL implementations. Furthermore, we analyze the usage of such metrics and FL implementations in the current literature of FL-enabled IDS. The main goal is to understand how IDS models are evaluated and compared in the FL ecosystem, taking into account the unique characteristics of FL and the associated challenges.

\subsection{Evaluation metrics}
\label{sec:metrics}
Evaluating classification scenarios involves the use of various metrics to assess the performance of the model. These metrics provide insights into how well a given model is performing and aid in comparing different models or tuning their hyperparameters. 

\subsubsection{Binary classification metrics}

\begin{itemize}
    \item True Positive (TP): number of positive instances correctly classified as positive
    \item True Negative (TN): number of negative instances correctly classified as negative
    \item False Positive (FP): number of negative instances incorrectly classified as positive 
    \item False Negative (FN): number of positive instances incorrectly classified as negative
    \item False Positive Rate (FPR): ratio of positive cases incorrectly classified as negative: 

            \[\text{False Positive Rate} = \frac{FP}{FP+TN} \]

    \item False Negative Rate (FNR): ratio of positive cases incorrectly classified as negative

            \[\text{False Negative Rate} = \frac{FN}{FN+TP} \] 

\end{itemize}

\subsubsection{General classification metrics}
\begin{itemize}

    \item Accuracy measures the overall correctness of the model's predictions and is calculated as the ratio of correct predictions to the total number of predictions. 
                \[ \text{Accuracy} = \frac{{TP + TN}}{{TP + TN + FP + FN}} \] 

    \item Precision is a measure that quantifies the accuracy of positive predictions made by the model. Precision is calculated by dividing the number of true positives by the sum of true positives and false positives. 
            \[ \text{Precision} = \frac{{TP}}{{TP + FP}} \]

    \item TP Rate (TPR) (also known as sensitivity, recall, or hit rate) is the ratio of true positives to the total number of actual positive instances.  It was used in 61\% of the surveyed studies. 

            \[\text{True Positive Rate / recall}  =  \frac{TP}{TP+FN} \]

    \item False Omission Rate (FOR) is a performance metric used in authentication tasks to assess the likelihood of incorrectly rejecting genuine users or samples. It measures the rate at which the system erroneously classifies genuine instances as impostors or unauthorized users. It appeared only in 1\% of the studies.

            \[\text{FOR} = \frac{FN}{FN+TN} \]

    \item F1-score combines the precision and TPR into a single value, providing a balanced measure of a model's performance. The F1-score is commonly used when there is an imbalance between the positive and negative classes or when both precision and TPR are equally important. 

            \[ \text{F1-score} = 2*\frac{TRP * Precision}{TRP + Precision} \]

    \item The Area Under the Curve (AUC) quantifies the overall quality of the model's predictions across different thresholds. To calculate the AUC, the ROC curve is generated by plotting the TPR on the y-axis against the FPR on the x-axis at different classification thresholds. The AUC is then computed by calculating the integral (area) under this curve; it ranges from 0 to 1. An AUC of 1 represents a perfect classifier that achieves a TPR of 1 while maintaining an FPR of 0. An AUC of 0.5 indicates a classifier that performs no better than random guessing. 

\item Matthew Correlation Coefficient (MCC) takes into account all four values in the confusion matrix (TP, TN, FP, FN):

\begin{equation} \label{eq:MCC}
	\small{MCC=\frac{TP\times TN - FN\times FP}{\sqrt{(TP+FP)(TP+FN)(TN+FP)(TN+FN)}}}
\end{equation}

MCC values have a range between -1 and 1. A value close to 1 means that both classes are predicted well, even if one class is disproportionately under- or over- represented. If $MCC=1$ then $FP = FN = 0$ which means that the model is perfect. When $MCC=-1$ then $TP = TN = 0$, indicating a negative correlation; consequently, the model produces the opposite of the true value. If $MCC=0$, it means that the model is totally random.

\item Cohen Kappa Score (CKS) takes values between -1 and 1, and is also used to measure the reliability of the classifier's predictions; nevertheless, it was not found in any of the reviewed works. Practically, CKS considers the possibility of agreeing by chance, and measures the number of predictions it makes that cannot be explained by a random guess. CKS is defined by the formula:

\begin{equation}
	\small{CKS=\frac{a-p_c}{1-p_c},}
\end{equation}

where $a$ is the accuracy of the model, and $p_c$ is the probability of having the correct prediction by chance. The calculation of $p_c$ is based on the confusion matrix. Let $p$ be the multiplication of the total sum of the elements of a column divided by the total number of elements, and the total sum of the elements of a row divided by the total number of elements. Then, $p_c$ is the sum of the previous $p$ applied to each column. 

\end{itemize}

\subsubsection{Regression metrics}

Unlike previous indicators, regression metrics are employed to evaluate the predictions of regression models. The most widespread metrics are:

\begin{itemize}

    \item Mean Square Error (MSE) is the average squared difference between the predicted values and the true values in a regression problem.
   \[\text{MSE} = \frac{1}{n} \sum_{i=1}^{n} (y_{i} - \hat{y}_{i})^2\]
    
    \item Mean Absolute Error (MAE) is used to calculate the average absolute difference between the predicted and actual values.
    
    \[\text{MAE} = \frac{1}{n} \sum_{i=1}^{n} |y_{i} - \hat{y}_{i}| \]
   
    \item Mean Absolute Percentage Error (MAPE) calculates the average percentage difference between the predicted and actual values.

    \[\text{MAPE} = \frac{1}{n} \sum_{i=1}^{n} \left|\frac{y_{i} - \hat{y}_{i}}{y_{i}}\right| \times 100 \]

    \item Root Mean Squared Error (RMSE) is used to measure the average magnitude of errors between the predicted and actual values, taking into account both the magnitude and the direction of the errors. RMSE is calculated by taking the square root of the average of the squared differences between the predicted and actual values.

    \[\text{RMSE} = \sqrt{MSE} \]
    
    \item Pearson correlation coefficient measures the strength and direction of the linear association between two variables, ranging from -1 (perfect negative correlation) to 1 (perfect positive correlation), with 0 indicating no correlation. 
\[\text{r} = \frac{{\sum_{i=1}^{n} (x_i - \bar{x})(y_i - \bar{y})}}{{\sqrt{{\sum_{i=1}^{n} (x_i - \bar{x})^2}} \sqrt{{\sum_{i=1}^{n} (y_i - \bar{y})^2}}}}, \]
where $n$ is the number of pairs of observations, $x_i$ and $y_i$ are the values of the first and second variables, respectively, for the i-th pair of observations, while $\bar{x}$ and $\bar{y}$ are the means of the variables $x$ and $y$, respectively.

\end{itemize}

\subsubsection{Additional considerations about metrics}

In addition to the described classification and regression metrics, the evaluation of FL systems, and therefore FL-enabled IDS, is characterized by some additional aspects that need to be considered. Specifically, as previously mentioned, the number of training rounds has a significant impact on model convergence and the communication overhead between the aggregator and the clients. Furthermore, the number of epochs determines the iterations a client performs on its local dataset in each training round before sending weight updates to the aggregator. Furthermore, the use of certain unsupervised approaches involves metrics such as Within-Cluster Sum of Squares (WCSS), which is used in clustering analysis. Specifically, WCSS quantifies the compactness or tightness of clusters by calculating the sum of the squared distances between each data point and the centroid of its assigned cluster. WCSS is commonly used alongside other metrics, such as the silhouette score or gap statistic, to evaluate and compare different clustering solutions, determine the optimal number of clusters, or assess the effectiveness of clustering algorithms in data partitioning. Additionally, although not explicitly mentioned, there are specific metrics for RL settings, such as rewards. A reward is often represented as a signal (negative or positive) received by the agent after taking an action in the environment to promote its learning ability.

\subsection{FL implementations}
\label{sec:implementations}

In this section, a brief overview of FL implementations is provided. The interested reader could also refer to~\cite{kholod2020open,fedml20} for more detailed descriptions and comparison of the various implementations. Regarding framework selection,~\cite{liu2022unifed} provides a unified benchmark for evaluating the most popular FL frameworks in terms of functionality, usability and performance.

\textbf{TensorFlow Federated (TFF)} is an open-source library developed by Google for training models in federated settings. TFF offers two interfaces to apply the included implementations of federated training on TensorFlow models~\cite{tensorflow}, as well as to define new algorithms to be used in federated environments. Additionally, it provides diverse implementations to reduce the communication overhead between devices and the server, as well as tools to apply DP techniques. As of June 2023, TFF is still in version 0.59, that is, in a pre-release status\footnote{https://github.com/tensorflow/federated/releases}.

\textbf{PySyft}~\cite{pysyft} is an open-source Python library designed for secure and private DL. PySyft decouples private data from model training, using FL, DP and SMPC. It was developed by the OpenMined~\cite{openMined} community and works mainly with DL frameworks, such as PyTorch~\cite{pytorch} and TensorFlow~\cite{tensorflow}. PySyft supports two types of computations: dynamic computations over hidden data, and static computations, that is, graphs of computations that can be executed later in a different environment. As PySyft does not support communication across networks, PyGrid\footnote{https://github.com/OpenMined/PySyft/tree/dev/packages/grid} is used to facilitate FL on web, mobile, and other types of devices. PySyft is not production-ready as it is still in a beta release in version 0.8.1, as of June 2023.

The \textbf{Federated AI Technology Enabler (FATE)}~\cite{fate} is an industrial grade open-source framework based on HE and SMPC. It provides various ML algorithms, including logistic regression, tree-based algorithms, and additional techniques for DL and TL. FATE can be installed on Linux or Mac platforms and supports standalone and cluster deployments. Unlike previous implementations, FATE can already be used in production environments; its current version is 1.11 (June 2023).

\textbf{Paddle Federated Learning (PaddleFL)}~\cite{paddlefl} is an open-source FL framework based on the PaddlePaddle deep learning platform~\cite{paddlepaddle}. PaddleFL is Python-based and can be used in production environments; currently, as of June 2023, it is in version 1.2. The data partitioning types currently supported by PaddleFL are horizontal and vertical FL, whereas support for federated TL will be developed in the future.

\textbf{IBM Federated Learning (IBMFL)}~\cite{ibmfl} is an open-source Python library designed to  support easy set up of FL in productive environments. IBMFL is an enterprise-level solution that provides a basic FL layer over which more advanced features could be added. It supports conventional ML algorithms, such as linear regression and k-means, as well as DNNs, while facilitating the easy implementation of new FL algorithms. In IBMFL, supervised and unsupervised, as well as RL approaches are included. The IBMFL code repository~\cite{ibmflrepo} is maintained and its latest version is 1.1 (as of June 2023).

\textbf{FedML}~\cite{fedml20} is an open research library and benchmark written in Python, which attempts to address limitations of other FL approaches and promote FL research. 
Its creators argue that FedML overcomes the main limitations of other FL libraries by: (i) supporting diverse FL computing paradigms and diverse FL configurations, (ii) providing standardised FL algorithm implementations and benchmarks, and (iii) being open and continuously evolving. It supports three main computing paradigms: on-device training for edge devices (for example, mobile and IoT), distributed computing, and standalone, single-machine simulation. At the time of writing this paper, it has a pre-release version (0.8)~\cite{fedmlrepo}.

\textbf{Flower}~\cite{flower} is a Python-based, open source FL framework focusing on large-scale experiments with heterogeneous devices. Indicatively, it can support experiments with up to 15M clients with a pair of high-end GPUs. Among Flower advantages are stability, wide support of languages, operating systems and ML frameworks, support of scenarios with diverse privacy requirements, as well as flexibility to enable the implementation of novel approaches with low engineering overhead. Moreover, it facilitates the extension of FL implementations to mobile devices with diverse hardware specifications and the transition of existing ML training setups to a FL environment. The code base of Flower~\cite{flowergit} is actively maintained with the latest release being version 1.5 in June 2023.

\textbf{FederatedScope}~\cite{fedscope} is a relatively novel platform for FL that was designed to be flexible enough to manage heterogeneous data, resources, behaviors and learning goals, using an event-driven architecture. Two further objectives of this platform are usability and extensibility. The former is fulfilled by providing different programming interfaces to users, whereas the latter by extensible $<$event, handler$>$ pairs thanks to its event-driven architecture. FederatedScope is open-source, Python-based and is intended for both research and production environments. Its source code is maintained~\cite{fedscopegit} but given its novelty there is no stable release yet; currently, its latest version is 0.3.0, as of June 2023. 

Apart from the above general-purpose FL frameworks, domain-specific FL implementations have emerged, such as Substra~\cite{substra}, NVIDIA Clara~\cite{clara} and Fed-BioMed~\cite{fedbiomed}, which focus on medical use cases. Substra is a Python-based implementation that is used by hospitals and biotech companies, and provides command line interfaces for administrators and graphical user interfaces for high-level users. Its operation is based on trusted execution environments for privacy, a distributed ledger for traceability and encryption for security. NVIDIA Clara is a platform designed to enable developers to quickly create and deploy healthcare AI applications that can be used in medical imaging, diagnostics, personalized medicine and accelerating drug discovery. Fed-BioMed is an open source, Python and PyTorch-based project for biomedical research through FL. The goal of the project is to provide a simplified framework for easy deployment and friendly user interface to foster research and collaboration. 

\subsection{Analysis}
\label{sec:implementations_analysis}

As shown in the analyzed studies in Table~\ref{tab:super_taxonomy_long_table}, the majority of metrics used to evaluate FL-enabled IDS correspond to the classification metrics described above. Specifically, Accuracy represents the most common metric, being used by 92 out of 104 studies (88.45\%). However, this metric alone does not provide a reliable measure, especially in imbalanced datasets, which are common in FL scenarios and in real-life datasets. 

Precision and recall are also widely used, in 56.2\% and 53.3\% of surveyed works, respectively. In practical terms, a high recall indicates that the model is effective at correctly identifying flows representing attacks. Moreover, high precision would indicate that when the model predicts an attack, it is likely to be correct. 

The F1-score is also relatively common (70.1\%) and it provides a single metric to evaluate the overall effectiveness of a classification model, considering both the correctness of positive predictions and the ability to capture positive instances. The AUC is a less frequent metric (3.8\%) but often preferred over others, such as accuracy, in situations where there is an imbalanced class distribution or a need to evaluate the model's performance across different threshold settings. In fact, as described by~\cite{belenguer2022review}, the use of ROC can provide a more accurate global view in FL-enabled IDS systems. It is also worth noting that only one study~\cite{popoola2021federated} uses the MCC metric, which relates the set of TP, TN, FP, and FN cases in a formula. This metric, along with the CKS metric, offers high reliability and simplicity of interpretation~\cite{belenguer2022review}. Other metrics employed in the different analyzed works are related to the efficiency of the proposed approach, including time or computation time (21\%), latency (2\%), overhead/communication overhead (11\%), delay (2\%), memory/CPU use (6\%), power (3\%), energy (3\%), and area on-chip (1\%).

The wide variety of metrics used, along with the issues mentioned in Section~\ref{sec:datasets} regarding the division of datasets among different parties, makes comparing the different proposed approaches a complex and non-intuitive task. Additionally, as shown in Table~\ref{tab:super_taxonomy_long_table}, it is noteworthy that some studies do not explicitly indicate the number of rounds and epochs considered in the training process. This is coupled with a lack of details about the hyperparameters used in the employed model. In fact, the process of tuning the hyperparameters of the models used is essential for improving the metrics~\cite{bello2021revisiting}. As described in~\cite{dehghani2021benchmark}, which is devoted to analyzing the factors that may lead to a method being perceived as superior,  \textit{``quantifiable progress necessitates the use of shared metrics, which are an essential part of a benchmark. Their choice requires great care, as at the end of the day, metrics reflect the progress and dictate future research directions.''} Based on the above discussion, we have shown that there is no agreement in FL research when it comes to reporting metrics, and in this sense, we believe that the full context of the problem needs to be provided in order to promote transparency and avoid hindering the progress of this area of knowledge. In fact, as described in~\cite{belenguer2022review}, future work should address this issue by defining a standardized set of metrics to facilitate the comparison and analysis of FL-enabled IDS.


\begin{figure}[h]
\begin{tikzpicture}
\begin{axis}[
xbar,
xmin=0,
bar width=12pt, 
/pgf/number format/.cd,
symbolic y coords={ 
   PaddleFL,
   FedML,
   FederatedScope,
   FATE,
   Flower,
   TFF,
   IBMFL,
   PySyft},
height=8cm,
width=7cm,
ytick=data,
     nodes near coords]
\addplot coordinates {
    (0,FedML)
    (0,FederatedScope)
    (0,PaddleFL)
    (1,FATE)
    (2,Flower)
    (3,TFF)
    (4,IBMFL)
    (12,PySyft)
};
\end{axis}
\end{tikzpicture}
\caption{Use of FL implementations in surveyed publications}
\label{fig:fl:libraries:usage}
\end{figure}
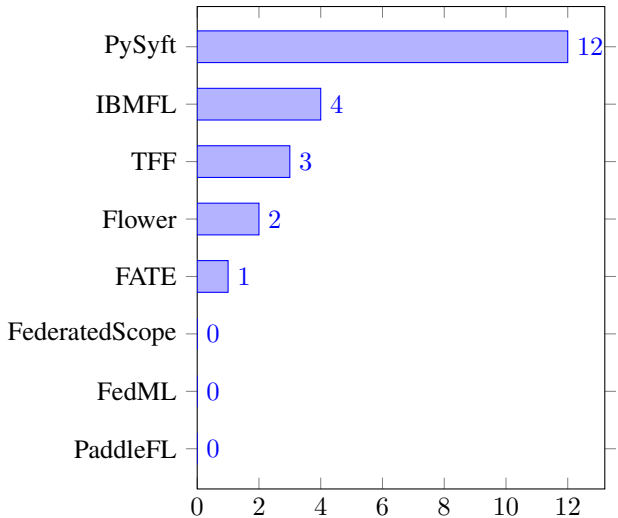

Regarding the implementations used in the analyzed works, Figure~\ref{fig:fl:libraries:usage} depicts the usage of different FL implementations previously described. Interestingly, only 22 out of 104 works employ libraries specifically designed for FL. Actually, some works do not provide details of the libraries used, while a significant portion references well-known ML libraries such as scikit-learn~\cite{kramer2016scikit}, TensorFlow~\cite{abadi2016tensorflow}, PyTorch~\cite{paszke2019pytorch}, or Keras~\cite{ketkar2017introduction}. Furthermore, it is worth noting that the majority of the analyzed frameworks offer preliminary versions that may contain bugs or implementation weaknesses. These aspects also do not facilitate the analysis and comparison among the analyzed FL-enabled IDS.
\section{Challenges}
\label{sec:challenges}
The analysis of the current literature on FL-enabled IDS approaches reveals significant challenges to be considered in the coming years. This section aims to provide an overview of such challenges, as well as current solutions and future paths to address them. The description of these challenges is based on the works analyzed in this survey, the analysis of current literature on FL~\cite{abdulrahman2021survey, li2020federated, aledhari2020federated}, as well as the authors' experience on the definition and implementation of FL-enabled IDS approaches~\cite{campos2021evaluating, ruzafa2021intrusion, matheu2022federated}. Additionally, to describe the current state of challenges related to the use of FL, particularly in the context of FL-enabled IDS, we attempt to answer three main questions:

\begin{enumerate}
    \item What is the nature of the challenge and what is its impact in the context of FL-enabled IDS?
    \item What solutions have been proposed in the context of FL-enabled IDS to address this challenge?
    \item What potential approaches should be considered in the future to address this challenge, and how can they be used in the context of FL-enabled IDS?
\end{enumerate}
    
Furthermore, Table~\ref{tab:challenges} provides a summary of the challenges considered by the FL-enabled IDS approaches analyzed in this study, as well as some of the key considerations to be addressed by future research. It should be noted that the same reference is repeated when it addresses more than one challenge. 

\subsection{Security}

The security aspects of FL settings have attracted a significant interest in recent years~\cite{blanco2021achieving, mothukuri2021survey}. Unlike centralized approaches, security in an FL scenario is typically compromised by \textit{model poisoning} attacks, where the weights/gradients sent by clients in each training round are deliberately/maliciously modified~\cite{bhagoji2019analyzing}. Indeed, since data is not shared, these attacks include \textit{data poisoning} attacks, such as \textit{dirty-label}, which results in a different value for the client's weights/gradients. Although these attacks are widely studied in the literature during the last few years, additional security threats can arise from \textit{backdoor} attacks, \textit{free-riding} attacks (where nodes only participate to obtain the resulting model from the training process), \textit{availability attacks} (where an attacker removes clients from the federated environment), or typical \textit{eavesdropping} attacks~\cite{mothukuri2021survey}.

The main impact in the context of IDS is represented by reduction in performance for detecting cyberattacks, which can have serious consequences depending on the environment where the IDS is deployed. Additionally, this could lead to false alarms resulting from misclassification during the training process. According to the analysis of the literature, 11/104 works (that is, 10.6\%) have proposed a security approach to address some of the mentioned issues in the context of FL-enabled IDS. On the one hand, some proposals have defined alternative aggregation functions to FedAvg (see Section~\ref{sec:aggregationMethods}) to provide a certain degree of resilience against malicious clients. For example,~\cite{shukla2021device} proposes a robust approach derived from FedProx as a defense mechanism against different malicious devices. Additionally,~\cite{rey2022federated} analyzes the impact of different robust aggregation functions against data and model poisoning attacks. In particular, some authors evaluate the use of well-known functions, such as trimmed mean~\cite{yin2018byzantine}, against attacks related to label-flipping and gradient modification. On the other hand, according to our analysis of the literature, the use of GANs has also been widely considered to enhance the robustness of FL-enabled IDS. Although this can be considered a double-edged sword, GANs can be used to generate synthetic weights and gradients that allow the identification of potential malicious nodes. For example,~\cite{singh2020collaborative} proposes the use of GANs to improve system robustness by training with data related to previously unseen attacks. These tools can be complemented by trust and reputation approaches to evaluate the level of trustworthiness offered by FL clients, as proposed by~\cite{lalouani2022robust}.

As defined in~\cite{blanco2021achieving}, there are different mechanisms to improve the security of FL environments that can be applied in the context of FL-enabled IDS. Although some works have considered more robust aggregation functions, there is a lack of comprehensive analysis considering additional well-known aggregation functions such as Krum/Multi-Krum~\cite{blanchard2017machine} or Bulyan~\cite{guerraoui2018hidden} to be deployed in FL-enabled IDS. Furthermore, most analyses do not consider the complexity of these functions, which could have a strong impact on IDS deployment due to the need to detect potential attacks as soon as possible. Moreover, there is also a lack of an exhaustive list of attacks to be considered in FL environments. This lack of consensus in the definition of attacks makes it challenging to compare different aggregation techniques and evaluate their robustness. Additionally, in most cases, this evaluation relies on dubious assumptions, where attackers are isolated nodes with limited knowledge of the system. In fact, this aspect is addressed by~\cite{shejwalkar2021manipulating}, which generates specific attacks for some of the previously mentioned aggregation functions. Beyond the definition of robust aggregation techniques, as mentioned by \cite{blanco2021achieving}, the use of statistical approaches can be essential to identify nodes that are sending forged updates. Likewise, the application of cryptographic approaches and tools like TEE~\cite{mo2021ppfl} is not widely considered, and needs to be evaluated in the context of FL-enabled IDS.

\subsection{Privacy}

One of the main advantages of using FL is related to privacy, as data does not need to be shared to train the global model. However, as extensively discussed in the literature, the exchange of weights/gradients during the training process raises serious privacy concerns, as information derived from the local dataset can be still inferred~\cite{mothukuri2021survey, abdulrahman2021survey, liu2022privacy}. As described in~\cite{wahab2021federated}, privacy threats can involve diverse types of actors, such as \textit{honest-but-curious aggregators}, which attempt to infer information from each client's dataset using the shared updates in each training round. Likewise, clients may also be able to infer information from the gradients of other clients using the information contained in the global models sent by the aggregator. Additionally, other attacks can be considered related to the inference of the participation of certain nodes in the training process, which can also have privacy implications~\cite{wahab2021federated}.

The impact of privacy concerns on the development of FL-enabled IDS depends largely on the nature of the data being used to detect attacks. For example, in the case of network data in a medical environment, the inference of health-related data of individuals can have a significant impact, especially regarding compliance with current data protection regulations, say, GDPR. According to our analysis, 9 out of 104 papers analyzed (8.7\%) propose the use of different techniques to address privacy concerns in FL-enabled IDS environments. Most of the presented works are based on the application of cryptographic techniques for privacy preservation, such as DP~\cite{dwork2014algorithmic}, SMPC~\cite{cramer2015secure}, and HE~\cite{acar2018survey}. DP involves adding noise to the weights exchanged in each training round and is used by~\cite{ruzafa2021intrusion}, considering different DP techniques following Gaussian and Laplacian distributions in the context of attack detection in IoT. Other works based on DP include~\cite{ibitoye2022differentially} and~\cite{kumar2021security}, as well as~\cite{hao2021secure}, which also integrates HE. This cryptographic approach allows computations to be performed on encrypted data. Moreover, SMPC refers to a cryptographic protocol where multiple parties perform a certain function without revealing their inputs. In this way, in the context of FL, the aggregator would not have access to the updates generated by each client~\cite{hsu2020privacy}.

Based on the analysis of current approaches, we perceive that most works rely on the application of well-known cryptographic techniques (especially based on DP) to prevent potential attackers from accessing model updates in each training round. However, the conducted analyses are insufficient to demonstrate the applicability of these techniques in different contexts, including the development of FL-enabled IDS. As described in~\cite{wahab2021federated}, additional analysis is required regarding the impact of different parameters in a federated environment, such as the number of clients or training rounds, on the leaked information. In fact, these aspects should be considered along with the type of data and the environment where the IDS is deployed, as they can have a crucial impact on finding trade-offs between privacy and system effectiveness in detecting attacks. Additionally, the described analyses do not encompass practical considerations, such as communication efficiency or overhead associated with the use of these privacy preservation approaches, as well as system heterogeneity. These aspects can pose a challenge for the deployed IDS to achieve adequate performance. Furthermore, other key aspects are inadequately addressed; for example, the use of GANs to generate synthetic data could be considered to avoid potential inference of real client data~\cite{xin2020private}. However, the challenge lies in generating data that accurately represents reality while still preserving client privacy. Beyond technical aspects, there is a lack of approaches assessing the level of privacy offered by FL environments in compliance with legal frameworks such as GDPR or the EU AI Act.

\subsection{Aggregator as a bottleneck}
\label{sec:challenges_aggregator_bottleneck}

Related to the previous challenges, a major issue of most current FL deployments is that they are carried out in a centralized mode where the aggregator can become a \textit{bottleneck}. Indeed, most of the security and privacy issues previously described are exacerbated in FL environments where a single entity is responsible for performing the entire aggregation process of client updates and generating updated global models in each training round; this entity also becomes a \textit{single point of failure}. Additionally, deploying a single aggregator has a clear impact on the scalability of the environment, as a large number of clients can cause an overwhelming computational and communication overhead. This problem can be further aggravated in synchronous FL approaches where the central entity performs the aggregation process only after all clients have sent their updates. Indeed, with reference to Section~\ref{sec:challenges_device_heterogeneity}, clients' heterogeneity can lead to convergence problems due to the presence of clients with low processing capabilities, a.k.a \textit{stragglers}~\cite{chai2020tifl}.

In the context of FL-enabled IDS, the consequences of the mentioned problems can limit the applicability and effectiveness of the developed IDS. For example, in an IoT environment like a smart city with a large number of deployed devices, the FL training process can be significantly affected. According to our analysis, only 6/104 (5.8\%) works address this problem in FL-enabled IDS approaches. The main solution is based on blockchain technology, which represents an immutable ledger managed by a set of nodes. The functionality of the blockchain is determined by a set of smart contracts executed by all nodes. Additionally, a certain consensus mechanism is used to ensure that nodes agree on the order and validity of transactions. We note that most proposed approaches utilize blockchain to maintain model updates throughout the training process (e.g.,~\cite{hei2020trusted, aliyu2021blockchain, preuveneers2018chained}). However, these proposed approaches do not provide some key specific details on how the system is designed and lack a comprehensive evaluation demonstrating the viability of the solution. Furthermore, although some approaches propose a distributed aggregation process~\cite{liu2021blockchain}, they do not provide details of the implemented smart contracts or evaluation of different consensus mechanisms. In fact, most approaches do not leverage the potential of smart contracts to perform a distributed process of aggregation and validation of updates computed in each training round.

Although the application of blockchain in FL environments has garnered significant interest in recent years~\cite{zhu2023blockchain}, some of the mentioned problems are common in current research beyond the application to FL-enabled IDS. We believe that additional research efforts are needed to provide evidence of the applicability of blockchain in FL to mitigate the problems associated with the aggregator as a centralized entity. Particularly, based on the experience of some authors in blockchain~\cite{bernabe2019privacy, hernandez2021sharing}, we advocate for comprehensive evaluations, considering the communication and computation requirements that can be crucial in IDS systems. Future efforts should harness the potential of designing robust aggregation functions using smart contracts. Additionally, we believe that the application of blockchain for attack detection can be especially useful in large-scale scenarios where different organizations (such as SOCs) without mutual trust can benefit from cybersecurity threat knowledge without sharing sensitive data. Beyond blockchain-based approaches, recent decentralized FL approaches also need to be considered in the future~\cite{lalitha2018fully}. In this scenario, clients themselves are responsible for performing the aggregation process based on the updates received from other participants. Although the deployment of decentralized FL environments has recently gained great interest~\cite{yuan2023decentralized}, additional efforts are still required to evaluate their applicability in specific contexts, especially considering the potential limitations of FL clients. In addition to this approach, other trends should consider asynchronous approaches in the aggregation process, where aggregation is performed as updates arrive~\cite{xu2021asynchronous}, aiming to alleviate the computational and communication overhead of traditional synchronous approaches.

\subsection{Data heterogeneity}
\label{sec:challenges_data_heterogeneity}

One of the main inherent characteristics of FL is represented by non-IID data distributions. This situation is common in real-world scenarios where different client devices may have unbalanced data. Indeed, such data heterogeneity (or statistical heterogeneity~\cite{abdulrahman2021survey}) is characterized by the presence of devices with different size and class distribution, so the data from a single device are not representative of the entire dataset~\cite{wahab2021federated}. Although these aspects also affect centralized ML environments, they are aggravated in FL settings. As demonstrated in previous works~\cite{li2019convergence}, statistical heterogeneity has a key impact on the convergence of the federated training process, especially when FedAvg is employed. Indeed, some of the main aggregation functions described in Section~\ref{sec:aggregationMethods}, such as FedProx or Fed+, were mainly proposed to address the limitations of FedAvg in non-IID settings.

In the context of IDS development, non-IID data distributions represent situations where devices may have a large number of samples of a certain type of attack. This is quite common in real-world scenarios where certain devices may be more vulnerable and serve as entry points to the system. According to our analysis, problems associated with data heterogeneity are widely considered in the proposed FL-enabled IDS approaches (26/104, i.e., 25\%). It is worth noting that some of these works (e.g.,~\cite{zhou2020federated, yuan2021towards, friha2022felids}) evaluate their approach in configurations with non-IID data distributions, although they do not propose any specific solution. As mentioned earlier, while the aggregation function is directly related to the convergence of the system in non-IID environments, according to Section~\ref{sec:aggregationMethods}, we note that most proposed schemes use aggregation approaches based on FedAvg, which has convergence issues~\cite{li2019convergence}. Actually, only a few approaches use alternative aggregation functions, such as Fed+ (~\cite{ruzafa2021intrusion}) or FedProx-based schemes (~\cite{su2022detection}), while other authors~\cite{liu2022intrusion} propose an alternative aggregation approach (FedBatch). The main goal of these solutions is to offer a certain level of personalization, so that the global model can adapt to the specific characteristics of the clients' data. In this regard, some works advocate the use of TL as an approach to address data heterogeneity. For example, TL has been used in scenarios where certain devices have access to a restricted set of training data~\cite{zhao2020network}, or a model built from labeled data is used to label other data~\cite{pei2022knowledge}. Another common practice to address convergence issues related to non-IID data distributions is the use of oversampling and undersampling techniques to balance the data distributions. While some of the proposed approaches are based on well-known techniques such as SMOTE-TOMEK~\cite{zeng2016effective} or SMOTE-ENN~\cite{xu2020hybrid}, a current trend is represented by generative approaches based on GANs~\cite{tabassum2022fedgan}.

While data heterogeneity has been widely considered, there are still aspects that require further attention. In particular, most proposed approaches do not address the heterogeneity associated with the representation of data considering different feature spaces (i.e., vertical FL). In fact, most proposed solutions are based on dividing a specific dataset into a certain number of parts acting as FL clients. Specifically, according to our analysis, only three works (~\cite{novikova2022federated, zhao2019multi, hsu2020privacy}) consider vertical FL scenarios. In real-world scenarios, the need for approaches addressing this level of heterogeneity will be required for the development of FL-enabled IDS. Otherwise, different organizations would be forced to represent their data using the same feature space, which may be unfeasible. Related to this issue, future approaches need to consider the heterogeneity of the model used for local training. On the opposite side, forcing all parties in a federated environment to use the same training model may limit its applicability in any context, particularly in the field of IDS. In this context, the potential of TL can be exploited as a potential approach. Additionally, a comprehensive theoretical and practical analysis of the impact of non-IID data distributions considering different aggregation functions and IDS datasets is needed. As mentioned, even though many works consider non-IID configurations, the reality is that most approaches still rely on using FedAvg as the aggregation approach. Furthermore, as mentioned in~\cite{wahab2021federated}, the conducted evaluations need to consider the tuning of FL hyperparameters to favor the convergence of the developed systems even in scenarios with a high degree of data heterogeneity.

\subsection{Device heterogeneity} \label{sec:challenges_device_heterogeneity}

According to the FL training process, a certain set of clients is selected in each training round to perform local inference and provide updated weights to the aggregator. However, FL scenarios are often characterized by the presence of diverse types of devices with varying processing capabilities, network connectivity, and battery life~\cite{li2020federated}. Additionally, depending on their local datasets, these devices may either benefit or hinder the convergence of the whole FL system. Furthermore, it is necessary to capture the dynamism of the environment where the devices are deployed. For example, depending on the considered scenario, certain clients may enter/leave the FL training process due to poor network capabilities, failures, or mobility factors. Moreover, certain devices may voluntarily choose not to participate in the training process due to critical tasks they need to perform at a given time. This heterogeneity of FL devices or clients significantly impacts the effectiveness and efficiency of FL systems. As described in~\cite{abdulrahman2021survey}, different processing capabilities can favor the presence of stragglers. With reference to Section~\ref{sec:challenges_aggregator_bottleneck}, this circumstance exacerbates the problem of the \textit{aggregator as a bottleneck} and requires asynchronous communication approaches to mitigate the impact in terms of processing time~\cite{li2020federated}.

In the context of IDS, most of the analyzed works do not address the problems associated to device heterogeneity. Specifically, only 8/104 of the analyzed works (7.7\%) tackle this issue. Most of them are based on client selection mechanisms considering different aspects, such as client contribution to the global model~\cite{mowla2019federated, sun2020adaptive}, attack type~\cite{ayed2021federated}, as well as communication capabilities and computational power~\cite{xia2022fed_adbn}. Communication aspects are also addressed by~\cite{agrawal2021temporal}, which considers an aggregation function to determine the waiting time of the aggregator before initiating a new training round. A common feature of the proposed approaches is that, beyond the contribution of a device to the global model, they do not consider the changing conditions of the environment and the devices themselves (e.g., devices entering and leaving the FL environment) to select specific clients in each training round. Additionally, the analyzed works do not offer a comprehensive analysis of the impact of these approaches on the overall performance of the system in terms of complexity and required delay, which could significantly affect the performance of IDS. Furthermore, as detailed in Section~\ref{sec:challenges_data_heterogeneity}, client selection approaches should also consider the heterogeneity of data associated with the different devices involved. In this regard, client selection approaches should rely on solutions that do not require accessing the data or related information, e.g., labels. In the context of IDS, this could provide information about which devices are being targeted and by what types of attacks.

As widely recognized, device heterogeneity is a fundamental problem in FL environments~\cite{aledhari2020federated, zhang2021survey}. Although in the context of IDS, this issue has been partially addressed, further research efforts are required to limit its effects. Firstly, a more comprehensive analysis is needed regarding which aspects and characteristics of the devices can affect the effectiveness and efficiency of the FL training process. While it is commonly accepted that computing and network capabilities have a significant impact, the extent to which each of these characteristics can affect the training process is unclear. As described by~\cite{lim2020federated}, a potential approach is the deployment of edge nodes that homogenize the resulting system. However, the addition of edge nodes can render management more complex and introduce communication delays. Secondly, device heterogeneity should be addressed by considering the dynamism of the environment and the devices themselves throughout their lifecycle. In the context of IDS, we observe that only very few works (e.g.,~\cite{qin2021fnel}) address dynamism during the FL training process. In fact, almost all the analyzed works consider static scenarios where devices already have their entire local dataset before the training process begins. In real-world scenarios, certain devices may receive new data while participating in FL training. Therefore, the design and implementation of resource management approaches are essential in the near future~\cite{lim2020federated, wahab2021federated}. Specifically, RL-based solutions~\cite{zhang2022multi} could help capture device properties through interaction with the environment to select the most appropriate clients. Additionally, the use of SDNs could facilitate such resource management by allowing clients and the underlying network to adapt to changing environmental conditions~\cite{balasubramanian2021intelligent}.

\subsection{Deployment in constrained scenarios}

Related to the previous challenge, an additional aspect is represented by the deployment of FL systems in scenarios with certain constraints. In fact, this aspect is inherent to FL since, unlike centralized ML approaches, end devices must bear a higher processing load. Firstly, depending on the scenario, certain devices may be unable to execute certain ML/DL algorithms and techniques. Even if they can be executed, a training process that requires a significant number of epochs or rounds may be infeasible depending on the ML model being considered. Besides, it has been shown that a key factor is the overhead required for model exchange between the aggregator and clients in each training round~\cite{luping2019cmfl}. This aspect also impacts the overhead required during the training process, which can exacerbate the bottleneck problem of the aggregator.

According to our analysis, only 7 out of 104 works address the problems associated with deploying FL-enabled IDS in resource-constrained environments. Namely, our analysis reveals that while many works make use of datasets related to IoT, the practical deployment aspects are not truly addressed. In particular, only a few works consider lightweight ML approaches based on NNs~\cite{man2021intelligent, LocKedge, zakariyya2021memory}, such as BNNs~\cite{qin2020line} for reducing the memory and computation overhead. Other works are more focused on lessening the communication overhead during the training process through the application of encoding techniques~\cite{shukla2022rafel} and FD~\cite{zhao2022semi}.

This limited sample of works highlights the need for additional approaches in the coming years that offer more comprehensive analysis of the limits in applying FL-based approaches in collaborative IDS development. In particular, a potential research direction is determined by the use of lightweight ML techniques, such as BNNs or Random NNs, which currently are not considered widely. Additionally, the deployment of systems based on TinyML~\cite{banbury2020benchmarking, sanchez2020tinyml} needs to be considered in order to accommodate the performance requirements of FL systems in constrained devices and environments. Despite the rise of well-known frameworks (e.g., TensorFlow Lite~\cite{david2021tensorflow}), their application in FL is still limited~\cite{mathur2021device}. Indeed, these frameworks offer implementations of pruning and quantization techniques to reduce the size of models, which can be considered in each FL training round~\cite{jiang2022model}. Specifically, pruning is a technique to reduce models by eliminating redundant and unnecessary connections or parameters. In the case of quantization, it involves reducing the numerical precision of a model's parameters. However, these techniques come with a cost in the model's performance in terms of inference precision. Therefore, additional efforts are required to find trade-offs between efficiency and effectiveness in FL scenarios.

\begin{table*}[]
\label{tab:challenges}
\tiny
\centering
\begin{tabular}{C{1cm}C{0.8cm}C{7.5cm}C{8cm}}
\hline
\textbf{Challenge} & \textbf{Reference} & \textbf{Approach} & \textbf{Future aspects} \\ \hline \hline

\multirow{11}{*}{Security} & \cite{singh2020collaborative}  & An Auxiliary Classifier GAN is proposed for generating malware attacks to increase the system's robustness &
\multirow{11}{=}{\begin{itemize}
    \item Precise definition of a comprehensive list of attacks considering different configurations and attack parameters.
    \item Exhaustive evaluation of aggregation mechanisms to assess their resilience in federated environments with malicious nodes.
    \item Evaluation of trust and reputation approaches that allow weighting.
    \item Analysis of the impact of cryptographic approaches.
    \item Use of statistical approaches for identifying malicious nodes to ensure a training process based on legitimate clients.
\end{itemize}}
    \\
\cline{2-3}
&\cite{taheri2020fed}  &  A GAN is used to generate poisoning attacks and different aggregation functions are exploited to evaluate the robustness in the scope of Android platform\\ \cline{2-3}
&\cite{chen2020intrusion}  & Different aggregation functions are considered to measure their impact in scenarios with non-IID data distributions and against poisoning attacks \\ \cline{2-3}
&\cite{attota2021ensemble}   & Label-flipping attacks are simulated to analyze the impact on the proposed approach \\ \cline{2-3}
&\cite{yadav2021clustering}   & A reward-based mechanism is proposed to detect model poisoning attacks, so malicious nodes can be removed\\ \cline{2-3}

&\cite{li2020deepfed}   &  Proposed cryptographic system to encrypt model parameters with the aggregation server \\ \cline{2-3}
&\cite{shukla2021device}  & Proposed robust version of FedProx as aggregation function to detect malicious updates during the training process \\ \cline{2-3}
&\cite{lalouani2022robust}  & Implementation of an approach to assess the trust level of each node based on the variations of each local model throughout the training rounds \\ \cline{2-3}

&\cite{vy2021federated}  & Implementation of a mechanism to reject malicious update generated through label-flipping and GANs\\ \cline{2-3}
&\cite{rey2022federated}  & Analysis of the impact of different data/model poisoning attacks by considering different aggregation functions \\ \cline{2-3}
&\cite{zhang2022secfednids}  &  Implementation of a model-level defensive mechanism to address poisoning attacks \\ \hline

\multirow{9}{*}{Privacy} & \cite{al2020federated} & Use of mimic learning where a certain model is used to train sensitive data &
\multirow{9}{=}{\begin{itemize}
    \item Analysis of privacy-preserving techniques' impact regarding the compliance with existing data protection regulations, such as GDPR.
    \item Identification of the limits on the required level of privacy and effectiveness for a certain IDS depending on the deployment scenario.
    \item Exhaustive  analyses to find trade-offs between the privacy level, as well as the effectiveness and performance of the deployed IDS.
    \item Analysis of GAN-based approaches for federated training using synthetic data instead of clients' real data.
\end{itemize}}
    \\
\cline{2-3}

& \cite{hsu2020privacy} & Implementation of a SMPC scheme for privacy-preserving aggregation \\ \cline{2-3}
& \cite{duy2021federated}  & Use of DP during the training process in an SDN-enabled FL setting by using TensorFlow Privacy \\ \cline{2-3}
& \cite{dong2021towards} & Use of data masking to further protect the privacy of clients' data   \\ \cline{2-3}

& \cite{kumar2021security}  & Integration of HE and DP techniques for privacy-preserving model updates in an edge-based FL-enabled IDS  \\ \cline{2-3}

& \cite{hao2021secure} & Use of HE and DP techniques to protect the model updates to be aggregated in each training round   \\ \cline{2-3}

& \cite{ruzafa2021intrusion} &  Analysis of different DP techniques on the impact of the effectiveness for intrusion detection \\ \cline{2-3}

& \cite{kumar2021pefl} & Perturbation-based encoding combined with an LSTM-AE model to transform original data used for the federated training  \\ \cline{2-3}

& \cite{ibitoye2022differentially}  & Implementation of a DP mechanism by adding a noise layer in the proposed CNN model  \\ \hline

\multirow{6}{*}{\parbox{4cm}{Aggregator as\\ bottleneck}} &\cite{hei2020trusted} &  Clients' model updates are shared through a blockchain infrastructure to keep track of them &
\multirow{6}{=}{\begin{itemize}
    \item Comprehensive evaluations on the applicability of blockchain in FL settings, particularly in the context of IDS, considering different deployment options such as consensus mechanisms, number of nodes, and frequency of model updates.
    \item Definition and implementation of smart contracts for the aggregation and validation process of model updates produced by both the aggregator and FL clients.
    \item Analysis of asynchronous aggregation approaches to alleviate network and computation overhead, which could have negative implications on the IDS efficiency.
    \item Analysis of the applicability of fully decentralized approaches and comparison with hierarchical solutions where certain nodes are responsible for performing partial aggregations.
\end{itemize}}
    \\
\cline{2-3}
&\cite{aliyu2021blockchain}   &  Use of blockchain to store the models generated during the federated training process in a vehicular scenario\\ \cline{2-3}

&\cite{abdel2021federated}   & Capitalize on blockchain technology to come up with a decentralized validation process of clients' local updates and evaluation, considering different settings with malicious nodes\\ \cline{2-3}

& \cite{preuveneers2018chained} &  The authors integrate a blockchain infrastructure to add accountability to the model updates \\ \cline{2-3}

& \cite{chen2021trust} & Partial aggregation is carried out by intermediate nodes with the weights providing the best performance  \\ \cline{2-3}

&\cite{liu2021blockchain}  & Collaborative aggregation approach where several RSUs in a vehicular context are responsible of performing partial aggregations \\ \hline

\multirow{26}{*}{\parbox{4cm}{Data\\ heterogeneity}} &\cite{rahman2020internet} &  Analysis of the impact of different data distributions in the accuracy of the intrusion detection process &
\multirow{26}{=}{\begin{itemize}
    \item Analysis of the requirements and the impact of vertical FL scenarios for the development of FL-enabled IDS approaches.
    \item Evaluation on the impact of aggregation functions in FL-enabled IDS under non-IID settings.
    \item Analysis on the integration of TL approaches to deal with non-IID settings where certain devices are unable to access data.
\end{itemize}}
    \\
\cline{2-3}

&\cite{zhou2020federated} & Analysis of the impact of non-IID data distributions in the system's classification accuracy \\ \cline{2-3}

&\cite{ruzafa2021intrusion} & Comparison between FedAvg and Fed+ as aggregation functions considering a setting with a non-IID data distribution and a subset of clients with the highest entropy\\ \cline{2-3}

&\cite{zhao2020network}  & Use of a TL scheme to simulate scenarios where certain nodes suffer from data scarcity  \\ \cline{2-3}

& \cite{zhang2021flddos} & Use of a data resampling algorithm based on SMOTE~\cite{chawla2002smote} for dealing with highly unbalanced data distributions  \\ \cline{2-3}

& \cite{campos2021evaluating}  & A client selection approach based on entropy values and use of Fed+ to deal with non-IID data distributions \\ \cline{2-3}

& \cite{li2021fids} & An aggregation function based on FedAvg in which the difference between data distributions is taken into account \\ \cline{2-3}

& \cite{ferrag2022edge} &  Evaluation of a proposed dataset considering the impact of non-IID data distributions \\ \cline{2-3}

& \cite{li2021fids}  & An approach to expand feature spaces to deal with non-IID data distributions \\ \cline{2-3}

& \cite{yuan2021towards}  &  A system based on LGBM, which is evaluated in both IID non-IID settings \\ \cline{2-3}

& \cite{qin2021federated}  & Devices are grouped according to their targeting attack types to carry out the federated training by considering a certain feature set \\ \cline{2-3}

& \cite{zhao2021federated}  & Evaluation of an IDS for false data injection attacks considering both IID and non-IID settings\\ \cline{2-3}

& \cite{popoola2021federated} & Analysis of the impact of different aggregation functions on the performance of intrusion detection considering different datasets \\ \cline{2-3}

& \cite{lv2022ddos}  & A personalization approach to address the impact of non-IID data distributions \\ \cline{2-3}

& \cite{su2022detection}  & A FedProx-based aggregation approach to handle non-IID data distributions \\ \cline{2-3}

& \cite{wang2022feco}  & Analysis of the proposed FL-enabled contrastive learning approach by considering different data distributions \\ \cline{2-3}

& \cite{liu2022intrusion}  & An aggregation mechanism (FedBatch), which is analyzed and compared with FedAvg under non-IID data distributions \\ \cline{2-3}

& \cite{fan2020iotdefender} & A TL scheme to obtain personalized models by using different datasets  \\ \cline{2-3}

& \cite{chen2022privacy} & Implementation of a two-phase optimization strategy to deal with non-IID data distributions  \\ \cline{2-3}

& \cite{friha2022felids}  & The authors evaluate the proposed IDS by considering different non-IID data distributions and number of clients  \\ \cline{2-3}

& \cite{zhao2022semi} & An FD-oriented aggregation mechanism which is evaluated under different scenarios with non-IID data distributions \\ \cline{2-3}

& \cite{pei2022knowledge}  & A TL approach to build a semi-supervised approach where nodes leverage on the labelled data used by a cloud model that is used by nodes to train on unlabelled data \\ \cline{2-3}

& \cite{otoum2021federated} & Use of a TL approach to create customized models based on a global model previously obtained from a federated training process \\ \cline{2-3}

& \cite{saadat2021hierarchical}  & Analysis on the impact of including edge nodes between end devices and aggregator in settings with non-IID data distributions \\ \cline{2-3}

& \cite{kelli2021ids} &  Analysis on the impact of performing local training after the FL process to obtain personalized local models \\ \cline{2-3}

& \cite{tabassum2022fedgan} & A GAN is proposed to increase the number of samples especially from rare classes in order to deal with data imbalance \\ \hline

\multirow{3}{*}{\parbox{4cm}{Device\\ heterogeneity}} &\cite{mowla2019federated} &  Application of the Dempster-Shafer theory \cite{shafer1992dempster} for client prioritization &
\multirow{3}{=}{\begin{itemize}
    \item Precise definition of device characteristics influencing the FL training process.
    \item Study of reinforcement learning-based approaches to maximize the effectiveness and adaptability of the FL environment through appropriate client selection.
    \item Analysis of the impact of changing environmental conditions and device dynamics throughout their lifecycle.
    \item Definition of resource orchestration approaches to address device heterogeneity considering the inherent dynamism in IDS environments.
\end{itemize}}
    \\
\cline{2-3}

& \cite{sun2020adaptive}  & FL clients with similar performance are dynamically grouped throughout the training rounds \\ \cline{2-3}

& \cite{ayed2021federated} & The evaluation process considers different aspects (e.g., the type of attack) for client selection during the training process  \\ \cline{2-3}

& \cite{xia2022fed_adbn} & A client selection approach based on communication quality, computing power, and attack risk aspects \\ \cline{2-3}

& \cite{cheng2022federated} & An RL-based approach to implement a client selection mechanism towards increasing system's accuracy\\ \cline{2-3}

& \cite{pasdar2022train}  & A TL-based approach to create personalized models based on FL considering smartphone devices  \\ \cline{2-3}

& \cite{agrawal2021temporal} & An aggregation function to determine the waiting time for the server to initiate a new training round \\ \cline{2-3}

& \cite{ibitoye2022differentially}  & Analysis of a client selection approach by including adversarial samples  \\ \hline

\multirow{7}{*}{\parbox{4cm}{Computation and \\communication \\requirements}} & \cite{qin2020line} &  Use of BNN \cite{hubara2016binarized} to reduce communication and memory overhead &
\multirow{7}{=}{\begin{itemize}
    \item Analysis and evaluation of lightweight ML techniques in FL environments using TinyML-based implementations.
    \item Design and implementation of quantization and pruning schemes to address bandwidth limitations in specific scenarios and networks.
    \item Analysis of trade-offs between communication and processing overhead regarding training rounds and epochs and the achieved IDS performance.
\end{itemize}}
    \\
\cline{2-3}

&\cite{man2021intelligent}  & An FedAvg-based aggregation function to reduce the number of training rounds required for model convergence\\ \cline{2-3}

& \cite{LocKedge}  &  A lightweight NN structure based on the reduction of number of layers and neurons in hidden layers\\ \cline{2-3}

& \cite{zakariyya2021memory}  & Use of a NN to build a memory efficient version destined to constrained scenarios \\ \cline{2-3}

& \cite{dong2021towards} & A data binning approach to group traffic statistics so that the communication overhead is reduced \\ \cline{2-3}

& \cite{shukla2022rafel} & Design of a bit-wise encoding technique to reduce the communication cost during the training process \\ \cline{2-3}

& \cite{zhao2022semi}  & Use of a FD mechanism to reduce the communication overhead during the training process \\ \hline

\end{tabular}
\caption{Analysis of challenges and future trends for FL-enabled IDS approaches}
\label{tab:challenges}
\end{table*}






\section{Conclusions}
\label{sec:conclusions}

During the last few years, FL has proliferated throughout the IDS domain thanks to its decentralized, collaborative, and privacy-preserving intrinsic traits. Characteristically, with reference to Figure~\ref{fig:fl:pubs:by:year}, the number of peer-review publications in FL-powered IDS has mushroomed from less than 5 in 2018 to more than 150 in 2022. Nevertheless, as explained in Section~\ref{sec:introduction}, so far, no survey work provides a full-fledged, systematic overview of this rapidly evolving topic. The article at hand aspires to address this noticeable research gap by offering a pluralistic understanding of this composite ecosystem spanning several key axes. That is, apart from providing a contemporary taxonomy of the FL-enabled IDS approaches in the literature from 2018 to 2022, we detail the major ML models, datasets, aggregation functions, and implementation libraries, which are used in the context of such works. Additionally, we offer a thorough view of the current state of affairs on this topic, and identify the main challenges and future directions. Overall, due to its holistic orientation, this work is envisioned to not only shed more light upon this interesting research branch but also serve as a source of reference for future work.

\section*{Acknowledgment}
This work has been funded by the European Commission through the HORIZON-MSCA-2021-PF-01-01 project INCENTIVE (g.a. 101065524). This study also forms part of the ThinkInAzul programme and was partially supported by MCIN with funding from European Union NextGenerationEU (PRTR-C17.I1) and by Comunidad Autónoma de la Región de Murcia - Fundación Séneca. It was partially funded by the HORIZON-MSCA-2021-SE-01-01 project Cloudstars (g.a. 101086248) as well.

\bibliographystyle{IEEEtran}
\bibliography{biblio}

\begin{thebibliography}{100}
\providecommand{\url}[1]{#1}
\csname url@samestyle\endcsname
\providecommand{\newblock}{\relax}
\providecommand{\bibinfo}[2]{#2}
\providecommand{\BIBentrySTDinterwordspacing}{\spaceskip=0pt\relax}
\providecommand{\BIBentryALTinterwordstretchfactor}{4}
\providecommand{\BIBentryALTinterwordspacing}{\spaceskip=\fontdimen2\font plus
\BIBentryALTinterwordstretchfactor\fontdimen3\font minus
  \fontdimen4\font\relax}
\providecommand{\BIBforeignlanguage}[2]{{%
\expandafter\ifx\csname l@#1\endcsname\relax
\typeout{** WARNING: IEEEtran.bst: No hyphenation pattern has been}%
\typeout{** loaded for the language `#1'. Using the pattern for}%
\typeout{** the default language instead.}%
\else
\language=\csname l@#1\endcsname
\fi
#2}}
\providecommand{\BIBdecl}{\relax}
\BIBdecl

\bibitem{jordan2015machine}
M.~I. Jordan and T.~M. Mitchell, ``Machine learning: Trends, perspectives, and
  prospects,'' \emph{Science}, vol. 349, no. 6245, pp. 255--260, 2015.

\bibitem{act2021proposal}
{European Commission}, ``Proposal for a regulation of the {E}uropean
  {P}arliament and the {C}ouncil laying down harmonised rules on {A}rtificial
  {I}ntelligence ({A}rtificial {I}ntelligence {A}ct) and amending certain
  {U}nion legislative acts. {COM}/2021/206 final,'' \emph{EUR-Lex-52021PC0206},
  April 2021.

\bibitem{taddeo2019trusting}
M.~Taddeo, T.~McCutcheon, and L.~Floridi, ``Trusting artificial intelligence in
  cybersecurity is a double-edged sword,'' \emph{Nature Machine Intelligence},
  vol.~1, no.~12, pp. 557--560, 2019.

\bibitem{mcmahan2017communication}
B.~McMahan, E.~Moore, D.~Ramage, S.~Hampson, and B.~A. y~Arcas,
  ``Communication-efficient learning of deep networks from decentralized
  data,'' in \emph{Artificial intelligence and statistics}.\hskip 1em plus
  0.5em minus 0.4em\relax PMLR, 2017, pp. 1273--1282.

\bibitem{lu2020low}
Y.~Lu, X.~Huang, K.~Zhang, S.~Maharjan, and Y.~Zhang, ``Low-latency federated
  learning and blockchain for edge association in digital twin empowered 6g
  networks,'' \emph{IEEE Transactions on Industrial Informatics}, vol.~17,
  no.~7, pp. 5098--5107, 2020.

\bibitem{nguyen2021federated2}
D.~C. Nguyen, M.~Ding, P.~N. Pathirana, A.~Seneviratne, J.~Li, and H.~V. Poor,
  ``Federated learning for internet of things: A comprehensive survey,''
  \emph{IEEE Communications Surveys \& Tutorials}, 2021.

\bibitem{xia2021survey}
Q.~Xia, W.~Ye, Z.~Tao, J.~Wu, and Q.~Li, ``A survey of federated learning for
  edge computing: Research problems and solutions,'' \emph{High-Confidence
  Computing}, vol.~1, no.~1, p. 100008, 2021.

\bibitem{nguyen2022federated}
D.~C. Nguyen, Q.-V. Pham, P.~N. Pathirana, M.~Ding, A.~Seneviratne, Z.~Lin,
  O.~Dobre, and W.-J. Hwang, ``Federated learning for smart healthcare: A
  survey,'' \emph{ACM Computing Surveys (CSUR)}, vol.~55, no.~3, pp. 1--37,
  2022.

\bibitem{du2020federated}
Z.~Du, C.~Wu, T.~Yoshinaga, K.-L.~A. Yau, Y.~Ji, and J.~Li, ``Federated
  learning for vehicular internet of things: Recent advances and open issues,''
  \emph{IEEE Open Journal of the Computer Society}, vol.~1, pp. 45--61, 2020.

\bibitem{alazab2021federated}
M.~Alazab, S.~P. RM, M.~Parimala, P.~K.~R. Maddikunta, T.~R. Gadekallu, and
  Q.-V. Pham, ``Federated learning for cybersecurity: Concepts, challenges, and
  future directions,'' \emph{IEEE Transactions on Industrial Informatics},
  vol.~18, no.~5, pp. 3501--3509, 2021.

\bibitem{ghimire2022recent}
B.~Ghimire and D.~B. Rawat, ``Recent advances on federated learning for
  cybersecurity and cybersecurity for federated learning for internet of
  things,'' \emph{IEEE Internet of Things Journal}, 2022.

\bibitem{khraisat2019survey}
A.~Khraisat, I.~Gondal, P.~Vamplew, and J.~Kamruzzaman, ``Survey of intrusion
  detection systems: techniques, datasets and challenges,''
  \emph{Cybersecurity}, vol.~2, no.~1, pp. 1--22, 2019.

\bibitem{ring2019survey}
M.~Ring, S.~Wunderlich, D.~Scheuring, D.~Landes, and A.~Hotho, ``A survey of
  network-based intrusion detection data sets,'' \emph{Computers \& Security},
  vol.~86, pp. 147--167, 2019.

\bibitem{thakkar2020review}
A.~Thakkar and R.~Lohiya, ``A review of the advancement in intrusion detection
  datasets,'' \emph{Procedia Computer Science}, vol. 167, pp. 636--645, 2020.

\bibitem{lavaur2022evolution}
L.~Lavaur, M.-O. Pahl, Y.~Busnel, and F.~Autrel, ``The evolution of federated
  learning-based intrusion detection and mitigation: a survey,'' \emph{IEEE
  Transactions on Network and Service Management}, 2022.

\bibitem{agrawal2021federated}
S.~Agrawal, S.~Sarkar, O.~Aouedi, G.~Yenduri, K.~Piamrat, S.~Bhattacharya,
  P.~K.~R. Maddikunta, and T.~R. Gadekallu, ``Federated learning for intrusion
  detection system: Concepts, challenges and future directions,'' \emph{arXiv
  preprint arXiv:2106.09527}, 2021.

\bibitem{campos2021evaluating}
E.~M. Campos, P.~F. Saura, A.~Gonz{\'a}lez-Vidal, J.~L. Hern{\'a}ndez-Ramos,
  J.~B. Bernabe, G.~Baldini, and A.~Skarmeta, ``Evaluating federated learning
  for intrusion detection in internet of things: Review and challenges,''
  \emph{Computer Networks}, p. 108661, 2021.

\bibitem{ijisec/KoliasKK17}
\BIBentryALTinterwordspacing
C.~Kolias, V.~Kolias, and G.~Kambourakis, ``Termid: a distributed swarm
  intelligence-based approach for wireless intrusion detection,'' \emph{Int. J.
  Inf. Sec.}, vol.~16, no.~4, pp. 401--416, 2017. [Online]. Available:
  \url{https://doi.org/10.1007/s10207-016-0335-z}
\BIBentrySTDinterwordspacing

\bibitem{yang2019federated}
Q.~Yang, Y.~Liu, T.~Chen, and Y.~Tong, ``Federated machine learning: Concept
  and applications,'' \emph{ACM Transactions on Intelligent Systems and
  Technology (TIST)}, vol.~10, no.~2, pp. 1--19, 2019.

\bibitem{zhang2021survey}
C.~Zhang, Y.~Xie, H.~Bai, B.~Yu, W.~Li, and Y.~Gao, ``A survey on federated
  learning,'' \emph{Knowledge-Based Systems}, vol. 216, p. 106775, 2021.

\bibitem{abdulrahman2021survey}
S.~AbdulRahman, H.~Tout, H.~Ould-Slimane, A.~Mourad, C.~Talhi, and M.~Guizani,
  ``A survey on federated learning: The journey from centralized to distributed
  on-site learning and beyond,'' \emph{IEEE Internet Things J}, vol.~8, no.~7,
  pp. 5476--5497, 2021.

\bibitem{wahab2021federated}
O.~A. Wahab, A.~Mourad, H.~Otrok, and T.~Taleb, ``Federated machine learning:
  Survey, multi-level classification, desirable criteria and future directions
  in communication and networking systems,'' \emph{IEEE Communications Surveys
  \& Tutorials}, vol.~23, no.~2, pp. 1342--1397, 2021.

\bibitem{aledhari2020federated}
M.~Aledhari, R.~Razzak, R.~M. Parizi, and F.~Saeed, ``Federated learning: A
  survey on enabling technologies, protocols, and applications,'' \emph{IEEE
  Access}, vol.~8, pp. 140\,699--140\,725, 2020.

\bibitem{li2020review}
L.~Li, Y.~Fan, M.~Tse, and K.-Y. Lin, ``A review of applications in federated
  learning,'' \emph{Computers \& Industrial Engineering}, vol. 149, p. 106854,
  2020.

\bibitem{liu2022distributed}
J.~Liu, J.~Huang, Y.~Zhou, X.~Li, S.~Ji, H.~Xiong, and D.~Dou, ``From
  distributed machine learning to federated learning: A survey,''
  \emph{Knowledge and Information Systems}, pp. 1--33, 2022.

\bibitem{lim2020federated}
W.~Y.~B. Lim, N.~C. Luong, D.~T. Hoang, Y.~Jiao, Y.-C. Liang, Q.~Yang,
  D.~Niyato, and C.~Miao, ``Federated learning in mobile edge networks: A
  comprehensive survey,'' \emph{IEEE Communications Surveys \& Tutorials},
  vol.~22, no.~3, pp. 2031--2063, 2020.

\bibitem{imteaj2021survey}
A.~Imteaj, U.~Thakker, S.~Wang, J.~Li, and M.~H. Amini, ``A survey on federated
  learning for resource-constrained iot devices,'' \emph{IEEE Internet of
  Things Journal}, 2021.

\bibitem{khan2021federated}
L.~U. Khan, W.~Saad, Z.~Han, E.~Hossain, and C.~S. Hong, ``Federated learning
  for internet of things: Recent advances, taxonomy, and open challenges,''
  \emph{IEEE Communications Surveys \& Tutorials}, 2021.

\bibitem{zhu2023blockchain}
J.~Zhu, J.~Cao, D.~Saxena, S.~Jiang, and H.~Ferradi, ``Blockchain-empowered
  federated learning: Challenges, solutions, and future directions,'' \emph{ACM
  Computing Surveys}, vol.~55, no.~11, pp. 1--31, 2023.

\bibitem{issa2023blockchain}
W.~Issa, N.~Moustafa, B.~Turnbull, N.~Sohrabi, and Z.~Tari, ``Blockchain-based
  federated learning for securing internet of things: A comprehensive survey,''
  \emph{ACM Computing Surveys}, vol.~55, no.~9, pp. 1--43, 2023.

\bibitem{mothukuri2021survey}
V.~Mothukuri, R.~M. Parizi, S.~Pouriyeh, Y.~Huang, A.~Dehghantanha, and
  G.~Srivastava, ``A survey on security and privacy of federated learning,''
  \emph{Future Generation Computer Systems}, vol. 115, pp. 619--640, 2021.

\bibitem{blanco2021achieving}
A.~Blanco-Justicia, J.~Domingo-Ferrer, S.~Mart{\'\i}nez, D.~S{\'a}nchez,
  A.~Flanagan, and K.~E. Tan, ``Achieving security and privacy in federated
  learning systems: Survey, research challenges and future directions,''
  \emph{Engineering Applications of Artificial Intelligence}, vol. 106, p.
  104468, 2021.

\bibitem{yin2021comprehensive}
X.~Yin, Y.~Zhu, and J.~Hu, ``A comprehensive survey of privacy-preserving
  federated learning: A taxonomy, review, and future directions,'' \emph{ACM
  Computing Surveys (CSUR)}, vol.~54, no.~6, pp. 1--36, 2021.

\bibitem{tariq2023trustworthy}
A.~Tariq, M.~A. Serhani, F.~Sallabi, T.~Qayyum, E.~S. Barka, and K.~A. Shuaib,
  ``Trustworthy federated learning: A survey,'' \emph{arXiv preprint
  arXiv:2305.11537}, 2023.

\bibitem{ferrag2021federated}
M.~A. Ferrag, O.~Friha, L.~Maglaras, H.~Janicke, and L.~Shu, ``Federated deep
  learning for cyber security in the internet of things: Concepts,
  applications, and experimental analysis,'' \emph{IEEE Access}, vol.~9, pp.
  138\,509--138\,542, 2021.

\bibitem{fedorchenko2022comparative}
E.~Fedorchenko, E.~Novikova, and A.~Shulepov, ``Comparative review of the
  intrusion detection systems based on federated learning: Advantages and open
  challenges,'' \emph{Algorithms}, vol.~15, no.~7, p. 247, 2022.

\bibitem{belenguer2022review}
A.~Belenguer, J.~Navaridas, and J.~A. Pascual, ``A review of federated learning
  in intrusion detection systems for iot,'' \emph{arXiv preprint
  arXiv:2204.12443}, 2022.

\bibitem{arisdakessian2022survey}
S.~Arisdakessian, O.~A. Wahab, A.~Mourad, H.~Otrok, and M.~Guizani, ``A survey
  on iot intrusion detection: Federated learning, game theory, social
  psychology, and explainable ai as future directions,'' \emph{IEEE Internet of
  Things Journal}, vol.~10, no.~5, pp. 4059--4092, 2022.

\bibitem{venkatasubramanian2023iot}
M.~Venkatasubramanian, A.~H. Lashkari, and S.~Hakak, ``Iot malware analysis
  using federated learning: A comprehensive survey,'' \emph{IEEE Access}, 2023.

\bibitem{keele2007guidelines}
S.~Keele \emph{et~al.}, ``Guidelines for performing systematic literature
  reviews in software engineering,'' Technical report, ver. 2.3 ebse technical
  report. ebse, Tech. Rep., 2007.

\bibitem{ruzafa2021intrusion}
P.~Ruzafa-Alcazar, P.~Fernandez-Saura, E.~Marmol-Campos, A.~Gonzalez-Vidal,
  J.~L.~H. Ramos, J.~Bernal, and A.~F. Skarmeta, ``Intrusion detection based on
  privacy-preserving federated learning for the industrial iot,'' \emph{IEEE
  Transactions on Industrial Informatics}, 2021.

\bibitem{nist80094}
{NIST}, ``\BIBforeignlanguage{English}{Guide to {I}ntrusion {D}etection and
  {P}revention {S}ystems ({IDPS}) - {NIST} {S}pecial {P}ublication 800-94},''
  2007.

\bibitem{da2019internet}
K.~A. da~Costa, J.~P. Papa, C.~O. Lisboa, R.~Munoz, and V.~H.~C.
  de~Albuquerque, ``Internet of things: A survey on machine learning-based
  intrusion detection approaches,'' \emph{Computer Networks}, vol. 151, pp.
  147--157, 2019.

\bibitem{karop22}
\BIBentryALTinterwordspacing
G.~Karopoulos, G.~Kambourakis, E.~Chatzoglou, J.~L. Hernández-Ramos, and
  V.~Kouliaridis, ``Demystifying in-vehicle intrusion detection systems: A
  survey of surveys and a meta-taxonomy,'' \emph{Electronics}, vol.~11, no.~7,
  2022. [Online]. Available: \url{https://www.mdpi.com/2079-9292/11/7/1072}
\BIBentrySTDinterwordspacing

\bibitem{liu2019machine}
H.~Liu and B.~Lang, ``Machine learning and deep learning methods for intrusion
  detection systems: A survey,'' \emph{applied sciences}, vol.~9, no.~20, p.
  4396, 2019.

\bibitem{mahbooba2021explainable}
B.~Mahbooba, M.~Timilsina, R.~Sahal, and M.~Serrano, ``Explainable artificial
  intelligence (xai) to enhance trust management in intrusion detection systems
  using decision tree model,'' \emph{Complexity}, vol. 2021, 2021.

\bibitem{ahmim2019novel}
A.~Ahmim, L.~Maglaras, M.~A. Ferrag, M.~Derdour, and H.~Janicke, ``A novel
  hierarchical intrusion detection system based on decision tree and
  rules-based models,'' in \emph{2019 15th International Conference on
  Distributed Computing in Sensor Systems (DCOSS)}.\hskip 1em plus 0.5em minus
  0.4em\relax IEEE, 2019, pp. 228--233.

\bibitem{panigrahi2021consolidated}
R.~Panigrahi, S.~Borah, A.~K. Bhoi, M.~F. Ijaz, M.~Pramanik, Y.~Kumar, and
  R.~H. Jhaveri, ``A consolidated decision tree-based intrusion detection
  system for binary and multiclass imbalanced datasets,'' \emph{Mathematics},
  vol.~9, no.~7, p. 751, 2021.

\bibitem{mohammadi2021comprehensive}
M.~Mohammadi, T.~A. Rashid, S.~H.~T. Karim, A.~H.~M. Aldalwie, Q.~T. Tho,
  M.~Bidaki, A.~M. Rahmani, and M.~Hosseinzadeh, ``A comprehensive survey and
  taxonomy of the svm-based intrusion detection systems,'' \emph{Journal of
  Network and Computer Applications}, vol. 178, p. 102983, 2021.

\bibitem{drewek2021survey}
A.~Drewek-Ossowicka, M.~Pietro{\l}aj, and J.~Rumi{\'n}ski, ``A survey of neural
  networks usage for intrusion detection systems,'' \emph{Journal of Ambient
  Intelligence and Humanized Computing}, vol.~12, no.~1, pp. 497--514, 2021.

\bibitem{nisioti2018intrusion}
A.~Nisioti, A.~Mylonas, P.~D. Yoo, and V.~Katos, ``From intrusion detection to
  attacker attribution: A comprehensive survey of unsupervised methods,''
  \emph{IEEE Communications Surveys \& Tutorials}, vol.~20, no.~4, pp.
  3369--3388, 2018.

\bibitem{borkar2019novel}
G.~M. Borkar, L.~H. Patil, D.~Dalgade, and A.~Hutke, ``A novel clustering
  approach and adaptive svm classifier for intrusion detection in wsn: A data
  mining concept,'' \emph{Sustainable Computing: Informatics and Systems},
  vol.~23, pp. 120--135, 2019.

\bibitem{li2020enhancing}
W.~Li, W.~Meng, and M.~H. Au, ``Enhancing collaborative intrusion detection via
  disagreement-based semi-supervised learning in iot environments,''
  \emph{Journal of Network and Computer Applications}, vol. 161, p. 102631,
  2020.

\bibitem{gao2018novel}
Y.~Gao, Y.~Liu, Y.~Jin, J.~Chen, and H.~Wu, ``A novel semi-supervised learning
  approach for network intrusion detection on cloud-based robotic system,''
  \emph{IEEE Access}, vol.~6, pp. 50\,927--50\,938, 2018.

\bibitem{kim2020collaborative}
S.~Kim, H.~Cai, C.~Hua, P.~Gu, W.~Xu, and J.~Park, ``Collaborative anomaly
  detection for internet of things based on federated learning,'' in \emph{2020
  IEEE/CIC International Conference on Communications in China (ICCC)}.\hskip
  1em plus 0.5em minus 0.4em\relax IEEE, 2020, pp. 623--628.

\bibitem{nishio2019client}
T.~Nishio and R.~Yonetani, ``Client selection for federated learning with
  heterogeneous resources in mobile edge,'' in \emph{ICC 2019-2019 IEEE
  international conference on communications (ICC)}.\hskip 1em plus 0.5em minus
  0.4em\relax IEEE, 2019, pp. 1--7.

\bibitem{cho2020client}
Y.~J. Cho, J.~Wang, and G.~Joshi, ``Client selection in federated learning:
  Convergence analysis and power-of-choice selection strategies,'' \emph{arXiv
  preprint arXiv:2010.01243}, 2020.

\bibitem{sarhan2021cyber}
M.~Sarhan, S.~Layeghy, N.~Moustafa, and M.~Portmann, ``A cyber threat
  intelligence sharing scheme based on federated learning for network intrusion
  detection,'' \emph{arXiv preprint arXiv:2111.02791}, 2021.

\bibitem{hernandez2019toward}
J.~L. Hern{\'a}ndez-Ramos, D.~Geneiatakis, I.~Kounelis, G.~Steri, and I.~N.
  Fovino, ``Toward a data-driven society: A technological perspective on the
  development of cybersecurity and data-protection policies,'' \emph{IEEE
  Security \& Privacy}, vol.~18, no.~1, pp. 28--38, 2019.

\bibitem{hernandez2021challenges}
J.~L. Hernandez-Ramos, S.~N. Matheu, and A.~Skarmeta, ``The challenges of
  software cybersecurity certification [building security in],'' \emph{IEEE
  Security \& Privacy}, vol.~19, no.~1, pp. 99--102, 2021.

\bibitem{neisse2020interledger}
R.~Neisse, J.~L. Hern{\'a}ndez-Ramos, S.~N. Matheu-Garcia, G.~Baldini,
  A.~Skarmeta, V.~Siris, D.~Lagutin, and P.~Nikander, ``An interledger
  blockchain platform for cross-border management of cybersecurity
  information,'' \emph{IEEE Internet Computing}, vol.~24, no.~3, pp. 19--29,
  2020.

\bibitem{li2021survey}
Q.~Li, Z.~Wen, Z.~Wu, S.~Hu, N.~Wang, Y.~Li, X.~Liu, and B.~He, ``A survey on
  federated learning systems: vision, hype and reality for data privacy and
  protection,'' \emph{IEEE Transactions on Knowledge and Data Engineering},
  2021.

\bibitem{JATAIN2021}
\BIBentryALTinterwordspacing
D.~Jatain, V.~Singh, and N.~Dahiya, ``A contemplative perspective on federated
  machine learning: Taxonomy, threats \& vulnerability assessment and
  challenges,'' \emph{Journal of King Saud University - Computer and
  Information Sciences}, 2021. [Online]. Available:
  \url{https://www.sciencedirect.com/science/article/pii/S1319157821001312}
\BIBentrySTDinterwordspacing

\bibitem{sultana2019survey}
N.~Sultana, N.~Chilamkurti, W.~Peng, and R.~Alhadad, ``Survey on sdn based
  network intrusion detection system using machine learning approaches,''
  \emph{Peer-to-Peer Networking and Applications}, vol.~12, no.~2, pp.
  493--501, 2019.

\bibitem{gamage2020deep}
S.~Gamage and J.~Samarabandu, ``Deep learning methods in network intrusion
  detection: A survey and an objective comparison,'' \emph{Journal of Network
  and Computer Applications}, vol. 169, p. 102767, 2020.

\bibitem{rey2022federated}
V.~Rey, P.~M.~S. S{\'a}nchez, A.~H. Celdr{\'a}n, and G.~Bovet, ``Federated
  learning for malware detection in iot devices,'' \emph{Computer Networks},
  vol. 204, p. 108693, 2022.

\bibitem{sharafaldin2018toward}
I.~Sharafaldin, A.~H. Lashkari, and A.~A. Ghorbani, ``Toward generating a new
  intrusion detection dataset and intrusion traffic characterization.''
  \emph{ICISSp}, vol.~1, pp. 108--116, 2018.

\bibitem{lashkari2017cicflowmeter}
A.~H. Lashkari, Y.~Zang, G.~Owhuo, M.~Mamun, and G.~Gil, ``Cicflowmeter,''
  2017.

\bibitem{ring2017flow}
M.~Ring, S.~Wunderlich, D.~Gr{\"u}dl, D.~Landes, and A.~Hotho, ``Flow-based
  benchmark data sets for intrusion detection,'' in \emph{Proceedings of the
  16th European Conference on Cyber Warfare and Security. ACPI}, 2017, pp.
  361--369.

\bibitem{ring2017creation}
------, ``Creation of flow-based data sets for intrusion detection,''
  \emph{Journal of Information Warfare}, vol.~16, no.~4, pp. 41--54, 2017.

\bibitem{jazi2017detecting}
H.~H. Jazi, H.~Gonzalez, N.~Stakhanova, and A.~A. Ghorbani, ``Detecting
  http-based application layer dos attacks on web servers in the presence of
  sampling,'' \emph{Computer Networks}, vol. 121, pp. 25--36, 2017.

\bibitem{shiravi2012toward}
A.~Shiravi, H.~Shiravi, M.~Tavallaee, and A.~A. Ghorbani, ``Toward developing a
  systematic approach to generate benchmark datasets for intrusion detection,''
  \emph{computers \& security}, vol.~31, no.~3, pp. 357--374, 2012.

\bibitem{sharafaldin2019developing}
I.~Sharafaldin, A.~H. Lashkari, S.~Hakak, and A.~A. Ghorbani, ``Developing
  realistic distributed denial of service (ddos) attack dataset and taxonomy,''
  in \emph{2019 International Carnahan Conference on Security Technology
  (ICCST)}.\hskip 1em plus 0.5em minus 0.4em\relax IEEE, 2019, pp. 1--8.

\bibitem{sharma2018new}
R.~Sharma, R.~Singla, and A.~Guleria, ``A new labeled flow-based dns dataset
  for anomaly detection: Puf dataset,'' \emph{Procedia computer science}, vol.
  132, pp. 1458--1466, 2018.

\bibitem{viegas2017toward}
E.~K. Viegas, A.~O. Santin, and L.~S. Oliveira, ``Toward a reliable
  anomaly-based intrusion detection in real-world environments,''
  \emph{Computer Networks}, vol. 127, pp. 200--216, 2017.

\bibitem{turcotte2019unified}
M.~J. Turcotte, A.~D. Kent, and C.~Hash, ``Unified host and network data set,''
  in \emph{Data Science for Cyber-Security}.\hskip 1em plus 0.5em minus
  0.4em\relax World Scientific, 2019, pp. 1--22.

\bibitem{chatzoglou2021empirical}
E.~Chatzoglou, G.~Kambourakis, and C.~Kolias, ``Empirical evaluation of attacks
  against ieee 802.11 enterprise networks: The awid3 dataset,'' \emph{IEEE
  Access}, vol.~9, pp. 34\,188--34\,205, 2021.

\bibitem{kolias2015intrusion}
C.~Kolias, G.~Kambourakis, A.~Stavrou, and S.~Gritzalis, ``Intrusion detection
  in 802.11 networks: empirical evaluation of threats and a public dataset,''
  \emph{IEEE Communications Surveys \& Tutorials}, vol.~18, no.~1, pp.
  184--208, 2015.

\bibitem{H23Q}
\BIBentryALTinterwordspacing
E.~Chatzoglou, V.~Kouliaridis, G.~Kambourakis, G.~Karopoulos, and S.~Gritzalis,
  ``A hands-on gaze on http/3 security through the lens of http/2 and a public
  dataset,'' \emph{Computers \& Security}, vol. 125, p. 103051, 2023. [Online].
  Available:
  \url{https://www.sciencedirect.com/science/article/pii/S0167404822004436}
\BIBentrySTDinterwordspacing

\bibitem{QUIC:2022}
\BIBentryALTinterwordspacing
E.~Chatzoglou, V.~Kouliaridis, G.~Karopoulos, and G.~Kambourakis, ``Revisiting
  quic attacks: A comprehensive review on quic security and a hands-on study,''
  \emph{International Journal of Information Security}, 2022. [Online].
  Available: \url{https://link.springer.com/article/10.1007/s10207-022-00630-6}
\BIBentrySTDinterwordspacing

\bibitem{preuveneers2018chained}
D.~Preuveneers, V.~Rimmer, I.~Tsingenopoulos, J.~Spooren, W.~Joosen, and
  E.~Ilie-Zudor, ``Chained anomaly detection models for federated learning: An
  intrusion detection case study,'' \emph{Applied Sciences}, vol.~8, no.~12, p.
  2663, 2018.

\bibitem{zhao2019multi}
Y.~Zhao, J.~Chen, D.~Wu, J.~Teng, and S.~Yu, ``Multi-task network anomaly
  detection using federated learning,'' in \emph{Proceedings of the tenth
  international symposium on information and communication technology}, 2019,
  pp. 273--279.

\bibitem{chen2020intrusion}
Z.~Chen, N.~Lv, P.~Liu, Y.~Fang, K.~Chen, and W.~Pan, ``Intrusion detection for
  wireless edge networks based on federated learning,'' \emph{IEEE Access},
  vol.~8, pp. 217\,463--217\,472, 2020.

\bibitem{qin2020line}
Q.~Qin, K.~Poularakis, K.~K. Leung, and L.~Tassiulas, ``Line-speed and scalable
  intrusion detection at the network edge via federated learning,'' in
  \emph{2020 IFIP Networking Conference (Networking)}.\hskip 1em plus 0.5em
  minus 0.4em\relax IEEE, 2020, pp. 352--360.

\bibitem{fan2020iotdefender}
Y.~Fan, Y.~Li, M.~Zhan, H.~Cui, and Y.~Zhang, ``Iotdefender: A federated
  transfer learning intrusion detection framework for 5g iot,'' in \emph{2020
  IEEE 14th International Conference on Big Data Science and Engineering
  (BigDataSE)}.\hskip 1em plus 0.5em minus 0.4em\relax IEEE, 2020, pp. 88--95.

\bibitem{qin2021fnel}
T.~Qin, G.~Cheng, W.~Chen, and X.~Lei, ``Fnel: An evolving intrusion detection
  system based on federated never-ending learning,'' in \emph{2021 17th
  International Conference on Mobility, Sensing and Networking (MSN)}.\hskip
  1em plus 0.5em minus 0.4em\relax IEEE, 2021, pp. 239--246.

\bibitem{kwon2022anomaly}
J.~Kwon, B.~Jung, H.~Lee, and S.~Lee, ``Anomaly detection in multi-host
  environment based on federated hypersphere classifier,'' \emph{Electronics},
  vol.~11, no.~10, p. 1529, 2022.

\bibitem{otoum2021federated}
Y.~Otoum, Y.~Wan, and A.~Nayak, ``Federated transfer learning-based ids for the
  internet of medical things (iomt),'' in \emph{2021 IEEE Globecom Workshops
  (GC Wkshps)}.\hskip 1em plus 0.5em minus 0.4em\relax IEEE, 2021, pp. 1--6.

\bibitem{yadav2021unsupervised}
K.~Yadav, B.~B. Gupta, C.-H. Hsu, and K.~T. Chui, ``Unsupervised federated
  learning based iot intrusion detection,'' in \emph{2021 IEEE 10th Global
  Conference on Consumer Electronics (GCCE)}.\hskip 1em plus 0.5em minus
  0.4em\relax IEEE, 2021, pp. 298--301.

\bibitem{ayed2021federated}
M.~A. Ayed and C.~Talhi, ``Federated learning for anomaly-based intrusion
  detection,'' in \emph{2021 International Symposium on Networks, Computers and
  Communications (ISNCC)}.\hskip 1em plus 0.5em minus 0.4em\relax IEEE, 2021,
  pp. 1--8.

\bibitem{hao2021secure}
H.~N. Hao, H.~M. Chu, V.-H. Pham \emph{et~al.}, ``A secure and privacy
  preserving federated learning approach for iot intrusion detection system,''
  in \emph{International Conference on Network and System Security}.\hskip 1em
  plus 0.5em minus 0.4em\relax Springer, 2021, pp. 353--368.

\bibitem{wettlaufer2021property}
J.~Wettlaufer, ``Property inference-based federated learning groups for
  collaborative network anomaly detection,'' \emph{Electronic Communications of
  the EASST}, vol.~80, 2021.

\bibitem{mcosker2021architecture}
C.~L. McOsker, M.~S. Handlin, L.~Li, H.~Shahriar, and L.~Zhao, ``An
  architecture for federated learning enabled collaborative intrusion detection
  system,'' 2021.

\bibitem{wei2021federated}
Y.~Wei, S.~Zhou, S.~Leng, S.~Maharjan, and Y.~Zhang, ``Federated learning
  empowered end-edge-cloud cooperation for 5g hetnet security,'' \emph{IEEE
  Network}, vol.~35, no.~2, pp. 88--94, 2021.

\bibitem{otoum2022feasibility}
S.~Otoum, N.~Guizani, and H.~Mouftah, ``On the feasibility of split learning,
  transfer learning and federated learning for preserving security in its
  systems,'' \emph{IEEE Transactions on Intelligent Transportation Systems},
  2022.

\bibitem{zhang2021flddos}
J.~Zhang, P.~Yu, L.~Qi, S.~Liu, H.~Zhang, and J.~Zhang, ``Flddos: Ddos attack
  detection model based on federated learning,'' in \emph{2021 IEEE 20th
  International Conference on Trust, Security and Privacy in Computing and
  Communications (TrustCom)}.\hskip 1em plus 0.5em minus 0.4em\relax IEEE,
  2021, pp. 635--642.

\bibitem{otoum2021federatedreinforcement}
S.~Otoum, N.~Guizani, and H.~Mouftah, ``Federated reinforcement
  learning-supported ids for iot-steered healthcare systems,'' in \emph{ICC
  2021-IEEE International Conference on Communications}.\hskip 1em plus 0.5em
  minus 0.4em\relax IEEE, 2021, pp. 1--6.

\bibitem{tang2022federated}
Z.~Tang, H.~Hu, and C.~Xu, ``A federated learning method for network intrusion
  detection,'' \emph{Concurrency and Computation: Practice and Experience},
  vol.~34, no.~10, p. e6812, 2022.

\bibitem{markovic2022random}
T.~Markovic, M.~Leon, D.~Buffoni, and S.~Punnekkat, ``Random forest based on
  federated learning for intrusion detection,'' in \emph{IFIP International
  Conference on Artificial Intelligence Applications and Innovations}.\hskip
  1em plus 0.5em minus 0.4em\relax Springer, 2022, pp. 132--144.

\bibitem{chen2022privacy}
Y.~Chen, J.~Zhang, and C.~K. Yeo, ``Privacy-preserving knowledge transfer for
  intrusion detection with federated deep autoencoding gaussian mixture
  model,'' \emph{Information Sciences}, vol. 609, pp. 1204--1220, 2022.

\bibitem{popoola2021federated}
S.~I. Popoola, G.~Gui, B.~Adebisi, M.~Hammoudeh, and H.~Gacanin, ``Federated
  deep learning for collaborative intrusion detection in heterogeneous
  networks,'' in \emph{2021 IEEE 94th Vehicular Technology Conference
  (VTC2021-Fall)}.\hskip 1em plus 0.5em minus 0.4em\relax IEEE, 2021, pp. 1--6.

\bibitem{shi2021data}
J.~Shi, B.~Ge, Y.~Liu, Y.~Yan, and S.~Li, ``Data privacy security guaranteed
  network intrusion detection system based on federated learning,'' in
  \emph{IEEE INFOCOM 2021-IEEE Conference on Computer Communications Workshops
  (INFOCOM WKSHPS)}.\hskip 1em plus 0.5em minus 0.4em\relax IEEE, 2021, pp.
  1--6.

\bibitem{friha2022felids}
O.~Friha, M.~A. Ferrag, L.~Shu, L.~Maglaras, K.-K.~R. Choo, and M.~Nafaa,
  ``Felids: Federated learning-based intrusion detection system for
  agricultural internet of things,'' \emph{Journal of Parallel and Distributed
  Computing}, vol. 165, pp. 17--31, 2022.

\bibitem{zhang2022secfednids}
Z.~Zhang, Y.~Zhang, D.~Guo, L.~Yao, and Z.~Li, ``Secfednids: Robust defense for
  poisoning attack against federated learning-based network intrusion detection
  system,'' \emph{Future Generation Computer Systems}, vol. 134, pp. 154--169,
  2022.

\bibitem{li2021fids}
J.~Li, Z.~Zhang, Y.~Li, X.~Guo, and H.~Li, ``Fids: Detecting ddos through
  federated learning based method,'' in \emph{2021 IEEE 20th International
  Conference on Trust, Security and Privacy in Computing and Communications
  (TrustCom)}.\hskip 1em plus 0.5em minus 0.4em\relax IEEE, 2021, pp. 856--862.

\bibitem{yuan2021towards}
S.~Yuan, H.~Li, R.~Zhang, M.~Hao, Y.~Li, and R.~Lu, ``Towards lightweight and
  efficient distributed intrusion detection framework,'' in \emph{2021 IEEE
  Global Communications Conference (GLOBECOM)}.\hskip 1em plus 0.5em minus
  0.4em\relax IEEE, 2021, pp. 1--6.

\bibitem{duy2021federated}
P.~T. Duy, T.~Van~Hung, N.~H. Ha, H.~Do~Hoang, and V.-H. Pham, ``Federated
  learning-based intrusion detection in sdn-enabled iiot networks,'' in
  \emph{2021 8th NAFOSTED Conference on Information and Computer Science
  (NICS)}.\hskip 1em plus 0.5em minus 0.4em\relax IEEE, 2021, pp. 424--429.

\bibitem{dong2021towards}
T.~Dong, H.~Qiu, J.~Lu, M.~Qiu, and C.~Fan, ``Towards fast network intrusion
  detection based on efficiency-preserving federated learning,'' in \emph{2021
  IEEE Intl Conf on Parallel \& Distributed Processing with Applications, Big
  Data \& Cloud Computing, Sustainable Computing \& Communications, Social
  Computing \& Networking (ISPA/BDCloud/SocialCom/SustainCom)}.\hskip 1em plus
  0.5em minus 0.4em\relax IEEE, 2021, pp. 468--475.

\bibitem{lv2022ddos}
D.~Lv, X.~Cheng, J.~Zhang, W.~Zhang, W.~Zhao, and H.~Xu, ``Ddos attack
  detection based on cnn and federated learning,'' in \emph{2021 Ninth
  International Conference on Advanced Cloud and Big Data (CBD)}.\hskip 1em
  plus 0.5em minus 0.4em\relax IEEE, 2022, pp. 236--241.

\bibitem{BestFeaturesAWID3}
E.~Chatzoglou, G.~Kambourakis, C.~Kolias, and C.~Smiliotopoulos, ``Pick quality
  over quantity: Expert feature selection and data preprocessing for 802.11
  intrusion detection systems,'' \emph{IEEE Access}, vol.~10, pp.
  64\,761--64\,784, 2022.

\bibitem{AWID3:BestFeatureSelection2}
\BIBentryALTinterwordspacing
E.~Chatzoglou, G.~Kambourakis, C.~Smiliotopoulos, and C.~Kolias, ``Best of both
  worlds: Detecting application layer attacks through 802.11 and non-802.11
  features,'' \emph{Sensors}, vol.~22, no.~15, 2022. [Online]. Available:
  \url{https://www.mdpi.com/1424-8220/22/15/5633}
\BIBentrySTDinterwordspacing

\bibitem{meidan2018n}
Y.~Meidan, M.~Bohadana, Y.~Mathov, Y.~Mirsky, A.~Shabtai, D.~Breitenbacher, and
  Y.~Elovici, ``N-baiot—network-based detection of iot botnet attacks using
  deep autoencoders,'' \emph{IEEE Pervasive Computing}, vol.~17, no.~3, pp.
  12--22, 2018.

\bibitem{Mirai-Kolias}
\BIBentryALTinterwordspacing
C.~Kolias, G.~Kambourakis, A.~Stavrou, and J.~M. Voas, ``Ddos in the iot: Mirai
  and other botnets,'' \emph{Computer}, vol.~50, no.~7, pp. 80--84, 2017.
  [Online]. Available: \url{https://doi.org/10.1109/MC.2017.201}
\BIBentrySTDinterwordspacing

\bibitem{mirsky2018kitsune}
Y.~Mirsky, T.~Doitshman, Y.~Elovici, and A.~Shabtai, ``Kitsune: an ensemble of
  autoencoders for online network intrusion detection,'' \emph{arXiv preprint
  arXiv:1802.09089}, 2018.

\bibitem{koroniotis2019towards}
N.~Koroniotis, N.~Moustafa, E.~Sitnikova, and B.~Turnbull, ``Towards the
  development of realistic botnet dataset in the internet of things for network
  forensic analytics: Bot-iot dataset,'' \emph{Future Generation Computer
  Systems}, vol. 100, pp. 779--796, 2019.

\bibitem{zolanvari2019machine}
M.~Zolanvari, M.~A. Teixeira, L.~Gupta, K.~M. Khan, and R.~Jain, ``Machine
  learning-based network vulnerability analysis of industrial internet of
  things,'' \emph{IEEE Internet of Things Journal}, vol.~6, no.~4, pp.
  6822--6834, 2019.

\bibitem{q70p-q449-19}
\BIBentryALTinterwordspacing
H.~Kang, D.~H. Ahn, G.~M. Lee, J.~D. Yoo, K.~H. Park, and H.~K. Kim, ``Iot
  network intrusion dataset,'' 2019. [Online]. Available:
  \url{https://dx.doi.org/10.21227/q70p-q449}
\BIBentrySTDinterwordspacing

\bibitem{ullah2020scheme}
I.~Ullah and Q.~H. Mahmoud, ``A scheme for generating a dataset for anomalous
  activity detection in iot networks.'' in \emph{Canadian Conference on AI},
  2020, pp. 508--520.

\bibitem{iot23}
\BIBentryALTinterwordspacing
G.~Sebastian, P.~Agustin, and M.~J. Erquiaga, ``Iot-23 dataset,'' 2020.
  [Online]. Available: \url{https://www.stratosphereips.org/datasets-iot23}
\BIBentrySTDinterwordspacing

\bibitem{0s0p-s959-21}
\BIBentryALTinterwordspacing
F.~Hussain, S.~G. Abbas, M.~Husnain, U.~U. Fayyaz, F.~Shahzad, and G.~A. Shah,
  ``Iot dos and ddos attack dataset,'' 2021. [Online]. Available:
  \url{https://dx.doi.org/10.21227/0s0p-s959}
\BIBentrySTDinterwordspacing

\bibitem{bhxy-ep04-20}
\BIBentryALTinterwordspacing
H.~Hindy, C.~Tachtatzis, R.~Atkinson, E.~Bayne, and X.~Bellekens,
  ``Mqtt-iot-ids2020: Mqtt internet of things intrusion detection dataset,''
  2020. [Online]. Available: \url{https://dx.doi.org/10.21227/bhxy-ep04}
\BIBentrySTDinterwordspacing

\bibitem{vaccari2020mqttset}
I.~Vaccari, G.~Chiola, M.~Aiello, M.~Mongelli, and E.~Cambiaso, ``Mqttset, a
  new dataset for machine learning techniques on mqtt,'' \emph{Sensors},
  vol.~20, no.~22, p. 6578, 2020.

\bibitem{9w13-2t13-21}
\BIBentryALTinterwordspacing
F.~Hussain, S.~G. Abbas, G.~A.~Shah, I.~M. Pires, U.~U. Fayyaz, F.~Shahzad,
  N.~M. Garcia, and E.~Zdravevski, ``Iot healthcare security dataset,'' 2021.
  [Online]. Available: \url{https://dx.doi.org/10.21227/9w13-2t13}
\BIBentrySTDinterwordspacing

\bibitem{liu2021using}
Z.~Liu, N.~Thapa, A.~Shaver, K.~Roy, M.~Siddula, X.~Yuan, and A.~Yu, ``Using
  embedded feature selection and cnn for classification on ccd-inid-v1—a new
  iot dataset,'' \emph{Sensors}, vol.~21, no.~14, p. 4834, 2021.

\bibitem{al2021x}
M.~Al-Hawawreh, E.~Sitnikova, and N.~Aboutorab, ``X-iiotid: A connectivity-and
  device-agnostic intrusion dataset for industrial internet of things,''
  \emph{IEEE Internet of Things Journal}, 2021.

\bibitem{ferrag2022edge}
M.~A. Ferrag, O.~Friha, D.~Hamouda, L.~Maglaras, and H.~Janicke,
  ``Edge-iiotset: A new comprehensive realistic cyber security dataset of iot
  and iiot applications for centralized and federated learning,'' \emph{IEEE
  Access}, 2022.

\bibitem{almaraz2022toward}
J.~G. Almaraz-Rivera, J.~A. Perez-Diaz, J.~A. Cantoral-Ceballos, J.~F. Botero,
  and L.~A. Trejo, ``Toward the protection of iot networks: Introducing the
  latam-ddos-iot dataset,'' \emph{IEEE Access}, vol.~10, pp.
  106\,909--106\,920, 2022.

\bibitem{dadkhah2022towards}
S.~Dadkhah, H.~Mahdikhani, P.~K. Danso, A.~Zohourian, K.~A. Truong, and A.~A.
  Ghorbani, ``Towards the development of a realistic multidimensional iot
  profiling dataset,'' in \emph{2022 19th Annual International Conference on
  Privacy, Security \& Trust (PST)}.\hskip 1em plus 0.5em minus 0.4em\relax
  IEEE, 2022, pp. 1--11.

\bibitem{alsaedi2020ton_iot}
A.~Alsaedi, N.~Moustafa, Z.~Tari, A.~Mahmood, and A.~Anwar, ``Ton\_iot
  telemetry dataset: A new generation dataset of iot and iiot for data-driven
  intrusion detection systems,'' \emph{IEEE Access}, vol.~8, pp.
  165\,130--165\,150, 2020.

\bibitem{guerra2020medbiot}
A.~Guerra-Manzanares, J.~Medina-Galindo, H.~Bahsi, and S.~N{\~o}mm, ``Medbiot:
  Generation of an iot botnet dataset in a medium-sized iot network.'' in
  \emph{ICISSP}, 2020, pp. 207--218.

\bibitem{zhao2022semi}
R.~Zhao, Y.~Wang, Z.~Xue, T.~Ohtsuki, B.~Adebisi, and G.~Gui, ``Semi-supervised
  federated learning based intrusion detection method for internet of things,''
  \emph{IEEE Internet of Things Journal}, 2022.

\bibitem{khoa2020collaborative}
T.~V. Khoa, Y.~M. Saputra, D.~T. Hoang, N.~L. Trung, D.~Nguyen, N.~V. Ha, and
  E.~Dutkiewicz, ``Collaborative learning model for cyberattack detection
  systems in iot industry 4.0,'' in \emph{2020 IEEE Wireless Communications and
  Networking Conference (WCNC)}.\hskip 1em plus 0.5em minus 0.4em\relax IEEE,
  2020, pp. 1--6.

\bibitem{popoola2021federatedzero}
S.~I. Popoola, R.~Ande, B.~Adebisi, G.~Gui, M.~Hammoudeh, and O.~Jogunola,
  ``Federated deep learning for zero-day botnet attack detection in iot edge
  devices,'' \emph{IEEE Internet of Things Journal}, 2021.

\bibitem{zakariyya2021memory}
I.~Zakariyya, H.~Kalutarage, and M.~O. Al-Kadri, ``Memory efficient federated
  deep learning for intrusion detection in iot networks.''\hskip 1em plus 0.5em
  minus 0.4em\relax CEUR Workshop Proceedings, 2021.

\bibitem{vy2021federated}
N.~C. Vy, N.~H. Quyen, V.-H. Pham \emph{et~al.}, ``Federated learning-based
  intrusion detection in the context of iiot networks: Poisoning attack and
  defense,'' in \emph{International Conference on Network and System
  Security}.\hskip 1em plus 0.5em minus 0.4em\relax Springer, 2021, pp.
  131--147.

\bibitem{LocKedge}
T.~Huong, P.~B. Ta, D.~Long, B.~Thang, N.~Binh, T.~Luong, and K.~P. TRAN,
  ``Lockedge: Low-complexity cyberattack detection in iot edge computing,''
  \emph{IEEE Access}, vol.~PP, 2021.

\bibitem{attota2021ensemble}
D.~C. Attota, V.~Mothukuri, R.~M. Parizi, and S.~Pouriyeh, ``An ensemble
  multi-view federated learning intrusion detection for iot,'' \emph{IEEE
  Access}, vol.~9, pp. 117\,734--117\,745, 2021.

\bibitem{kumar2021pefl}
P.~Kumar, G.~P. Gupta, and R.~Tripathi, ``Pefl: Deep privacy-encoding-based
  federated learning framework for smart agriculture,'' \emph{IEEE Micro},
  vol.~42, no.~1, pp. 33--40, 2021.

\bibitem{de2022improving}
G.~de~Carvalho~Bertoli, L.~A. Pereira~J{\'u}nior, and O.~Saotome, ``Improving
  detection of scanning attacks on heterogeneous networks with federated
  learning,'' \emph{ACM SIGMETRICS Performance Evaluation Review}, vol.~49,
  no.~4, pp. 118--123, 2022.

\bibitem{singh2022dew}
P.~Singh, G.~S. Gaba, A.~Kaur, M.~Hedabou, and A.~Gurtov, ``Dew-cloud-based
  hierarchical federated learning for intrusion detection in iomt,'' \emph{IEEE
  Journal of Biomedical and Health Informatics}, 2022.

\bibitem{elsayed2020insdn}
M.~S. Elsayed, N.-A. Le-Khac, and A.~D. Jurcut, ``Insdn: A novel sdn intrusion
  dataset,'' \emph{IEEE Access}, vol.~8, pp. 165\,263--165\,284, 2020.

\bibitem{lee2017otids}
H.~Lee, S.~H. Jeong, and H.~K. Kim, ``Otids: A novel intrusion detection system
  for in-vehicle network by using remote frame,'' in \emph{2017 15th Annual
  Conference on Privacy, Security and Trust (PST)}.\hskip 1em plus 0.5em minus
  0.4em\relax IEEE, 2017, pp. 57--5709.

\bibitem{heijden2018veremi}
R.~W. Heijden, T.~Lukaseder, and F.~Kargl, ``Veremi: A dataset for comparable
  evaluation of misbehavior detection in vanets,'' in \emph{International
  conference on security and privacy in communication systems}.\hskip 1em plus
  0.5em minus 0.4em\relax Springer, 2018, pp. 318--337.

\bibitem{00dg-0d12-20}
\BIBentryALTinterwordspacing
J.~Whelan, T.~Sangarapillai, O.~Minawi, A.~Almehmadi, and K.~El-Khatib, ``Uav
  attack dataset,'' 2020. [Online]. Available:
  \url{https://dx.doi.org/10.21227/00dg-0d12}
\BIBentrySTDinterwordspacing

\bibitem{24m9-a446-19}
\BIBentryALTinterwordspacing
M.~Sami, ``Intrusion detection in can bus,'' 2019. [Online]. Available:
  \url{https://dx.doi.org/10.21227/24m9-a446}
\BIBentrySTDinterwordspacing

\bibitem{song2020vehicle}
H.~M. Song, J.~Woo, and H.~K. Kim, ``In-vehicle network intrusion detection
  using deep convolutional neural network,'' \emph{Vehicular Communications},
  vol.~21, p. 100198, 2020.

\bibitem{kang2021car}
H.~Kang, B.~I. Kwak, Y.~H. Lee, H.~Lee, H.~Lee, and H.~K. Kim, ``Car hacking
  and defense competition on in-vehicle network,'' in \emph{Workshop on
  Automotive and Autonomous Vehicle Security (AutoSec)}, vol. 2021, 2021,
  p.~25.

\bibitem{aliyu2021blockchain}
I.~Aliyu, M.~C. Feliciano, S.~Van~Engelenburg, D.~O. Kim, and C.~G. Lim, ``A
  blockchain-based federated forest for sdn-enabled in-vehicle network
  intrusion detection system,'' \emph{IEEE Access}, vol.~9, pp.
  102\,593--102\,608, 2021.

\bibitem{uprety2021privacy}
A.~Uprety, D.~B. Rawat, and J.~Li, ``Privacy preserving misbehavior detection
  in iov using federated machine learning,'' in \emph{2021 IEEE 18th Annual
  Consumer Communications \& Networking Conference (CCNC)}.\hskip 1em plus
  0.5em minus 0.4em\relax IEEE, 2021, pp. 1--6.

\bibitem{whelan2020novelty}
J.~Whelan, T.~Sangarapillai, O.~Minawi, A.~Almehmadi, and K.~El-Khatib,
  ``Novelty-based intrusion detection of sensor attacks on unmanned aerial
  vehicles,'' in \emph{Proceedings of the 16th ACM symposium on QoS and
  security for wireless and mobile networks}, 2020, pp. 23--28.

\bibitem{abdel2021federated}
M.~Abdel-Basset, N.~Moustafa, H.~Hawash, I.~Razzak, K.~M. Sallam, and O.~M.
  Elkomy, ``Federated intrusion detection in blockchain-based smart
  transportation systems,'' \emph{IEEE Transactions on Intelligent
  Transportation Systems}, vol.~23, no.~3, pp. 2523--2537, 2021.

\bibitem{driss2022federated}
M.~Driss, I.~Almomani, J.~Ahmad \emph{et~al.}, ``A federated learning framework
  for cyberattack detection in vehicular sensor networks,'' \emph{Complex \&
  Intelligent Systems}, pp. 1--15, 2022.

\bibitem{tavallaee2009detailed}
M.~Tavallaee, E.~Bagheri, W.~Lu, and A.~A. Ghorbani, ``A detailed analysis of
  the kdd cup 99 data set,'' in \emph{2009 IEEE symposium on computational
  intelligence for security and defense applications}.\hskip 1em plus 0.5em
  minus 0.4em\relax IEEE, 2009, pp. 1--6.

\bibitem{bolon2011feature}
V.~Bolon-Canedo, N.~Sanchez-Marono, and A.~Alonso-Betanzos, ``Feature selection
  and classification in multiple class datasets: An application to kdd cup 99
  dataset,'' \emph{Expert Systems with Applications}, vol.~38, no.~5, pp.
  5947--5957, 2011.

\bibitem{al2020federated}
N.~A. A.-A. Al-Marri, B.~S. Ciftler, and M.~M. Abdallah, ``Federated mimic
  learning for privacy preserving intrusion detection,'' in \emph{2020 IEEE
  International Black Sea Conference on Communications and Networking
  (BlackSeaCom)}.\hskip 1em plus 0.5em minus 0.4em\relax IEEE, 2020, pp. 1--6.

\bibitem{rahman2020internet}
S.~A. Rahman, H.~Tout, C.~Talhi, and A.~Mourad, ``Internet of things intrusion
  detection: Centralized, on-device, or federated learning?'' \emph{IEEE
  Network}, vol.~34, no.~6, pp. 310--317, 2020.

\bibitem{mirzaee2021fids}
P.~H. Mirzaee, M.~Shojafar, Z.~Pooranian, P.~Asefy, H.~Cruickshank, and
  R.~Tafazolli, ``Fids: A federated intrusion detection system for 5g smart
  metering network,'' in \emph{2021 17th International Conference on Mobility,
  Sensing and Networking (MSN)}.\hskip 1em plus 0.5em minus 0.4em\relax IEEE,
  2021, pp. 215--222.

\bibitem{wang2022feco}
N.~Wang, Y.~Chen, Y.~Hu, W.~Lou, and Y.~T. Hou, ``Feco: Boosting intrusion
  detection capability in iot networks via contrastive learning,'' in
  \emph{IEEE INFOCOM 2022-IEEE Conference on Computer Communications}.\hskip
  1em plus 0.5em minus 0.4em\relax IEEE, 2022, pp. 1409--1418.

\bibitem{shahid2021detecting}
O.~Shahid, V.~Mothukuri, S.~Pouriyeh, R.~M. Parizi, and H.~Shahriar,
  ``Detecting network attacks using federated learning for iot devices,'' in
  \emph{2021 IEEE 29th International Conference on Network Protocols
  (ICNP)}.\hskip 1em plus 0.5em minus 0.4em\relax IEEE, 2021, pp. 1--6.

\bibitem{saadat2021hierarchical}
H.~Saadat, A.~Aboumadi, A.~Mohamed, A.~Erbad, and M.~Guizani, ``Hierarchical
  federated learning for collaborative ids in iot applications,'' in \emph{2021
  10th Mediterranean Conference on Embedded Computing (MECO)}.\hskip 1em plus
  0.5em minus 0.4em\relax IEEE, 2021, pp. 1--6.

\bibitem{liu2022intrusion}
W.~Liu, X.~Xu, L.~Wu, L.~Qi, A.~Jolfaei, W.~Ding, and M.~R. Khosravi,
  ``Intrusion detection for maritime transportation systems with batch
  federated aggregation,'' \emph{IEEE Transactions on Intelligent
  Transportation Systems}, 2022.

\bibitem{luo2022federation}
J.~Luo, X.~Yang, and M.~N. Mohammed, ``Federation learning for intrusion
  detection methods by parse convolutional neural network,'' in \emph{2022
  Second International Conference on Advances in Electrical, Computing,
  Communication and Sustainable Technologies (ICAECT)}.\hskip 1em plus 0.5em
  minus 0.4em\relax IEEE, 2022, pp. 1--7.

\bibitem{yang2022federation}
X.~Yang, J.~Luo, and M.~N. Mohammed, ``Federation learning of optimized
  convolutional neural network structure for intrusion detection,'' in
  \emph{2022 Second International Conference on Advances in Electrical,
  Computing, Communication and Sustainable Technologies (ICAECT)}.\hskip 1em
  plus 0.5em minus 0.4em\relax IEEE, 2022, pp. 1--7.

\bibitem{gaussian2020network}
F.~D.~A. Gaussian, ``Network anomaly detection using federated deep
  autoencoding gaussian mixture model,'' in \emph{Machine Learning for
  Networking: Second IFIP TC 6 International Conference, MLN 2019, Paris,
  France, December 3-5, 2019, Revised Selected Papers}, vol. 12081.\hskip 1em
  plus 0.5em minus 0.4em\relax Springer Nature, 2020, p.~1.

\bibitem{hei2020trusted}
X.~Hei, X.~Yin, Y.~Wang, J.~Ren, and L.~Zhu, ``A trusted feature aggregator
  federated learning for distributed malicious attack detection,''
  \emph{Computers \& Security}, vol.~99, p. 102033, 2020.

\bibitem{liu2021blockchain}
H.~Liu, S.~Zhang, P.~Zhang, X.~Zhou, X.~Shao, G.~Pu, and Y.~Zhang, ``Blockchain
  and federated learning for collaborative intrusion detection in vehicular
  edge computing,'' \emph{IEEE Transactions on Vehicular Technology}, 2021.

\bibitem{su2022detection}
D.~Su and Z.~Qu, ``Detection ddos of attacks based on federated learning with
  digital twin network,'' in \emph{International Conference on Knowledge
  Science, Engineering and Management}.\hskip 1em plus 0.5em minus 0.4em\relax
  Springer, 2022, pp. 153--164.

\bibitem{almomani2016wsn}
I.~Almomani, B.~Al-Kasasbeh, and M.~Al-Akhras, ``Wsn-ds: A dataset for
  intrusion detection systems in wireless sensor networks,'' \emph{Journal of
  Sensors}, vol. 2016, 2016.

\bibitem{beigi2014towards}
E.~B. Beigi, H.~H. Jazi, N.~Stakhanova, and A.~A. Ghorbani, ``Towards effective
  feature selection in machine learning-based botnet detection approaches,'' in
  \emph{2014 IEEE Conference on Communications and Network Security}.\hskip 1em
  plus 0.5em minus 0.4em\relax IEEE, 2014, pp. 247--255.

\bibitem{moustafa2015unsw}
N.~Moustafa and J.~Slay, ``Unsw-nb15: a comprehensive data set for network
  intrusion detection systems (unsw-nb15 network data set),'' in \emph{2015
  military communications and information systems conference (MilCIS)}.\hskip
  1em plus 0.5em minus 0.4em\relax IEEE, 2015, pp. 1--6.

\bibitem{zhao2020network}
Y.~Zhao, J.~Chen, Q.~Guo, J.~Teng, and D.~Wu, ``Network anomaly detection using
  federated learning and transfer learning,'' in \emph{International Conference
  on Security and Privacy in Digital Economy}.\hskip 1em plus 0.5em minus
  0.4em\relax Springer, 2020, pp. 219--231.

\bibitem{aouedi2022fluids}
O.~Aouedi, K.~Piamrat, G.~Muller, and K.~Singh, ``Fluids: Federated learning
  with semi-supervised approach for intrusion detection system,'' in \emph{2022
  IEEE 19th Annual Consumer Communications \& Networking Conference
  (CCNC)}.\hskip 1em plus 0.5em minus 0.4em\relax IEEE, 2022, pp. 523--524.

\bibitem{lalouani2022robust}
W.~Lalouani and M.~Younis, ``A robust distributed intrusion detection system
  for collusive attacks on edge of things,'' in \emph{2022 IEEE Wireless
  Communications and Networking Conference (WCNC)}.\hskip 1em plus 0.5em minus
  0.4em\relax IEEE, 2022, pp. 1004--1009.

\bibitem{cheng2022federated}
Y.~Cheng, J.~Lu, D.~Niyato, B.~Lyu, J.~Kang, and S.~Zhu, ``Federated transfer
  learning with client selection for intrusion detection in mobile edge
  computing,'' \emph{IEEE Communications Letters}, vol.~26, no.~3, pp.
  552--556, 2022.

\bibitem{iot-ddos}
\BIBentryALTinterwordspacing
``Ddos botnet attack on iot devices,'' 2019. [Online]. Available:
  \url{https://www.kaggle.com/datasets/siddharthm1698/ddos-botnet-attack-on-iot-devices}
\BIBentrySTDinterwordspacing

\bibitem{teixeira2018scada}
M.~A. Teixeira, T.~Salman, M.~Zolanvari, R.~Jain, N.~Meskin, and M.~Samaka,
  ``Scada system testbed for cybersecurity research using machine learning
  approach,'' \emph{Future Internet}, vol.~10, no.~8, p.~76, 2018.

\bibitem{zhou2020federated}
P.~Zhou, ``Federated deep payload classification for industrial internet with
  cloud-edge architecture,'' in \emph{2020 16th International Conference on
  Mobility, Sensing and Networking (MSN)}.\hskip 1em plus 0.5em minus
  0.4em\relax IEEE, 2020, pp. 228--235.

\bibitem{igbe2017deterministic}
O.~Igbe, I.~Darwish, and T.~Saadawi, ``Deterministic dendritic cell algorithm
  application to smart grid cyber-attack detection,'' in \emph{2017 IEEE 4th
  International Conference on Cyber Security and Cloud Computing
  (CSCloud)}.\hskip 1em plus 0.5em minus 0.4em\relax IEEE, 2017, pp. 199--204.

\bibitem{fovino2009design}
I.~N. Fovino, A.~Carcano, M.~Masera, and A.~Trombetta, ``Design and
  implementation of a secure modbus protocol,'' in \emph{International
  conference on critical infrastructure protection}.\hskip 1em plus 0.5em minus
  0.4em\relax Springer, 2009, pp. 83--96.

\bibitem{morris2014industrial}
T.~Morris and W.~Gao, ``Industrial control system traffic data sets for
  intrusion detection research,'' in \emph{International conference on critical
  infrastructure protection}.\hskip 1em plus 0.5em minus 0.4em\relax Springer,
  2014, pp. 65--78.

\bibitem{east2009taxonomy}
S.~East, J.~Butts, M.~Papa, and S.~Shenoi, ``A taxonomy of attacks on the dnp3
  protocol,'' in \emph{International Conference on Critical Infrastructure
  Protection}.\hskip 1em plus 0.5em minus 0.4em\relax Springer, 2009, pp.
  67--81.

\bibitem{li2020deepfed}
B.~Li, Y.~Wu, J.~Song, R.~Lu, T.~Li, and L.~Zhao, ``Deepfed: Federated deep
  learning for intrusion detection in industrial cyber--physical systems,''
  \emph{IEEE Transactions on Industrial Informatics}, vol.~17, no.~8, pp.
  5615--5624, 2020.

\bibitem{aouedi2022federated}
O.~Aouedi, K.~Piamrat, G.~Muller, and K.~Singh, ``Federated semi-supervised
  learning for attack detection in industrial internet of things,'' \emph{IEEE
  Transactions on Industrial Informatics}, 2022.

\bibitem{kelli2021ids}
V.~Kelli, V.~Argyriou, T.~Lagkas, G.~Fragulis, E.~Grigoriou, and
  P.~Sarigiannidis, ``Ids for industrial applications: a federated learning
  approach with active personalization,'' \emph{Sensors}, vol.~21, no.~20, p.
  6743, 2021.

\bibitem{mothukuri2021federated}
V.~Mothukuri, P.~Khare, R.~M. Parizi, S.~Pouriyeh, A.~Dehghantanha, and
  G.~Srivastava, ``Federated learning-based anomaly detection for iot security
  attacks,'' \emph{IEEE Internet of Things Journal}, 2021.

\bibitem{frazao2018denial}
I.~Fraz{\~a}o, P.~H. Abreu, T.~Cruz, H.~Ara{\'u}jo, and P.~Sim{\~o}es, ``Denial
  of service attacks: Detecting the frailties of machine learning algorithms in
  the classification process,'' in \emph{International Conference on Critical
  Information Infrastructures Security}.\hskip 1em plus 0.5em minus 0.4em\relax
  Springer, 2018, pp. 230--235.

\bibitem{novikova2022federated}
E.~Novikova, E.~Doynikova, and S.~Golubev, ``Federated learning for intrusion
  detection in the critical infrastructures: Vertically partitioned data use
  case,'' \emph{Algorithms}, vol.~15, no.~4, p. 104, 2022.

\bibitem{goh2016dataset}
J.~Goh, S.~Adepu, K.~N. Junejo, and A.~Mathur, ``A dataset to support research
  in the design of secure water treatment systems,'' in \emph{International
  conference on critical information infrastructures security}.\hskip 1em plus
  0.5em minus 0.4em\relax Springer, 2016, pp. 88--99.

\bibitem{ibitoye2022differentially}
O.~Ibitoye, M.~O. Shafiq, and A.~Matrawy, ``Differentially private
  self-normalizing neural networks for adversarial robustness in federated
  learning,'' \emph{Computers \& Security}, vol. 116, p. 102631, 2022.

\bibitem{garcia2014empirical}
S.~Garcia, M.~Grill, J.~Stiborek, and A.~Zunino, ``An empirical comparison of
  botnet detection methods,'' \emph{computers \& security}, vol.~45, pp.
  100--123, 2014.

\bibitem{draper2016characterization}
G.~Draper-Gil, A.~H. Lashkari, M.~S.~I. Mamun, and A.~A. Ghorbani,
  ``Characterization of encrypted and vpn traffic using time-related,'' in
  \emph{Proceedings of the 2nd international conference on information systems
  security and privacy (ICISSP)}, 2016, pp. 407--414.

\bibitem{lashkari2017characterization}
A.~H. Lashkari, G.~Draper-Gil, M.~S.~I. Mamun, A.~A. Ghorbani \emph{et~al.},
  ``Characterization of tor traffic using time based features.'' in
  \emph{ICISSp}, 2017, pp. 253--262.

\bibitem{singh2020collaborative}
N.~Singh, H.~Kasyap, and S.~Tripathy, ``Collaborative learning based effective
  malware detection system,'' in \emph{Joint European Conference on Machine
  Learning and Knowledge Discovery in Databases}.\hskip 1em plus 0.5em minus
  0.4em\relax Springer, 2020, pp. 205--219.

\bibitem{ronen2018microsoft}
R.~Ronen, M.~Radu, C.~Feuerstein, E.~Yom-Tov, and M.~Ahmadi, ``Microsoft
  malware classification challenge,'' \emph{arXiv preprint arXiv:1802.10135},
  2018.

\bibitem{arp2014drebin}
D.~Arp, M.~Spreitzenbarth, M.~Hubner, H.~Gascon, K.~Rieck, and C.~Siemens,
  ``Drebin: Effective and explainable detection of android malware in your
  pocket.'' in \emph{Ndss}, vol.~14, 2014, pp. 23--26.

\bibitem{zhou2012dissecting}
Y.~Zhou and X.~Jiang, ``Dissecting android malware: Characterization and
  evolution,'' in \emph{2012 IEEE symposium on security and privacy}.\hskip 1em
  plus 0.5em minus 0.4em\relax IEEE, 2012, pp. 95--109.

\bibitem{contagiodataset}
\BIBentryALTinterwordspacing
``Contagio dataset.'' [Online]. Available:
  \url{https://www.impactcybertrust.org/dataset_view?idDataset=1273}
\BIBentrySTDinterwordspacing

\bibitem{taheri2020fed}
R.~Taheri, M.~Shojafar, M.~Alazab, and R.~Tafazolli, ``Fed-iiot: A robust
  federated malware detection architecture in industrial iot,'' \emph{IEEE
  Transactions on Industrial Informatics}, vol.~17, no.~12, pp. 8442--8452,
  2020.

\bibitem{pasdar2022train}
A.~Pasdar, Y.~C. Lee, T.~Liu, and S.-H. Hong, ``Train me to fight:
  Machine-learning based on-device malware detection for mobile devices,'' in
  \emph{2022 22nd IEEE International Symposium on Cluster, Cloud and Internet
  Computing (CCGrid)}.\hskip 1em plus 0.5em minus 0.4em\relax IEEE, 2022, pp.
  239--248.

\bibitem{mahdavifar2020dynamic}
S.~Mahdavifar, A.~F.~A. Kadir, R.~Fatemi, D.~Alhadidi, and A.~A. Ghorbani,
  ``Dynamic android malware category classification using semi-supervised deep
  learning,'' in \emph{2020 IEEE Intl Conf on Dependable, Autonomic and Secure
  Computing, Intl Conf on Pervasive Intelligence and Computing, Intl Conf on
  Cloud and Big Data Computing, Intl Conf on Cyber Science and Technology
  Congress (DASC/PiCom/CBDCom/CyberSciTech)}.\hskip 1em plus 0.5em minus
  0.4em\relax IEEE, 2020, pp. 515--522.

\bibitem{allix2016androzoo}
K.~Allix, T.~F. Bissyand{\'e}, J.~Klein, and Y.~Le~Traon, ``Androzoo:
  Collecting millions of android apps for the research community,'' in
  \emph{2016 IEEE/ACM 13th Working Conference on Mining Software Repositories
  (MSR)}.\hskip 1em plus 0.5em minus 0.4em\relax IEEE, 2016, pp. 468--471.

\bibitem{pei2022knowledge}
X.~Pei, X.~Deng, S.~Tian, L.~Zhang, and K.~Xue, ``A knowledge transfer-based
  semi-supervised federated learning for iot malware detection,'' \emph{IEEE
  Transactions on Dependable and Secure Computing}, 2022.

\bibitem{virusshare}
V.~VirusShare, ``Virusshare. com--because sharing is caring.''

\bibitem{mowla2019federated}
N.~I. Mowla, N.~H. Tran, I.~Doh, and K.~Chae, ``Federated learning-based
  cognitive detection of jamming attack in flying ad-hoc network,'' \emph{IEEE
  Access}, vol.~8, pp. 4338--4350, 2019.

\bibitem{crawdad}
O.~Puñal, C.~Pereira, A.~Aguiar, and J.~Gross, ``The
  uportorwthaachen/vanetjamming2014 dataset,'' 2004.

\bibitem{zhao2020intelligent}
R.~Zhao, Y.~Yin, Y.~Shi, and Z.~Xue, ``Intelligent intrusion detection based on
  federated learning aided long short-term memory,'' \emph{Physical
  Communication}, vol.~42, p. 101157, 2020.

\bibitem{schonlau2001computer}
M.~Schonlau, W.~DuMouchel, W.-H. Ju, A.~F. Karr, M.~Theus, and Y.~Vardi,
  ``Computer intrusion: Detecting masquerades,'' \emph{Statistical science},
  pp. 58--74, 2001.

\bibitem{nguyen2019diot}
T.~D. Nguyen, S.~Marchal, M.~Miettinen, H.~Fereidooni, N.~Asokan, and A.-R.
  Sadeghi, ``D{\"i}ot: A federated self-learning anomaly detection system for
  iot,'' in \emph{2019 IEEE 39th International Conference on Distributed
  Computing Systems (ICDCS)}.\hskip 1em plus 0.5em minus 0.4em\relax IEEE,
  2019, pp. 756--767.

\bibitem{li2020distributed}
K.~Li, H.~Zhou, Z.~Tu, W.~Wang, and H.~Zhang, ``Distributed network intrusion
  detection system in satellite-terrestrial integrated networks using federated
  learning,'' \emph{IEEE Access}, vol.~8, pp. 214\,852--214\,865, 2020.

\bibitem{sun2020intrusion}
Y.~Sun, H.~Ochiai, and H.~Esaki, ``Intrusion detection with segmented federated
  learning for large-scale multiple lans,'' in \emph{2020 International Joint
  Conference on Neural Networks (IJCNN)}.\hskip 1em plus 0.5em minus
  0.4em\relax IEEE, 2020, pp. 1--8.

\bibitem{hsu2020privacy}
R.-H. Hsu, Y.-C. Wang, C.-I. Fan, B.~Sun, T.~Ban, T.~Takahashi, T.-W. Wu, and
  S.-W. Kao, ``A privacy-preserving federated learning system for android
  malware detection based on edge computing,'' in \emph{2020 15th Asia Joint
  Conference on Information Security (AsiaJCIS)}.\hskip 1em plus 0.5em minus
  0.4em\relax IEEE, 2020, pp. 128--136.

\bibitem{rehman2022federated}
A.~Rehman, I.~Razzak, and G.~Xu, ``Federated learning for privacy preservation
  of healthcare data from smartphone-based side-channel attacks,'' \emph{IEEE
  Journal of Biomedical and Health Informatics}, 2022.

\bibitem{zhao2021federated}
L.~Zhao, J.~Li, Q.~Li, and F.~Li, ``A federated learning framework for
  detecting false data injection attacks in solar farms,'' \emph{IEEE
  Transactions on Power Electronics}, vol.~37, no.~3, pp. 2496--2501, 2021.

\bibitem{zhang2022grid}
Y.~Zhang, Y.~Liu, N.~Zhang, D.~Wang, S.~Zhang, and Y.~Wu, ``Grid false data
  intrusion detection method based on edge computing and federated learning,''
  in \emph{3D Imaging—Multidimensional Signal Processing and Deep
  Learning}.\hskip 1em plus 0.5em minus 0.4em\relax Springer, 2022, pp.
  179--188.

\bibitem{shukla2021device}
S.~Shukla, P.~S. Manoj, G.~Kolhe, and S.~Rafatirad, ``On-device malware
  detection using performance-aware and robust collaborative learning,'' in
  \emph{2021 58th ACM/IEEE Design Automation Conference (DAC)}.\hskip 1em plus
  0.5em minus 0.4em\relax IEEE, 2021, pp. 967--972.

\bibitem{peng2019opening}
P.~Peng, L.~Yang, L.~Song, and G.~Wang, ``Opening the blackbox of virustotal:
  Analyzing online phishing scan engines,'' in \emph{Proceedings of the
  Internet Measurement Conference}, 2019, pp. 478--485.

\bibitem{shukla2022rafel}
S.~Shukla, G.~Kolhe, H.~Homayoun, S.~Rafatirad, and S.~M. PD, ``Rafel-robust
  and data-aware federated learning-inspired malware detection in
  internet-of-things (iot) networks,'' in \emph{Proceedings of the Great Lakes
  Symposium on VLSI 2022}, 2022, pp. 153--157.

\bibitem{nguyen2021federated1}
T.~G. Nguyen, T.~V. Phan, D.~T. Hoang, T.~N. Nguyen, and C.~So-In, ``Federated
  deep reinforcement learning for traffic monitoring in sdn-based iot
  networks,'' \emph{IEEE Transactions on Cognitive Communications and
  Networking}, vol.~7, no.~4, pp. 1048--1065, 2021.

\bibitem{alazzam2022federated}
M.~B. Alazzam, F.~Alassery, and A.~Almulihi, ``Federated deep learning
  approaches for the privacy and security of iot systems,'' \emph{Wireless
  Communications and Mobile Computing}, vol. 2022, 2022.

\bibitem{nreldataset}
\BIBentryALTinterwordspacing
``Nrel national renewable energy laboratory.'' [Online]. Available:
  \url{https://www.nrel.gov/solar/}
\BIBentrySTDinterwordspacing

\bibitem{javed2020alphalogger}
A.~R. Javed, M.~O. Beg, M.~Asim, T.~Baker, and A.~H. Al-Bayatti, ``Alphalogger:
  Detecting motion-based side-channel attack using smartphone keystrokes,''
  \emph{Journal of Ambient Intelligence and Humanized Computing}, pp. 1--14,
  2020.

\bibitem{kddcup_dataset}
\BIBentryALTinterwordspacing
``Kddcup99 dataset,'' 1999. [Online]. Available:
  \url{http://kdd.ics.uci.edu/databases/kddcup99/kddcup99.html}
\BIBentrySTDinterwordspacing

\bibitem{apruzzese2022role}
G.~Apruzzese, P.~Laskov, E.~M. de~Oca, W.~Mallouli, L.~B. Rapa, A.~V.
  Grammatopoulos, and F.~Di~Franco, ``The role of machine learning in
  cybersecurity,'' \emph{arXiv preprint arXiv:2206.09707}, 2022.

\bibitem{wright1995logistic}
R.~E. Wright, ``Logistic regression.'' 1995.

\bibitem{kotsiantis2013decision}
S.~B. Kotsiantis, ``Decision trees: a recent overview,'' \emph{Artificial
  Intelligence Review}, vol.~39, pp. 261--283, 2013.

\bibitem{breiman2001random}
L.~Breiman, ``Random forests,'' \emph{Machine learning}, vol.~45, pp. 5--32,
  2001.

\bibitem{anderson1995introduction}
J.~A. Anderson, \emph{An introduction to neural networks}.\hskip 1em plus 0.5em
  minus 0.4em\relax MIT press, 1995.

\bibitem{hearst1998support}
M.~A. Hearst, S.~T. Dumais, E.~Osuna, J.~Platt, and B.~Scholkopf, ``Support
  vector machines,'' \emph{IEEE Intelligent Systems and their applications},
  vol.~13, no.~4, pp. 18--28, 1998.

\bibitem{svozil1997introduction}
D.~Svozil, V.~Kvasnicka, and J.~Pospichal, ``Introduction to multi-layer
  feed-forward neural networks,'' \emph{Chemometrics and intelligent laboratory
  systems}, vol.~39, no.~1, pp. 43--62, 1997.

\bibitem{salehinejad2017recent}
H.~Salehinejad, S.~Sankar, J.~Barfett, E.~Colak, and S.~Valaee, ``Recent
  advances in recurrent neural networks,'' \emph{arXiv preprint
  arXiv:1801.01078}, 2017.

\bibitem{gu2018recent}
J.~Gu, Z.~Wang, J.~Kuen, L.~Ma, A.~Shahroudy, B.~Shuai, T.~Liu, X.~Wang,
  G.~Wang, J.~Cai \emph{et~al.}, ``Recent advances in convolutional neural
  networks,'' \emph{Pattern recognition}, vol.~77, pp. 354--377, 2018.

\bibitem{murtagh1991multilayer}
F.~Murtagh, ``Multilayer perceptrons for classification and regression,''
  \emph{Neurocomputing}, vol.~2, no. 5-6, pp. 183--197, 1991.

\bibitem{gonzalez2022parking}
A.~Gonzalez-Vidal, F.~Terroso-S{\'a}enz, and A.~Skarmeta, ``Parking
  availability prediction with coarse-grained human mobility data,''
  \emph{CMC-COMPUTERS MATERIALS \& CONTINUA}, vol.~71, no.~3, pp. 4355--4375,
  2022.

\bibitem{shafer1992dempster}
G.~Shafer, ``Dempster-shafer theory,'' \emph{Encyclopedia of artificial
  intelligence}, vol.~1, pp. 330--331, 1992.

\bibitem{mitchell2018never}
T.~Mitchell, W.~Cohen, E.~Hruschka, P.~Talukdar, B.~Yang, J.~Betteridge,
  A.~Carlson, B.~Dalvi, M.~Gardner, B.~Kisiel \emph{et~al.}, ``Never-ending
  learning,'' \emph{Communications of the ACM}, vol.~61, no.~5, pp. 103--115,
  2018.

\bibitem{khosla2020supervised}
P.~Khosla, P.~Teterwak, C.~Wang, A.~Sarna, Y.~Tian, P.~Isola, A.~Maschinot,
  C.~Liu, and D.~Krishnan, ``Supervised contrastive learning,'' \emph{Advances
  in neural information processing systems}, vol.~33, pp. 18\,661--18\,673,
  2020.

\bibitem{singh2019detailed}
A.~Singh, P.~Vepakomma, O.~Gupta, and R.~Raskar, ``Detailed comparison of
  communication efficiency of split learning and federated learning,''
  \emph{arXiv preprint arXiv:1909.09145}, 2019.

\bibitem{hubara2016binarized}
I.~Hubara, M.~Courbariaux, D.~Soudry, R.~El-Yaniv, and Y.~Bengio, ``Binarized
  neural networks,'' \emph{Advances in neural information processing systems},
  vol.~29, 2016.

\bibitem{xu2021learning}
Y.~Xu, K.~Han, C.~Xu, Y.~Tang, C.~Xu, and Y.~Wang, ``Learning frequency domain
  approximation for binary neural networks,'' \emph{Advances in Neural
  Information Processing Systems}, vol.~34, pp. 25\,553--25\,565, 2021.

\bibitem{zhang2021surveyMTL}
Y.~Zhang and Q.~Yang, ``A survey on multi-task learning,'' \emph{IEEE
  Transactions on Knowledge and Data Engineering}, vol.~34, no.~12, pp.
  5586--5609, 2021.

\bibitem{crawshaw2020multi}
M.~Crawshaw, ``Multi-task learning with deep neural networks: A survey,''
  \emph{arXiv preprint arXiv:2009.09796}, 2020.

\bibitem{zhang2019incremental}
C.~Zhang, Y.~Zhang, X.~Shi, G.~Almpanidis, G.~Fan, and X.~Shen, ``On
  incremental learning for gradient boosting decision trees,'' \emph{Neural
  Processing Letters}, vol.~50, pp. 957--987, 2019.

\bibitem{ke2017lightgbm}
G.~Ke, Q.~Meng, T.~Finley, T.~Wang, W.~Chen, W.~Ma, Q.~Ye, and T.-Y. Liu,
  ``Lightgbm: A highly efficient gradient boosting decision tree,''
  \emph{Advances in neural information processing systems}, vol.~30, 2017.

\bibitem{bracewell1986fourier}
R.~N. Bracewell and R.~N. Bracewell, \emph{The Fourier transform and its
  applications}.\hskip 1em plus 0.5em minus 0.4em\relax McGraw-Hill New York,
  1986, vol. 31999.

\bibitem{howard2017mobilenets}
A.~G. Howard, M.~Zhu, B.~Chen, D.~Kalenichenko, W.~Wang, T.~Weyand,
  M.~Andreetto, and H.~Adam, ``Mobilenets: Efficient convolutional neural
  networks for mobile vision applications,'' \emph{arXiv preprint
  arXiv:1704.04861}, 2017.

\bibitem{lin2022survey}
T.~Lin, Y.~Wang, X.~Liu, and X.~Qiu, ``A survey of transformers,'' \emph{AI
  Open}, 2022.

\bibitem{xu2015comprehensive}
D.~Xu and Y.~Tian, ``A comprehensive survey of clustering algorithms,''
  \emph{Annals of Data Science}, vol.~2, pp. 165--193, 2015.

\bibitem{sorzano2014survey}
C.~O.~S. Sorzano, J.~Vargas, and A.~P. Montano, ``A survey of dimensionality
  reduction techniques,'' \emph{arXiv preprint arXiv:1403.2877}, 2014.

\bibitem{ahmed2020k}
M.~Ahmed, R.~Seraj, and S.~M.~S. Islam, ``The k-means algorithm: A
  comprehensive survey and performance evaluation,'' \emph{Electronics},
  vol.~9, no.~8, p. 1295, 2020.

\bibitem{xie2021improved}
B.~Xie, X.~Dong, and C.~Wang, ``An improved-means clustering intrusion
  detection algorithm for wireless networks based on federated learning,''
  \emph{Wireless Communications and Mobile Computing}, vol. 2021, 2021.

\bibitem{senoussaoui2013study}
M.~Senoussaoui, P.~Kenny, T.~Stafylakis, and P.~Dumouchel, ``A study of the
  cosine distance-based mean shift for telephone speech diarization,''
  \emph{IEEE/ACM Transactions on Audio, Speech, and Language Processing},
  vol.~22, no.~1, pp. 217--227, 2013.

\bibitem{wang2018ce3}
P.~Wang and Y.~Yao, ``Ce3: A three-way clustering method based on mathematical
  morphology,'' \emph{Knowledge-based systems}, vol. 155, pp. 54--65, 2018.

\bibitem{yadav2021clustering}
K.~Yadav and B.~Gupta, ``Clustering based rewarding algorithm to detect
  adversaries in federated machine learning based iot environment,'' in
  \emph{2021 IEEE International Conference on Consumer Electronics
  (ICCE)}.\hskip 1em plus 0.5em minus 0.4em\relax IEEE, 2021, pp. 1--6.

\bibitem{bank2020autoencoders}
D.~Bank, N.~Koenigstein, and R.~Giryes, ``Autoencoders,'' \emph{arXiv preprint
  arXiv:2003.05991}, 2020.

\bibitem{zong2018deep}
B.~Zong, Q.~Song, M.~R. Min, W.~Cheng, C.~Lumezanu, D.~Cho, and H.~Chen, ``Deep
  autoencoding gaussian mixture model for unsupervised anomaly detection,'' in
  \emph{International conference on learning representations}, 2018.

\bibitem{cetin2019federated}
B.~Cetin, A.~Lazar, J.~Kim, A.~Sim, and K.~Wu, ``Federated wireless network
  intrusion detection,'' in \emph{2019 IEEE International Conference on Big
  Data (Big Data)}.\hskip 1em plus 0.5em minus 0.4em\relax IEEE, 2019, pp.
  6004--6006.

\bibitem{jahromi2021deep}
A.~N. Jahromi, H.~K. Schulich, and A.~Dehghantanha, ``Deep federated
  learning-based cyber-attack detection in industrial control systems,'' in
  \emph{2021 18th International Conference on Privacy, Security and Trust
  (PST)}.\hskip 1em plus 0.5em minus 0.4em\relax IEEE, 2021, pp. 1--6.

\bibitem{wu2022fl}
D.~Wu, Y.~Deng, and M.~Li, ``Fl-mgvn: Federated learning for anomaly detection
  using mixed gaussian variational self-encoding network,'' \emph{Information
  Processing \& Management}, vol.~59, no.~2, p. 102839, 2022.

\bibitem{gui2021review}
J.~Gui, Z.~Sun, Y.~Wen, D.~Tao, and J.~Ye, ``A review on generative adversarial
  networks: Algorithms, theory, and applications,'' \emph{IEEE transactions on
  knowledge and data engineering}, 2021.

\bibitem{melchior2017gaussian}
J.~Melchior, N.~Wang, and L.~Wiskott, ``Gaussian-binary restricted boltzmann
  machines for modeling natural image statistics,'' \emph{PloS one}, vol.~12,
  no.~2, p. e0171015, 2017.

\bibitem{xia2022fed_adbn}
Z.~Xia, Y.~Chen, B.~Yin, H.~Liang, H.~Zhou, K.~Gu, and F.~Yu, ``Fed\_adbn: An
  efficient intrusion detection framework based on client selection in ami
  network,'' \emph{Expert Systems}, 2022.

\bibitem{qu2021survey}
X.~Qu, L.~Yang, K.~Guo, L.~Ma, M.~Sun, M.~Ke, and M.~Li, ``A survey on the
  development of self-organizing maps for unsupervised intrusion detection,''
  \emph{Mobile networks and applications}, vol.~26, pp. 808--829, 2021.

\bibitem{hariri2019extended}
S.~Hariri, M.~C. Kind, and R.~J. Brunner, ``Extended isolation forest,''
  \emph{IEEE Transactions on Knowledge and Data Engineering}, vol.~33, no.~4,
  pp. 1479--1489, 2019.

\bibitem{ashrafuzzaman2020elliptic}
M.~Ashrafuzzaman, S.~Das, A.~A. Jillepalli, Y.~Chakhchoukh, and F.~T. Sheldon,
  ``Elliptic envelope based detection of stealthy false data injection attacks
  in smart grid control systems,'' in \emph{2020 IEEE Symposium Series on
  Computational Intelligence (SSCI)}.\hskip 1em plus 0.5em minus 0.4em\relax
  IEEE, 2020, pp. 1131--1137.

\bibitem{qin2021federated}
Y.~Qin and M.~Kondo, ``Federated learning-based network intrusion detection
  with a feature selection approach,'' in \emph{2021 International Conference
  on Electrical, Communication, and Computer Engineering (ICECCE)}.\hskip 1em
  plus 0.5em minus 0.4em\relax IEEE, 2021, pp. 1--6.

\bibitem{liang2006fast}
N.-Y. Liang, G.-B. Huang, P.~Saratchandran, and N.~Sundararajan, ``A fast and
  accurate online sequential learning algorithm for feedforward networks,''
  \emph{IEEE Transactions on neural networks}, vol.~17, no.~6, pp. 1411--1423,
  2006.

\bibitem{kaelbling1996reinforcement}
L.~P. Kaelbling, M.~L. Littman, and A.~W. Moore, ``Reinforcement learning: A
  survey,'' \emph{Journal of artificial intelligence research}, vol.~4, pp.
  237--285, 1996.

\bibitem{lopez2020application}
M.~Lopez-Martin, B.~Carro, and A.~Sanchez-Esguevillas, ``Application of deep
  reinforcement learning to intrusion detection for supervised problems,''
  \emph{Expert Systems with Applications}, vol. 141, p. 112963, 2020.

\bibitem{clifton2020q}
J.~Clifton and E.~Laber, ``Q-learning: Theory and applications,'' \emph{Annual
  Review of Statistics and Its Application}, vol.~7, pp. 279--301, 2020.

\bibitem{mnih2015human}
V.~Mnih, K.~Kavukcuoglu, D.~Silver, A.~A. Rusu, J.~Veness, M.~G. Bellemare,
  A.~Graves, M.~Riedmiller, A.~K. Fidjeland, G.~Ostrovski \emph{et~al.},
  ``Human-level control through deep reinforcement learning,'' \emph{nature},
  vol. 518, no. 7540, pp. 529--533, 2015.

\bibitem{van2016deep}
H.~Van~Hasselt, A.~Guez, and D.~Silver, ``Deep reinforcement learning with
  double q-learning,'' in \emph{Proceedings of the AAAI conference on
  artificial intelligence}, vol.~30, no.~1, 2016.

\bibitem{nakip2021mirai}
M.~Nakip and E.~Gelenbe, ``Mirai botnet attack detection with auto-associative
  dense random neural network,'' in \emph{2021 IEEE Global Communications
  Conference (GLOBECOM)}.\hskip 1em plus 0.5em minus 0.4em\relax IEEE, 2021,
  pp. 01--06.

\bibitem{bertino2021ai}
E.~Bertino, M.~Kantarcioglu, C.~G. Akcora, S.~Samtani, S.~Mittal, and M.~Gupta,
  ``Ai for security and security for ai,'' in \emph{Proceedings of the Eleventh
  ACM Conference on Data and Application Security and Privacy}, 2021, pp.
  333--334.

\bibitem{arora2021survey}
S.~Arora and P.~Doshi, ``A survey of inverse reinforcement learning:
  Challenges, methods and progress,'' \emph{Artificial Intelligence}, vol. 297,
  p. 103500, 2021.

\bibitem{lalouani2021robust}
W.~Lalouani and M.~Younis, ``Robust distributed intrusion detection system for
  edge of things,'' in \emph{2021 IEEE Global Communications Conference
  (GLOBECOM)}.\hskip 1em plus 0.5em minus 0.4em\relax IEEE, 2021, pp. 01--06.

\bibitem{agrawal2021temporal}
S.~Agrawal, A.~Chowdhuri, S.~Sarkar, R.~Selvanambi, and T.~R. Gadekallu,
  ``Temporal weighted averaging for asynchronous federated intrusion detection
  systems,'' \emph{Computational Intelligence and Neuroscience}, vol. 2021,
  2021.

\bibitem{tahir2021experience}
B.~Tahir, A.~Jolfaei, and M.~Tariq, ``Experience driven attack design and
  federated learning based intrusion detection in industry 4.0,'' \emph{IEEE
  Transactions on Industrial Informatics}, 2021.

\bibitem{chen2021trust}
N.~Chen, Y.~Jin, Y.~Li, and L.~Cai, ``Trust-based federated learning for
  network anomaly detection,'' in \emph{Web Intelligence}, no. Preprint.\hskip
  1em plus 0.5em minus 0.4em\relax IOS Press, 2021, pp. 1--11.

\bibitem{man2021intelligent}
D.~Man, F.~Zeng, W.~Yang, M.~Yu, J.~Lv, and Y.~Wang, ``Intelligent intrusion
  detection based on federated learning for edge-assisted internet of things,''
  \emph{Security and Communication Networks}, vol. 2021, 2021.

\bibitem{sun2020adaptive}
Y.~Sun, H.~Esaki, and H.~Ochiai, ``Adaptive intrusion detection in the
  networking of large-scale lans with segmented federated learning,''
  \emph{IEEE Open Journal of the Communications Society}, vol.~2, pp. 102--112,
  2020.

\bibitem{tabassum2022fedgan}
A.~Tabassum, A.~Erbad, W.~Lebda, A.~Mohamed, and M.~Guizani, ``Fedgan-ids:
  Privacy-preserving ids using gan and federated learning,'' \emph{Computer
  Communications}, vol. 192, pp. 299--310, 2022.

\bibitem{sun2022hierarchical}
X.~Sun, Z.~Tang, M.~Du, C.~Deng, W.~Lin, J.~Chen, Q.~Qi, and H.~Zheng, ``A
  hierarchical federated learning-based intrusion detection system for 5g smart
  grids,'' \emph{Electronics}, vol.~11, no.~16, p. 2627, 2022.

\bibitem{kumar2021security}
K.~S. Kumar, S.~A.~H. Nair, D.~G. Roy, B.~Rajalingam, and R.~S. Kumar,
  ``Security and privacy-aware artificial intrusion detection system using
  federated machine learning,'' \emph{Computers \& Electrical Engineering},
  vol.~96, p. 107440, 2021.

\bibitem{reisizadeh2020fedpaq}
A.~Reisizadeh, A.~Mokhtari, H.~Hassani, A.~Jadbabaie, and R.~Pedarsani,
  ``Fedpaq: A communication-efficient federated learning method with periodic
  averaging and quantization,'' in \emph{International Conference on Artificial
  Intelligence and Statistics}.\hskip 1em plus 0.5em minus 0.4em\relax PMLR,
  2020, pp. 2021--2031.

\bibitem{gholami2021survey}
A.~Gholami, S.~Kim, Z.~Dong, Z.~Yao, M.~W. Mahoney, and K.~Keutzer, ``A survey
  of quantization methods for efficient neural network inference,'' \emph{arXiv
  preprint arXiv:2103.13630}, 2021.

\bibitem{alistarh2017qsgd}
D.~Alistarh, D.~Grubic, J.~Li, R.~Tomioka, and M.~Vojnovic, ``Qsgd:
  Communication-efficient sgd via gradient quantization and encoding,''
  \emph{Advances in neural information processing systems}, vol.~30, 2017.

\bibitem{wang2020federated}
H.~Wang, M.~Yurochkin, Y.~Sun, D.~Papailiopoulos, and Y.~Khazaeni, ``Federated
  learning with matched averaging,'' \emph{arXiv preprint arXiv:2002.06440},
  2020.

\bibitem{yurochkin2019bayesian}
M.~Yurochkin, M.~Agarwal, S.~Ghosh, K.~Greenewald, N.~Hoang, and Y.~Khazaeni,
  ``Bayesian nonparametric federated learning of neural networks,'' in
  \emph{International conference on machine learning}.\hskip 1em plus 0.5em
  minus 0.4em\relax PMLR, 2019, pp. 7252--7261.

\bibitem{fedprox}
T.~Li, A.~K. Sahu, M.~Zaheer, M.~Sanjabi, A.~Talwalkar, and V.~Smith,
  ``Federated optimization in heterogeneous networks,'' \emph{Proceedings of
  Machine Learning and Systems}, vol.~2, pp. 429--450, 2020.

\bibitem{yu2021fed}
P.~Yu, A.~Kundu, L.~Wynter, and S.~H. Lim, ``Fed+: A unified approach to robust
  personalized federated learning,'' 2021.

\bibitem{so2021turbo}
J.~So, B.~G{\"u}ler, and A.~S. Avestimehr, ``Turbo-aggregate: Breaking the
  quadratic aggregation barrier in secure federated learning,'' \emph{IEEE
  Journal on Selected Areas in Information Theory}, vol.~2, no.~1, pp.
  479--489, 2021.

\bibitem{bonawitz2017practical}
K.~Bonawitz, V.~Ivanov, B.~Kreuter, A.~Marcedone, H.~B. McMahan, S.~Patel,
  D.~Ramage, A.~Segal, and K.~Seth, ``Practical secure aggregation for
  privacy-preserving machine learning,'' in \emph{proceedings of the 2017 ACM
  SIGSAC Conference on Computer and Communications Security}, 2017, pp.
  1175--1191.

\bibitem{yu2019lagrange}
Q.~Yu, S.~Li, N.~Raviv, S.~M.~M. Kalan, M.~Soltanolkotabi, and S.~A.
  Avestimehr, ``Lagrange coded computing: Optimal design for resiliency,
  security, and privacy,'' in \emph{The 22nd International Conference on
  Artificial Intelligence and Statistics}.\hskip 1em plus 0.5em minus
  0.4em\relax PMLR, 2019, pp. 1215--1225.

\bibitem{kopparapu2020fedcd}
K.~Kopparapu, E.~Lin, and J.~Zhao, ``Fedcd: Improving performance in non-iid
  federated learning,'' \emph{arXiv preprint arXiv:2006.09637}, 2020.

\bibitem{blanchard2017machine}
P.~Blanchard, E.~M. El~Mhamdi, R.~Guerraoui, and J.~Stainer, ``Machine learning
  with adversaries: Byzantine tolerant gradient descent,'' \emph{Advances in
  Neural Information Processing Systems}, vol.~30, 2017.

\bibitem{wu2020safa}
W.~Wu, L.~He, W.~Lin, R.~Mao, C.~Maple, and S.~Jarvis, ``Safa: A
  semi-asynchronous protocol for fast federated learning with low overhead,''
  \emph{IEEE Transactions on Computers}, vol.~70, no.~5, pp. 655--668, 2020.

\bibitem{casado2022concept}
F.~E. Casado, D.~Lema, M.~F. Criado, R.~Iglesias, C.~V. Regueiro, and S.~Barro,
  ``Concept drift detection and adaptation for federated and continual
  learning,'' \emph{Multimedia Tools and Applications}, vol.~81, no.~3, pp.
  3397--3419, 2022.

\bibitem{lu2018learning}
J.~Lu, A.~Liu, F.~Dong, F.~Gu, J.~Gama, and G.~Zhang, ``Learning under concept
  drift: A review,'' \emph{IEEE transactions on knowledge and data
  engineering}, vol.~31, no.~12, pp. 2346--2363, 2018.

\bibitem{cheng2021secureboost}
K.~Cheng, T.~Fan, Y.~Jin, Y.~Liu, T.~Chen, D.~Papadopoulos, and Q.~Yang,
  ``Secureboost: A lossless federated learning framework,'' \emph{IEEE
  Intelligent Systems}, vol.~36, no.~6, pp. 87--98, 2021.

\bibitem{liang2004privacy}
G.~Liang and S.~S. Chawathe, ``Privacy-preserving inter-database operations,''
  in \emph{International Conference on Intelligence and Security
  Informatics}.\hskip 1em plus 0.5em minus 0.4em\relax Springer, 2004, pp.
  66--82.

\bibitem{chen2016xgboost}
T.~Chen and C.~Guestrin, ``Xgboost: A scalable tree boosting system,'' in
  \emph{Proceedings of the 22nd acm sigkdd international conference on
  knowledge discovery and data mining}, 2016, pp. 785--794.

\bibitem{paillier1999public}
P.~Paillier, ``Public-key cryptosystems based on composite degree residuosity
  classes,'' in \emph{International conference on the theory and applications
  of cryptographic techniques}.\hskip 1em plus 0.5em minus 0.4em\relax
  Springer, 1999, pp. 223--238.

\bibitem{luping2019cmfl}
W.~Luping, W.~Wei, and L.~Bo, ``Cmfl: Mitigating communication overhead for
  federated learning,'' in \emph{2019 IEEE 39th international conference on
  distributed computing systems (ICDCS)}.\hskip 1em plus 0.5em minus
  0.4em\relax IEEE, 2019, pp. 954--964.

\bibitem{chaudhuri2023dynamic}
A.~Chaudhuri, A.~Nandi, and B.~Pradhan, ``A dynamic weighted federated learning
  for android malware classification,'' in \emph{Soft Computing: Theories and
  Applications: Proceedings of SoCTA 2022}.\hskip 1em plus 0.5em minus
  0.4em\relax Springer, 2023, pp. 147--159.

\bibitem{nilsson2018performance}
A.~Nilsson, S.~Smith, G.~Ulm, E.~Gustavsson, and M.~Jirstrand, ``A performance
  evaluation of federated learning algorithms,'' in \emph{Proceedings of the
  second workshop on distributed infrastructures for deep learning}, 2018, pp.
  1--8.

\bibitem{zhu2021federated}
H.~Zhu, J.~Xu, S.~Liu, and Y.~Jin, ``Federated learning on non-iid data: A
  survey,'' \emph{Neurocomputing}, vol. 465, pp. 371--390, 2021.

\bibitem{yin2018byzantine}
D.~Yin, Y.~Chen, R.~Kannan, and P.~Bartlett, ``Byzantine-robust distributed
  learning: Towards optimal statistical rates,'' in \emph{International
  Conference on Machine Learning}.\hskip 1em plus 0.5em minus 0.4em\relax PMLR,
  2018, pp. 5650--5659.

\bibitem{kholod2020open}
I.~Kholod, E.~Yanaki, D.~Fomichev, E.~Shalugin, E.~Novikova, E.~Filippov, and
  M.~Nordlund, ``Open-source federated learning frameworks for iot: A
  comparative review and analysis,'' \emph{Sensors}, vol.~21, no.~1, p. 167,
  2020.

\bibitem{fedml20}
\BIBentryALTinterwordspacing
C.~He, S.~Li, J.~So, X.~Zeng, M.~Zhang, H.~Wang, X.~Wang, P.~Vepakomma,
  A.~Singh, H.~Qiu, X.~Zhu, J.~Wang, L.~Shen, P.~Zhao, Y.~Kang, Y.~Liu,
  R.~Raskar, Q.~Yang, M.~Annavaram, and S.~Avestimehr, ``Fedml: A research
  library and benchmark for federated machine learning,'' 2020. [Online].
  Available: \url{https://arxiv.org/abs/2007.13518}
\BIBentrySTDinterwordspacing

\bibitem{liu2022unifed}
X.~Liu, T.~Shi, C.~Xie, Q.~Li, K.~Hu, H.~Kim, X.~Xu, B.~Li, and D.~Song,
  ``Unifed: A benchmark for federated learning frameworks,'' \emph{arXiv
  preprint arXiv:2207.10308}, 2022.

\bibitem{tensorflow}
\BIBentryALTinterwordspacing
``{TensorFlow}.'' [Online]. Available: \url{https://www.tensorflow.org}
\BIBentrySTDinterwordspacing

\bibitem{pysyft}
\BIBentryALTinterwordspacing
{OpenMined}, ``{PySyft}.'' [Online]. Available:
  \url{https://github.com/OpenMined/PySyft}
\BIBentrySTDinterwordspacing

\bibitem{openMined}
\BIBentryALTinterwordspacing
``Openmined.'' [Online]. Available: \url{https://www.openmined.org/}
\BIBentrySTDinterwordspacing

\bibitem{pytorch}
\BIBentryALTinterwordspacing
``pytorch.'' [Online]. Available: \url{https://pytorch.org/}
\BIBentrySTDinterwordspacing

\bibitem{fate}
\BIBentryALTinterwordspacing
``{An Industrial Grade Federated Learning Framework},'' last visited 8/11/2022.
  [Online]. Available: \url{https://fate.fedai.org/}
\BIBentrySTDinterwordspacing

\bibitem{paddlefl}
\BIBentryALTinterwordspacing
``{Paddle Federated Learning},'' last visited 8/11/2022. [Online]. Available:
  \url{https://github.com/PaddlePaddle/PaddleFL}
\BIBentrySTDinterwordspacing

\bibitem{paddlepaddle}
\BIBentryALTinterwordspacing
``{PaddlePaddle},'' last visited 8/11/2022. [Online]. Available:
  \url{https://github.com/PaddlePaddle/Paddle}
\BIBentrySTDinterwordspacing

\bibitem{ibmfl}
\BIBentryALTinterwordspacing
H.~Ludwig, N.~Baracaldo, G.~Thomas, Y.~Zhou, A.~Anwar, S.~Rajamoni, Y.~Ong,
  J.~Radhakrishnan, A.~Verma, M.~Sinn, M.~Purcell, A.~Rawat, T.~Minh,
  N.~Holohan, S.~Chakraborty, S.~Whitherspoon, D.~Steuer, L.~Wynter, H.~Hassan,
  S.~Laguna, M.~Yurochkin, M.~Agarwal, E.~Chuba, and A.~Abay, ``Ibm federated
  learning: an enterprise framework white paper v0.1,'' 2020. [Online].
  Available: \url{https://arxiv.org/abs/2007.10987}
\BIBentrySTDinterwordspacing

\bibitem{ibmflrepo}
\BIBentryALTinterwordspacing
``{IBM Federated Learning},'' last visited 9/11/2022. [Online]. Available:
  \url{https://github.com/IBM/federated-learning-lib}
\BIBentrySTDinterwordspacing

\bibitem{fedmlrepo}
\BIBentryALTinterwordspacing
``{FedML: The Community Building Open and Collaborative AI Anywhere at Any
  Scale},'' last visited 5/12/2022. [Online]. Available:
  \url{https://github.com/FedML-AI/FedML}
\BIBentrySTDinterwordspacing

\bibitem{flower}
\BIBentryALTinterwordspacing
D.~J. Beutel, T.~Topal, A.~Mathur, X.~Qiu, T.~Parcollet, and N.~D. Lane,
  ``Flower: {A} friendly federated learning research framework,'' \emph{CoRR},
  vol. abs/2007.14390, 2020. [Online]. Available:
  \url{https://arxiv.org/abs/2007.14390}
\BIBentrySTDinterwordspacing

\bibitem{flowergit}
\BIBentryALTinterwordspacing
``{Flower - A Friendly Federated Learning Framework},'' last visited 7/12/2022.
  [Online]. Available: \url{https://github.com/adap/flower}
\BIBentrySTDinterwordspacing

\bibitem{fedscope}
\BIBentryALTinterwordspacing
Y.~Xie, Z.~Wang, D.~Gao, D.~Chen, L.~Yao, W.~Kuang, Y.~Li, B.~Ding, and
  J.~Zhou, ``Federatedscope: A flexible federated learning platform for
  heterogeneity,'' 2022. [Online]. Available:
  \url{https://arxiv.org/abs/2204.05011}
\BIBentrySTDinterwordspacing

\bibitem{fedscopegit}
\BIBentryALTinterwordspacing
``{FederatedScope},'' last visited 13/12/2022. [Online]. Available:
  \url{https://github.com/alibaba/FederatedScope}
\BIBentrySTDinterwordspacing

\bibitem{substra}
\BIBentryALTinterwordspacing
``{Substra},'' last visited 13/12/2022. [Online]. Available:
  \url{https://github.com/Substra/substra}
\BIBentrySTDinterwordspacing

\bibitem{clara}
\BIBentryALTinterwordspacing
``{What is NVIDIA Clara?}'' last visited 13/12/2022. [Online]. Available:
  \url{https://developer.nvidia.com/industries/healthcare}
\BIBentrySTDinterwordspacing

\bibitem{fedbiomed}
\BIBentryALTinterwordspacing
``{Fed-BioMed - An open-source federated learning framework},'' last visited
  13/12/2022. [Online]. Available:
  \url{https://fedbiomed.gitlabpages.inria.fr/}
\BIBentrySTDinterwordspacing

\bibitem{bello2021revisiting}
I.~Bello, W.~Fedus, X.~Du, E.~D. Cubuk, A.~Srinivas, T.-Y. Lin, J.~Shlens, and
  B.~Zoph, ``Revisiting resnets: Improved training and scaling strategies,''
  \emph{Advances in Neural Information Processing Systems}, vol.~34, pp.
  22\,614--22\,627, 2021.

\bibitem{dehghani2021benchmark}
M.~Dehghani, Y.~Tay, A.~A. Gritsenko, Z.~Zhao, N.~Houlsby, F.~Diaz, D.~Metzler,
  and O.~Vinyals, ``The benchmark lottery,'' \emph{arXiv preprint
  arXiv:2107.07002}, 2021.

\bibitem{kramer2016scikit}
O.~Kramer and O.~Kramer, ``Scikit-learn,'' \emph{Machine learning for evolution
  strategies}, pp. 45--53, 2016.

\bibitem{abadi2016tensorflow}
M.~Abadi, P.~Barham, J.~Chen, Z.~Chen, A.~Davis, J.~Dean, M.~Devin,
  S.~Ghemawat, G.~Irving, M.~Isard \emph{et~al.}, ``$\{$TensorFlow$\}$: a
  system for $\{$Large-Scale$\}$ machine learning,'' in \emph{12th USENIX
  symposium on operating systems design and implementation (OSDI 16)}, 2016,
  pp. 265--283.

\bibitem{paszke2019pytorch}
A.~Paszke, S.~Gross, F.~Massa, A.~Lerer, J.~Bradbury, G.~Chanan, T.~Killeen,
  Z.~Lin, N.~Gimelshein, L.~Antiga \emph{et~al.}, ``Pytorch: An imperative
  style, high-performance deep learning library,'' \emph{Advances in neural
  information processing systems}, vol.~32, 2019.

\bibitem{ketkar2017introduction}
N.~Ketkar and N.~Ketkar, ``Introduction to keras,'' \emph{Deep learning with
  python: a hands-on introduction}, pp. 97--111, 2017.

\bibitem{li2020federated}
T.~Li, A.~K. Sahu, A.~Talwalkar, and V.~Smith, ``Federated learning:
  Challenges, methods, and future directions,'' \emph{IEEE Signal Processing
  Magazine}, vol.~37, no.~3, pp. 50--60, 2020.

\bibitem{matheu2022federated}
S.~N. Matheu, E.~M{\'a}rmol, J.~L. Hern{\'a}ndez-Ramos, A.~Skarmeta, and
  G.~Baldini, ``Federated cyberattack detection for internet of things-enabled
  smart cities,'' \emph{Computer}, vol.~55, no.~12, pp. 65--73, 2022.

\bibitem{bhagoji2019analyzing}
A.~N. Bhagoji, S.~Chakraborty, P.~Mittal, and S.~Calo, ``Analyzing federated
  learning through an adversarial lens,'' in \emph{International Conference on
  Machine Learning}.\hskip 1em plus 0.5em minus 0.4em\relax PMLR, 2019, pp.
  634--643.

\bibitem{guerraoui2018hidden}
R.~Guerraoui, S.~Rouault \emph{et~al.}, ``The hidden vulnerability of
  distributed learning in byzantium,'' in \emph{International Conference on
  Machine Learning}.\hskip 1em plus 0.5em minus 0.4em\relax PMLR, 2018, pp.
  3521--3530.

\bibitem{shejwalkar2021manipulating}
V.~Shejwalkar and A.~Houmansadr, ``Manipulating the byzantine: Optimizing model
  poisoning attacks and defenses for federated learning,'' in \emph{NDSS},
  2021.

\bibitem{mo2021ppfl}
F.~Mo, H.~Haddadi, K.~Katevas, E.~Marin, D.~Perino, and N.~Kourtellis, ``Ppfl:
  privacy-preserving federated learning with trusted execution environments,''
  in \emph{Proceedings of the 19th Annual International Conference on Mobile
  Systems, Applications, and Services}, 2021, pp. 94--108.

\bibitem{liu2022privacy}
Z.~Liu, J.~Guo, W.~Yang, J.~Fan, K.-Y. Lam, and J.~Zhao, ``Privacy-preserving
  aggregation in federated learning: A survey,'' \emph{IEEE Transactions on Big
  Data}, 2022.

\bibitem{dwork2014algorithmic}
C.~Dwork, A.~Roth \emph{et~al.}, ``The algorithmic foundations of differential
  privacy,'' \emph{Foundations and Trends{\textregistered} in Theoretical
  Computer Science}, vol.~9, no. 3--4, pp. 211--407, 2014.

\bibitem{cramer2015secure}
R.~Cramer, I.~B. Damg{\aa}rd \emph{et~al.}, \emph{Secure multiparty
  computation}.\hskip 1em plus 0.5em minus 0.4em\relax Cambridge University
  Press, 2015.

\bibitem{acar2018survey}
A.~Acar, H.~Aksu, A.~S. Uluagac, and M.~Conti, ``A survey on homomorphic
  encryption schemes: Theory and implementation,'' \emph{ACM Computing Surveys
  (Csur)}, vol.~51, no.~4, pp. 1--35, 2018.

\bibitem{xin2020private}
B.~Xin, W.~Yang, Y.~Geng, S.~Chen, S.~Wang, and L.~Huang, ``Private fl-gan:
  Differential privacy synthetic data generation based on federated learning,''
  in \emph{ICASSP 2020-2020 IEEE International Conference on Acoustics, Speech
  and Signal Processing (ICASSP)}.\hskip 1em plus 0.5em minus 0.4em\relax IEEE,
  2020, pp. 2927--2931.

\bibitem{chai2020tifl}
Z.~Chai, A.~Ali, S.~Zawad, S.~Truex, A.~Anwar, N.~Baracaldo, Y.~Zhou,
  H.~Ludwig, F.~Yan, and Y.~Cheng, ``Tifl: A tier-based federated learning
  system,'' in \emph{Proceedings of the 29th international symposium on
  high-performance parallel and distributed computing}, 2020, pp. 125--136.

\bibitem{bernabe2019privacy}
J.~B. Bernabe, J.~L. Canovas, J.~L. Hernandez-Ramos, R.~T. Moreno, and
  A.~Skarmeta, ``Privacy-preserving solutions for blockchain: Review and
  challenges,'' \emph{IEEE Access}, vol.~7, pp. 164\,908--164\,940, 2019.

\bibitem{hernandez2021sharing}
J.~L. Hern{\'a}ndez-Ramos, G.~Karopoulos, D.~Geneiatakis, T.~Martin,
  G.~Kambourakis, and I.~N. Fovino, ``Sharing pandemic vaccination certificates
  through blockchain: Case study and performance evaluation,'' \emph{Wireless
  Communications and Mobile Computing}, vol. 2021, pp. 1--12, 2021.

\bibitem{lalitha2018fully}
A.~Lalitha, S.~Shekhar, T.~Javidi, and F.~Koushanfar, ``Fully decentralized
  federated learning,'' in \emph{Third workshop on bayesian deep learning
  (NeurIPS)}, vol.~2, 2018.

\bibitem{yuan2023decentralized}
L.~Yuan, L.~Sun, P.~S. Yu, and Z.~Wang, ``Decentralized federated learning: A
  survey and perspective,'' \emph{arXiv preprint arXiv:2306.01603}, 2023.

\bibitem{xu2021asynchronous}
C.~Xu, Y.~Qu, Y.~Xiang, and L.~Gao, ``Asynchronous federated learning on
  heterogeneous devices: A survey,'' \emph{arXiv preprint arXiv:2109.04269},
  2021.

\bibitem{li2019convergence}
X.~Li, K.~Huang, W.~Yang, S.~Wang, and Z.~Zhang, ``On the convergence of fedavg
  on non-iid data,'' \emph{arXiv preprint arXiv:1907.02189}, 2019.

\bibitem{zeng2016effective}
M.~Zeng, B.~Zou, F.~Wei, X.~Liu, and L.~Wang, ``Effective prediction of three
  common diseases by combining smote with tomek links technique for imbalanced
  medical data,'' in \emph{2016 IEEE International Conference of Online
  Analysis and Computing Science (ICOACS)}.\hskip 1em plus 0.5em minus
  0.4em\relax IEEE, 2016, pp. 225--228.

\bibitem{xu2020hybrid}
Z.~Xu, D.~Shen, T.~Nie, and Y.~Kou, ``A hybrid sampling algorithm combining
  m-smote and enn based on random forest for medical imbalanced data,''
  \emph{Journal of Biomedical Informatics}, vol. 107, p. 103465, 2020.

\bibitem{zhang2022multi}
S.~Q. Zhang, J.~Lin, and Q.~Zhang, ``A multi-agent reinforcement learning
  approach for efficient client selection in federated learning,'' in
  \emph{Proceedings of the AAAI Conference on Artificial Intelligence},
  vol.~36, no.~8, 2022, pp. 9091--9099.

\bibitem{balasubramanian2021intelligent}
V.~Balasubramanian, M.~Aloqaily, M.~Reisslein, and A.~Scaglione, ``Intelligent
  resource management at the edge for ubiquitous iot: An sdn-based federated
  learning approach,'' \emph{IEEE network}, vol.~35, no.~5, pp. 114--121, 2021.

\bibitem{banbury2020benchmarking}
C.~R. Banbury, V.~J. Reddi, M.~Lam, W.~Fu, A.~Fazel, J.~Holleman, X.~Huang,
  R.~Hurtado, D.~Kanter, A.~Lokhmotov \emph{et~al.}, ``Benchmarking tinyml
  systems: Challenges and direction,'' \emph{arXiv preprint arXiv:2003.04821},
  2020.

\bibitem{sanchez2020tinyml}
R.~Sanchez-Iborra and A.~F. Skarmeta, ``Tinyml-enabled frugal smart objects:
  Challenges and opportunities,'' \emph{IEEE Circuits and Systems Magazine},
  vol.~20, no.~3, pp. 4--18, 2020.

\bibitem{david2021tensorflow}
R.~David, J.~Duke, A.~Jain, V.~Janapa~Reddi, N.~Jeffries, J.~Li, N.~Kreeger,
  I.~Nappier, M.~Natraj, T.~Wang \emph{et~al.}, ``Tensorflow lite micro:
  Embedded machine learning for tinyml systems,'' \emph{Proceedings of Machine
  Learning and Systems}, vol.~3, pp. 800--811, 2021.

\bibitem{mathur2021device}
A.~Mathur, D.~J. Beutel, P.~P.~B. de~Gusmao, J.~Fernandez-Marques, T.~Topal,
  X.~Qiu, T.~Parcollet, Y.~Gao, and N.~D. Lane, ``On-device federated learning
  with flower,'' \emph{arXiv preprint arXiv:2104.03042}, 2021.

\bibitem{jiang2022model}
Y.~Jiang, S.~Wang, V.~Valls, B.~J. Ko, W.-H. Lee, K.~K. Leung, and
  L.~Tassiulas, ``Model pruning enables efficient federated learning on edge
  devices,'' \emph{IEEE Transactions on Neural Networks and Learning Systems},
  2022.

\bibitem{chawla2002smote}
N.~V. Chawla, K.~W. Bowyer, L.~O. Hall, and W.~P. Kegelmeyer, ``Smote:
  synthetic minority over-sampling technique,'' \emph{Journal of artificial
  intelligence research}, vol.~16, pp. 321--357, 2002.

\bibitem{stin_dataset}
\BIBentryALTinterwordspacing
``Stin-data-set,'' 2020. [Online]. Available:
  \url{https://github.com/kun9717/STIN-data-set/}
\BIBentrySTDinterwordspacing

\bibitem{qvr7-n418-21}
\BIBentryALTinterwordspacing
H.~Kang, B.~I. Kwak, Y.~H. Lee, H.~Lee, H.~Lee, and H.~K. Kim, ``Car hacking:
  Attack \& defense challenge 2020 dataset,'' 2021. [Online]. Available:
  \url{https://dx.doi.org/10.21227/qvr7-n418}
\BIBentrySTDinterwordspacing

\end{thebibliography}
\appendices
\section{List of works analyzed}\label{sec:appendix}
\onecolumn
\begin{center}
\tiny
\begin{longtable}{C{0.4cm}C{0.4cm}C{1.2cm}C{1.2cm}C{1.2cm}C{1.5cm}C{1.5cm}C{1.5cm}C{1.5cm}C{0.8cm}C{0.8cm}C{2.5cm}}
\caption{Classification of existing works on FL-enabled IDS. N/S stands for not specified}
\label{tab:super_taxonomy_long_table} \\

\hline 
\textbf{Ref.} & \textbf{Year} & \textbf{Network architecture} & \textbf{Data availability} & \textbf{Data partitioning} & \textbf{Dataset} & \textbf{ML model} & \textbf{Aggregation function} & \textbf{FL implementation}  & \textbf{Training parties} & \textbf{Training rounds} & \textbf{Evaluation metrics} 
\\
\hline 
\endfirsthead

\multicolumn{12}{c}%

{\tablename\ \thetable{} -- continued from previous page} \\
\hline 
\textbf{Ref.} & \textbf{Year} & \textbf{Network architecture} & \textbf{Data availability} & \textbf{Data partitioning} & \textbf{Dataset} & \textbf{ML model} & \textbf{Aggregation function} & \textbf{Implementation}  & \textbf{Training parties} & \textbf{Training rounds} & \textbf{Evaluation metrics} 
\\
\hline 
\endhead

\hline \multicolumn{12}{c}{{Continued on next page}} \\ \hline
\endfoot

\hline \hline
\endlastfoot

\rowcolor{black}\multicolumn{12}{c}{\textcolor{white}{Based on supervised models}}\\

\cite{nguyen2019diot} & 2019 & Cross-device & Horizontal & Centralized & Generated & GRU & FedAvg & Based on Flask and Keras  & 2-15 & 3/51 & FPR, TPR, time  \\ \hline

\cite{zhao2019multi} & 2019 & Cross-device & Vertical & Centralized & CIC-IDS2017, ISCXVPN2016, ISCXTor2016 & NN (multi-task)  & FedAvg & Based on Pytorch & N/S & N/S & Accuracy, precision, recall, training time \\ \hline

\cite{li2020distributed} & 2020 & Cross-device & Horizontal & Centralized & Generated \cite{stin_dataset} & CNN  & N/S & TensorFlow Federated & N/S & 1-8/1 & FPR, TPR, training time, CPU use   \\ \hline

\cite{chen2020intrusion} & 2020 & Cross-device & Horizontal & Centralized & WSN-DS, KDDCup99, CIC-IDS2017 & GRU-SVM  & FedAvg, CMFL and based on FedAvg & PySyft & 10-50 & 1-50/1 & Accuracy, FAR, recall   \\ \hline

\cite{zhao2020network} & 2020  & Cross-device & Transfer Learning & Centralized & UNSW-NB15 & DNN  & FedAvg & N/S & 10 & N/S-N/S& Accuracy, precision, recall   \\ \hline

\cite{mowla2019federated} & 2020 & Cross-device & Horizontal & Centralized & CRAWDAD \cite{crawdad} & NN  & FedAvg & N/S & 6 & 1-10/25 & Accuracy  \\ \hline

\cite{al2020federated} & 2020 & Cross-device & Horizontal & Centralized &  NSL-KDD & MLP  & FedAvg & TensorFlow, Keras & 10 & 20/10 & Accuracy, precision, recall, F1-score   \\ \hline

\cite{qin2020line} & 2020 & Cross-device & Horizontal & Centralized & CIC-IDS2017, ISCXBotnet2014 & BNN & signSGD & TensorFlow, Keras & 2-8 & N/S-N/S& Accuracy, precision, recall, F1-score, latency, overhead  \\ \hline

\cite{sun2020intrusion} & 2020 & Cross-device & Horizontal & Centralized & Generated & CNN & FedAvg & N/S & 20 & 1-60/1 & Accuracy, precision, recall, F1-score  \\ \hline

\cite{hsu2020privacy} & 2020 & Cross-device & Horizontal, Vertical & Centralized &  Generated & SVM & FedAvg & Based on Python & 2-30 & 10/30 & Accuracy, precision, recall, FPR, F1-score  \\ \hline

\cite{zhao2020intelligent} & 2020 & Cross-device & Horizontal & Centralized & SEA \cite{schonlau2001computer} & LSTM, CNN  & FedAvg & TensorFlow, Keras & 4 & 1-50/1 & Accuracy, precision, recall, F1-score, latency, overhead  \\ \hline

\cite{rahman2020internet} & 2020 & Cross-device & Horizontal & Centralized & NSL-KDD & NN  & FedAvg & N/S & 4 & 1-5/1 & Accuracy   \\ \hline

\cite{zhou2020federated} & 2020 & Cross-device & Horizontal & Centralized & \cite{igbe2017deterministic, morris2014industrial} & CNN  & FedAvg & Based on Pytorch & 2-10 & 5/20 & Accuracy\\ \hline

\cite{fan2020iotdefender} & 2020 & Cross-device & Transfer Learning & Centralized & CIC-IDS2017, NSL-KDD, Kitsune, IoT network intrusion dataset & CNN  & FedAvg & N/S & 4 & 1-20/1-10 & Accuracy, TPR, FPR   \\ \hline

\cite{hei2020trusted} & 2020 & Cross-device & Horizontal & Centralized & KDDCup99 & MLP, DT, SVM, RF  & FedAvg & N/S & 4 & N/S-N/S& Accuracy, Precision, Recall, F1-score, AUC, delay   \\ \hline

\cite{mirzaee2021fids} & 2021 & Cross-device & Horizontal & Centralized & NSL-KDD & DNN & FedAvg  & N/S & 16 & 200/50 & Accuracy, Precision, Recall, F1-score, AUC, FPR, delay   \\ \hline

\cite{zakariyya2021memory} & 2021 & Cross-device & Horizontal & Centralized & N-BaIoT, Kitsune, IoT-DDoS, WUSTL \cite{teixeira2018scada} & Fully connected NN & FedAvg & Pytorch, Pysyft & 3 & 30/4 & Accuracy, Precision, Recall, F1-score, memory use   \\ \hline

\cite{qin2021fnel} & 2021 & Cross-device & Transfer & Centralized & CIC-IDS2017, CIC-DDoS2019 & NN  & N/S & N/S & N/S & 1-200/1 & Precision, FPR, FNR   \\ \hline

\cite{zhang2021flddos} & 2021  & Cross-device & Horizontal & Centralized & NSL-KDD, CIC-IDS2017, CIC-DDoS2019 & RNN & Based on FedAvg & Pytorch, Pysyft & 21 & 1-8000/100 & Accuracy, Precision, Recall, F1-score  \\ \hline

\cite{li2021fids} & 2021 & Cross-device & Horizontal & Centralized & CIC-DDoS2019 & GRU & Based on FedAvg & Pytorch  & 5 & 1-10/200 & Accuracy, loss   \\ \hline

\cite{yuan2021towards} & 2021 & Cross-device & Horizontal & Centralized & CIC-DDoS2019 & DT & N/S & N/S  & 1-30 & 1-200/1 & Accuracy, Precision, Recall, F1-score, FPR, communication overhead, time \\ \hline

\cite{lalouani2021robust} & 2021 & Cross-device & Horizontal & Centralized & UNSW-NB15-v2 & NN & FedAvg & N/S  & 10 & 55/1 & Accuracy, Precision, Recall, F1-score, TP   \\ \hline

\cite{duy2021federated} & 2021 & Cross-device & Horizontal & Centralized & CIC-DDoS2019 & CNN & FedAvg & Flower, TensorFlow Privacy & 3 & 1-10/1 & Accuracy, Precision, Recall, F1-score, communication overhead   \\ \hline

\cite{otoum2021federated} & 2021 & Cross-device & Transfer & Centralized & CIC-IDS2017 & DNN & FedAvg & Pytorch, TensorFlow & N/S & N/S-5 & Accuracy, Precision, F1-score, time  \\ \hline

\cite{shahid2021detecting} & 2021 & Cross-device & Horizontal & Centralized & NSL-KDD & LR, DNN & FedAvg & scikit-learn, PySyft & 4-8 & 1-4/1 & Accuracy, Precision, Recall, F1-score   \\ \hline

\cite{dong2021towards} & 2021 & Cross-device & Horizontal & Centralized & CIC-DDoS2019 & MLP, GBDT & Based on FedAvg (DataBin) & Based on Python & 21,66 & 1000-2000/1 & Accuracy, FOR  \\ \hline

\cite{chen2021trust} & 2021 & Cross-device & Horizontal & Centralized & KDDCup99 & CNN & Based on FedAvg & Pytorch & 1-20 & 50-300/1 & Accuracy, time   \\ \hline

\cite{ayed2021federated} & Cross-device & Horizontal & Centralized & 2021 & CIC-IDS2017 & CNN & FedAvg & N/S & 10 & 7/10 & Accuracy, loss   \\ \hline

\cite{hao2021secure} & 2021 & Cross-device & Horizontal & Centralized & CIC-IDS2017 & LSTM, NN, CNN & FedAvg & Flask & 2-6 & 3-7/10 & Accuracy, precision, recall, F1   \\ \hline

\cite{vy2021federated} & 2021 & Cross-device & Horizontal & Centralized & Kitsune & CNN, GRU & FedAvg? & TensorFlow, Keras & 3-9 & 6-12/1 & Accuracy, precision, recall, F1   \\ \hline

\cite{popoola2021federated} & 2021 & Cross-silo & Horizontal & Centralized & Ton\_IoT, UNSW-NB15, BoT-IoT, CSE-CIC-IDS2018 & DNN & FedAvg, CM, Fed+ & N/S & 4 & 1-10/1-10 & Accuracy, precision, recall, F1-score, MCC   \\ \hline

\cite{agrawal2021temporal} & 2021 & Cross-device & Horizontal & Centralized & NSL-KDD & NN & Temporal Weighted Averaging (proposed) & TensorFlow & 30 & 15/1 & Accuracy, loss   \\ \hline

\cite{ruzafa2021intrusion} & 2021 & Cross-device & Horizontal & Centralized & Ton\_IoT & LR & FedAvg, Fed+ & IBMFL & 4 & 50/1 & Accuracy, Pearson correlation   \\ \hline

\cite{man2021intelligent} & 2021 & Cross-device & Horizontal & Centralized & NSL-KDD & CNN & Based on FedAvg (proposed) & Pytorch & 10 & 40-200/40-200 & Accuracy, precision, recall, F1-score   \\ \hline

\cite{attota2021ensemble} & 2021 & Cross-device & Horizontal & Centralized & MQTT dataset \cite{bhxy-ep04-20} & NN & FedAvg & PySyft & 10 & 1-10/1 & Accuracy, precision, recall, F1-score   \\ \hline

\cite{sun2020adaptive} & 2021 & Cross-device & Horizontal & Centralized & Generated & CNN & FedAvg & N/S& 20 & 1-60/1 & Accuracy, precision, recall, F1-score   \\ \hline

\cite{aliyu2021blockchain} & 2021 & Cross-device & Horizontal & Centralized & CAN-intrusion dataset (OTIDS) & RF & FedAvg & Based on sklearn & 5-20 & N/S-N/S & Accuracy, precision, recall, F1-score   \\ \hline
 
\cite{LocKedge} & 2021 & Cross-device & Horizontal & Centralized & BoT-IoT & NN & FedAvg & Based on Python & 4 & 1000/1 & Accuracy, Precision, Recall, F1, ROC curve, memory/CPU use   \\ \hline
 
\cite{uprety2021privacy} & 2021 & Cross-device & Horizontal & Centralized & Veremi & NN & FedAvg & TensorFlow & 10 & 1-500/1 & Accuracy, precision, recall, communication overhead   \\ \hline

\cite{shi2021data} & 2021 & Cross-device & Horizontal & Centralized & UNSW-NB15, CSE-CIC-IDS2018 & CNN & FedAvg & N/S & 10, 15 & 15,20/1 & Accuracy, loss \\ \hline

\cite{liu2021blockchain} & 2021  & Cross-device & Horizontal & Decentralized & KDDCup99 & MLP & FedAvg & Pytorch, Pysyft & 2-6 & 10-40/1 & Accuracy, precision, recall  \\ \hline

\cite{saadat2021hierarchical} & 2021 & Cross-device & Horizontal & Centralized & NSL-KDD & NN & FedSGD, FedAvg & N/S & 4-8 & 1-20/1 & Accuracy, loss \\ \hline

\cite{li2020deepfed} & 2021 & Cross-device & Horizontal & Centralized & \cite{morris2014industrial} & CNN, GRU, MLP & FedAvg & Flask, Keras & 3-7 & 2-10/1 & Accuracy, precision, recall, F1-score   \\ \hline

\cite{kelli2021ids} & 2021 & Cross-device & Horizontal & Centralized & Generated & NN & FedAvg & N/S & 3 & 1-3/1 & Accuracy, precision, F1-score   \\ \hline

\cite{kumar2021security} & 2021 & Cross-device & Horizontal & Centralized & KDDCup99 & RF, SVM, NB & FedAvg? & N/S & N/S & N/S-1 & Accuracy, overhead  \\ \hline

\cite{shukla2021device} & 2021 & Cross-device & Horizontal & Centralized & Generated from VirusTotal & CNN, RF, LR, KNN & Based on FedProx & N/S & 10-50 & 200/20-40 & Accuracy, time, power, energy  \\ \hline

\cite{pasdar2022train} & 2022 & Cross-silo & Transfer & Centralized & Maldroid, Drebin, AndroZoo & CNN & FedAvg & TensorFlow Lite, keras & 3 & 1-4000/1-8000 & Accuracy, loss, F1, time, CPU, memory  \\ \hline

\cite{lv2022ddos} & 2022 & Cross-device & Horizontal & Centralized & CIC-DDoS2019 & CNN & FedAvg & Pytorch & 50 & 1-1000/1 & Accuracy, precision, recall, F1-score   \\ \hline

\cite{zhang2022grid} & 2022 & Cross-device & Horizontal & Centralized & Generated & CNN-LSTM & FedAvg? & N/S & 4 & N/S-N/S & Accuracy    \\ \hline

\cite{su2022detection} & 2022 & Cross-device & Horizontal & Centralized & KDDCup99 & LSTM & FedAvg, Based on FedProx & Based on Python & N/S & 1-200/1-20 & Accuracy, precision, recall, F1  \\ \hline

\cite{singh2022dew} & 2022 & Cross-device & Horizontal & Centralized & NSL-KDD, Ton\_IoT & LSTM & Based on FedAvg (hierarchical) & Keras, Scikit-learn, TensorFlow, Numpy & N/S & 1-100/1 & Accuracy, precision, recall, F1   \\ \hline

\cite{luo2022federation} & 2022 & Cross-device & Horizontal & Centralized & NSL-KDD & CNN & FedAvg & Pytorch & N/S & N/S-N/S & Accuracy, Recall, FPR, time  \\ \hline

\cite{yang2022federation} & 2022 & Cross-device & Horizontal & Centralized & NSL-KDD & CNN & FedAvg & Pytorch & N/S & N/S-N/S & Accuracy, Recall, FPR, time   \\ \hline

\cite{markovic2022random} & 2022 & Cross-device & Horizontal & Centralized & KDDCup99, NSL-KDD, UNSW-NB15, CIC-IDS2017 & RF & FedAvg & N/S & 3-14 (depending on the dataset) & N/S & Accuracy  \\ \hline

\cite{wang2022feco} & 2022 & Cross-device & Horizontal & Centralized & NSL-KDD & NN & FedAvg & Pytorch & 50 & 15/4 & Accuracy, recall, precision, F1, FPR, time  \\ \hline

\cite{liu2022intrusion} & 2022 & Cross-device & Horizontal & Centralized & NSL-KDD & CNN-MLP & FedBatch (proposed) & Pytorch & 100 & N/S & Accuracy, precision, recall F1-score   \\ \hline

\cite{driss2022federated} & 2022 & Cross-device & Horizontal & Centralized & Car Hacking \cite{qvr7-n418-21} & RF, GRU & FedAvg & TensorFlow, Keras & N/S & N/S-100 & Accuracy, precision, recall F1-score, time  \\ \hline

\cite{lalouani2022robust} & 2021 & Horizontal & Centralized &  Cross-device& UNSW-NB15-v2 & NN & FedAvg & N/S  & 10 & 1-18/1 & Accuracy, Precision, Recall, F1-score\\ \hline

\cite{rehman2022federated} & 2022 & Cross-silo & Horizontal & Centralized & Generated (Based on \cite{javed2020alphalogger}) & DNN & FedAvg & N/S & 2 & 1-3/20 & Accuracy, precision, recall, F1, loss, ROC   \\ \hline

\cite{otoum2022feasibility} & 2022 & Cross-device & Horizontal & Centralized & CIC-IDS2017 & NN & FedAvg & sklearn & 1-100 & 1-10/1 & Accuracy, DR, power consumption  \\ \hline

\cite{ferrag2022edge} & 2022 & Cross-device & Horizontal & Centralized & Edge-IIoTset (proposed) & DT, RF, KNN, SVM, DNN & FedAvg & Based on sklearn  & 5-15 & 1-10/1 & Accuracy \\ \hline

\cite{kumar2021pefl} & 2022 & Cross-device & Horizontal & Centralized & Ton\_IoT & GRU & FedAvg? & PySyft  & 2 & 1-100/1-100 & Accuracy, precision, F1, detection rate  \\ \hline

\cite{campos2021evaluating} & 2022 & Cross-device & Horizontal & Centralized  & Ton\_IoT & LR & FedAvg, Fed+ & IBMFL  & 4-10 & 1-300/1 & Accuracy, precision, recall, F1, FPR    \\ \hline

\cite{mothukuri2021federated} & 2022 & Cross-device & Horizontal & Centralized & \cite{frazao2018denial} & LSTM, GRU & FedAvg & Pytorch, Pysyft  & 1-40 & N/S-N/S & Accuracy, precision, recall, F1, time   \\ \hline

\cite{de2022improving} & 2022 & Cross-device & Horizontal & Centralized & Ton\_IoT & LR & FedAvg & sklearn  & 13 & 50/10 & F1  \\ \hline

\cite{cheng2022federated} & 2022 & Cross-device & Transfer & Centralized & NSL-KDD, UNSW-NB15 & CNN & FedSGD & Pytorch, Pysyft  & 10 & 1-40/1 & Accuracy  \\ \hline

\cite{abdel2021federated} & 2022 & Cross-device & Horizontal & Centralized & Car Hacking \cite{qvr7-n418-21} & Transformer & FedAvg & PySyft  & 20 & 30/50 & Accuracy, precision, recall, F1   \\ \hline

\cite{zhao2021federated} & 2022 & Cross-device & Horizontal & Centralized  & Generated & LSTM & FedAvg & Flower, Pytorch  & 3 & 300/1 & Accuracy, precision, recall, F1, loss, time, communication overhead   \\ \hline

\cite{popoola2021federatedzero} & 2022 & Cross-device & Horizontal & Centralized & BoT-IoT, N-BaIoT & DNN & FedAvg & IBMFL, TensorFlow, Keras  & 5 & 1-8/5 & Accuracy, precision, recall, F1, time, memory  \\ \hline

\cite{novikova2022federated} & 2022 & Cross-device & Vertical & Centralized & SWaT \cite{goh2016dataset} & GBDT & SecureBoost & FATE & 6 & N/S-N/S & Accuracy, time   \\ \hline

\cite{ibitoye2022differentially} & 2022 & Cross-device & Horizontal & Centralized & MNIST, CTU-13 & CNN & FedAvg & TensorFlow, IBMFL & 200 & N/S-N/S & Accuracy, time   \\ \hline

\cite{tang2022federated} & 2022 & Cross-device & Horizontal & Centralized & CIC-IDS2017 & GRU & FedAvg & Pytorch, Pysyft & 2 & 1-500/1 & Accuracy, precision, recall, F1   \\ \hline

\cite{shukla2022rafel} & 2022 & Cross-device & Horizontal & Centralized & Generated & CNN, Mobile-Net, RF, LR, KNN, MLP & Based on FedProx & N/S & 10-50 & 200/20-40 & Accuracy, power, energy, and area on-chip  \\ \hline

\cite{friha2022felids} & 2022 & Cross-device & Horizontal & Centralized & CSE-CIC-IDS2018, InSDN \cite{elsayed2020insdn}, MQTTset \cite{vaccari2020mqttset}
& DNN, CNN, RNN & FedAvg & Sherpa.ai, TensorFlow, Keras & 5-15 & 1-50/1 & Accuracy, precision, recall, F1, time, energy   \\ \hline

\cite{tabassum2022fedgan} & 2022 & Cross-device & Horizontal & Centralized & NSL-KDD, KDDCup99, UNSW-NB15 & CNN & FedAvg & N/S & N/S & 1-30/1-200 & Accuracy, precision, recall, F1, AUC  \\ \hline
  
\cite{zhang2022secfednids} & 2022 & Cross-device & Horizontal & Centralized & UNSW-NB15, CSE-CIC-IDS2018 & CNN & FedAvg? & Pytorch & 100 & 1-50/1 & Accuracy, recall   \\ \hline

\cite{tahir2021experience} & 2022 & Cross-device & Horizontal & Centralized & Based on \cite{nreldataset} & GRU & FedAvg, Fed-AA, DeepFed-Avg, DeepFed-AA & Matpower, Pytorch & N/S & 1-50/1 & Accuracy, precision, recall, F1, time   \\ \hline

\cite{sun2022hierarchical} & 2022 & Cross-device & Horizontal & Centralized & NSL-KDD & MLP, CNN & FedAvg & TensorFlow & 10 & 1-100/1-10 & Accuracy, precision, F1, overhead   \\ \hline

\rowcolor{black}\multicolumn{12}{c}{\textcolor{white}{Based on unsupervised models}}\\
\cite{preuveneers2018chained} & 2018 & Cross-device & Horizontal & Centralized & CIC-IDS2017 & AE  &  FedAvg & Based on Keras, Deeplearning4j and Spring Boot 2.0 & 2-12 & 1/1-50 & Accuracy, loss, time  \\ \hline

\cite{cetin2019federated} & 2019 & Cross-device & Horizontal & Centralized &  AWID & Stacked AE & FedAvg & LEAF & 933 & 1-20/1 & Accuracy, communication overhead   \\ \hline

\cite{taheri2020fed} & 2020 & Cross-device & Horizontal & Centralized & Drebin \cite{arp2014drebin}, \cite{zhou2012dissecting}, \cite{contagiodataset} & GAN based on CNN &  FedAvg, Krum & TensorFlow, Keras & 5-15 & 1-300/1-300 & Accuracy   \\ \hline

\cite{gaussian2020network} & 2020 & Cross-silo & Horizontal & Centralized & KDDCup99 & Deep AE-GMM &  FedAvg & N/S & 2 & 1-60/1 & Precision, recall, F1-score  \\ \hline

\cite{khoa2020collaborative} & 2020 & Cross-device & Horizontal & Centralized & KDDCup99, NSL-KDD, UNSW-NB15, N-BaIoT & DBN  & FedAvg & N/S & 2, 3 & N/S-N/S & Accuracy, precision, TPR   \\ \hline

\cite{jahromi2021deep} & 2021 & Cross-device & Horizontal & Centralized & SWaT & Stacked AE & FedAvg & N/S & N/S & N/S-N/S & Accuracy, precision, recall, F1   \\ \hline

\cite{yadav2021unsupervised} & 2021 & Cross-device & Horizontal & Centralized & CIC-IDS2017 & AE, NN & FedAvg & N/S  & N/S & 1-100/1-100 & Accuracy, MSE   \\ \hline

\cite{wettlaufer2021property} & 2021 & Cross-device & Horizontal & Centralized & CIC-IDS2017 & Elliptic Envelope, Isolation Forest & FedAvg? & N/S & N/S & 5/1 & Accuracy, F1-score   \\ \hline

\cite{mcosker2021architecture} & 2021 & Cross-device & Horizontal & Centralized & CIC-IDS2017 & SOM & FedAvg? & PySyft & N/S & N/S-N/S & Accuracy, precision, recall F1-score   \\ \hline

\cite{xie2021improved} & 2021 & Cross-device & Horizontal & Centralized & AWID & K-means clustering & FedAvg & Based on Python & 10 & 2-10/1 & Accuracy, FAR   \\ \hline

\cite{yadav2021clustering} & 2021 & Cross-device & Horizontal & Centralized & NSL-KDD & Based on K-means & FedAvg & Based on Python & 10 & 1-100/1 & Accuracy, WCSS   \\ \hline

\cite{alazzam2022federated} & 2022 & Cross-device & Horizontal & Centralized & Generated & AE & FedAvg? & Pytorch & N/S & N/S-N/S & Accuracy, precision, recall, F1, FPR, FNR, memory use   \\ \hline

\cite{xia2022fed_adbn} & 2022 & Cross-device & Horizontal & Centralized & NSL-KDD & DBN & FedAvg & TensorFlow  & 100 & 1-30/30-50 & Accuracy, precision, recall, F1, time \\ \hline

\cite{rey2022federated} & 2022 & Cross-device & Horizontal & Centralized & N-BaIoT & MLP, AE & FedAvg, Coordinate-wise median/trimmed mean & Own library  & 8 & 1-29/1-4 & Accuracy, F1, FPR, TPR   \\ \hline

\cite{kwon2022anomaly} & 2022  & Cross-device & Horizontal & Centralized & MNIST, CIFAR-10, CIC-IDS2017, TON\_IoT & K-means clustering & FedAvg & Based on Pytorch & N/S & N/S-N/S & AUC, F1   \\ \hline

\cite{chen2022privacy} & 2022  & Cross-silo & Horizontal & Centralized & KDDCup99, NSL-KDD, CIC-IDS2017, CSE-CIC-IDS2018  & AE & FedAvg & N/S & 2 & 1-120/200 & Precision, recall, F1  \\ \hline

\cite{wu2022fl} & 2022  & Cross-silo & Horizontal & Centralized & NSL-KDD  & Variational AE & FedAvg & TensorFlow, Keras & 5-1000 & 100/1 & Accuracy, Precision, recall, F1  \\ \hline

\rowcolor{black}\multicolumn{12}{c}{\textcolor{white}{Based on semi-supervised models}}
\\ \hline
\cite{qin2021federated} & 2021 & Cross-device & Horizontal & Centralized & NSL-KDD & AE & FedAvg & N/S & 8 & N/S-N/S & Accuracy  \\ \hline

\cite{singh2020collaborative} & 2020 & Cross-device & Horizontal & Centralized & Microsoft Malware \cite{ronen2018microsoft} & CNN, Auxiliary Classifier GAN & FedAvg & TensorFlow Federated  & 5 & 50-100/10 & Accuracy, precision, recall, F1   \\ \hline

\cite{aouedi2022fluids} & 2022 & Cross-device & Horizontal & Centralized & UNSW-NB15 & AE-NN  & FedAvg & Pytorch & 100 & 6/5 & F1   \\ \hline

\cite{aouedi2022federated} & 2022 & Cross-device & Horizontal & Centralized & \cite{morris2014industrial} & AE-NN  & FedAvg & Pytorch & 100 & 18/20-100 & Accuracy, precision, recall, F1, communication overhead   \\ \hline

\cite{pei2022knowledge} & 2022 & Cross-device & Transfer & Centralized & Drebin, MalDroid, AndroZoo, VirusShare & SACN & FedAvg & Keras & 100-500 &N/S-N/S & FPR, TPR, ROC curve, F1-score  \\ \hline

\cite{zhao2022semi} & 2022 & Cross-device & Horizontal & Centralized & N-BaIoT & CNN & FedAvg & Pytorch & 27, 89 & 1-200/100 & Accuracy, precision, F1-score, communication overhead  \\ \hline

\rowcolor{black}\multicolumn{12}{c}{\textcolor{white}{Based on reinforcement models}} 
\\ \hline

\cite{wei2021federated} & 2021 & Cross-device & Horizontal & Centralized & CIC-IDS2017 & DQN  & Based on FedAvg & Pytorch & 1-10 & 1-4/1-1000 & Error rate, reward  \\ \hline

\cite{nguyen2021federated1} & 2021 & Cross-device & Horizontal & Centralized & N/S & DDQN, SOM  & FedAvg & TensorFlow Federated & 6 & 1-200/1 & Reward, loss, communication overhead  \\ \hline

\cite{otoum2021federatedreinforcement} & 2021 & Cross-device & Horizontal & Centralized & CIC-IDS2017 & Q-learning  & FedAvg & MATLAB/Simulink & 50 & 1-10/1-10 & Accuracy, DR, FPR, FNR  \\ \hline

\end{longtable}
\end{center}
\twocolumn

\begin{IEEEbiographynophoto}{José L. Hernández-Ramos} received the  Ph.D. degree in computer science from the University of Murcia (UMU), Spain. He is currently a Marie Sklodowska-Curie Postdoc Fellow at UMU working on the topic of federated learning and intrusion detection. Before that, he was a Scientific Project Officer with the Joint Research Centre, European Commission. He has participated in different European research projects, such as SocIoTal, SMARTIE, and SerIoT. He has published more than 60 peer-reviewed papers. His research interests include application of security and privacy mechanisms in the Internet of Things and transport systems scenarios, including blockchain and machine learning.
\end{IEEEbiographynophoto}

\begin{IEEEbiographynophoto}{Aurora González-Vidal} graduated in Mathematics from the University of Murcia in 2014. In 2015 she got a fellowship to work in the Statistical Division of the Research Support Service, where she specialized in Statistics and Data Analysis. Afterward, she studied a Big Data Master. In 2019, she got a Ph.D. in Computer Science. Currently, she is a postdoctoral researcher at the University of Murcia. She has collaborated in several national and European projects such as ENTROPY, IoTCrawler, and DEMETER. Her research covers machine learning in IoT-based environments, missing values imputation, and time-series segmentation. She is the president of the R Users Association UMUR.
\end{IEEEbiographynophoto}

\begin{IEEEbiographynophoto}{Georgios Karopoulos} received the Ph.D. degree in computer network security from the University of the Aegean, Greece. He is currently a Scientific Officer at the Joint Research Centre, European Commission. In the past, he was a Marie Curie Fellow Researcher at the University of Athens, Greece, and an ERCIM Fellow at IIT-CNR, Italy. His research interests include network security, VoIP security and privacy, smart grid security, and critical infrastructure protection. He has published and is a frequent reviewer in conferences and scientific journals in the above areas.
\end{IEEEbiographynophoto}

\begin{IEEEbiographynophoto}{Efstratios Chatzoglou} received the M.Sc. degree in security of information and communication systems from the University of Aegean, Samos, Greece. He previously worked as a Web Developer and a Penetration Tester at the Hellenic National Defense General Stuff. He is currently a Penetration Tester at TwelveSec and a research associate at Info-Sec-Lab with the University of Aegean. His research interests lie in the fields of wireless and cellular networks security, IoT networks security, Android application security, Web application security, and Machine Learning.
\end{IEEEbiographynophoto}

\begin{IEEEbiographynophoto}{Vasileios Kouliaridis} received his Ph.D. from the dept. of Information and Communication Systems Engineering, University of the Aegean, Greece. He is currently working as a Scientific Project Officer for the Joint Research Center of the European Commission. His research interests are in the fields of mobile and network security, privacy and machine learning. He has served as a guest editor and is a frequent reviewer in  scientific journals on the above fields.
\end{IEEEbiographynophoto}

\begin{IEEEbiographynophoto}{Enrique Mármol Campos} is a Ph.D. Student at the university of Murcia. He graduated in Mathematics in 2018. Then, in 2019, he finished the M.S. in advanced math, in the specialty of operative research and statistic, at the University of Murcia. He is currently researching on federated learning applied to cybersecurity in IoT devices.
\end{IEEEbiographynophoto}

\begin{IEEEbiographynophoto}{Georgios Kambourakis} is a Full Professor with the Department of Information and Communication Systems Engineering, University of the Aegean, Greece. He was the Head of the Department from Sept. 2018 to Oct. 2019, and was the Director of Info-Sec-Laboratory from Sept. 2014 to Dec. 2018. He has more than 160 refereed publications in his research areas. His research interests are in fields of mobile and wireless networks security and privacy, VoIP security, IoT security and privacy, DNS security, and security education. More info at: http://www.icsd.aegean.gr/gkamb
\end{IEEEbiographynophoto}

\end{document}